\documentclass{aa}  

\usepackage[dvipsnames]{xcolor}
\usepackage{color}
\usepackage{ulem}
\usepackage{yfonts}

\newcommand{\teff}{$T_{\rm eff}$}
\newcommand{\logrhk}{$\log R'_{\rm HK}$}
\newcommand{\prot}{$P_{\rm rot}$}

\usepackage{graphicx}
\usepackage{txfonts}
\usepackage{natbib}
\begin{document}

\title{Extending the FIP bias sample to magnetically active stars
}

  \subtitle{Challenging the FIP bias paradigm?      
  }
   \author{B. Seli\inst{1,2}
          \and
          K. Ol\'ah\inst{1}
          \and
          L. Kriskovics\inst{1}
          \and          
          Zs. K\H{o}v\'ari\inst{1}
          \and
          K. Vida\inst{1}
          \and
          L. G. Bal\'azs\inst{1,2}
          \and
          J. M. Laming\inst{3}
          \and
          L. van Driel-Gesztelyi\inst{1,4,5}
          \and
          D. Baker\inst{4}
          }
   \institute{Konkoly Observatory, Research Centre for Astronomy and Earth Sciences (ELKH), Budapest, Hungary
   \and
    E\"otv\"os University, Department of Astronomy, Pf. 32, 1518 Budapest, Hungary
    \and
    Space Science Division, Code 7684, Naval Research Laboratory, Washington DC 20375, USA
    \and
    University College London, Mullard Space Science Laboratory, Holmbury St.  Mary, Dorking, Surrey, RH5 6NT, UK
    \and
    LESIA, Observatoire de Paris, Universit\'e PSL, CNRS, Sorbonne Universit\'e, Univ. Paris Diderot, Sorbonne Paris Cit\'e, 5 place Jules Janssen, 92195 Meudon, France\\
          \email{seli.balint@csfk.org}
             }

   \date{Received ...; accepted ...}


  \abstract
   {The different elemental abundances of the photosphere and the corona are striking features of not only the Sun, but other stars as well. This phenomenon is known as the FIP effect (FIP stands for first ionization potential), and its strength can be characterized by the FIP bias, the logarithmic abundance difference between low- and high-FIP elements in the corona, compared to the photosphere. The FIP bias was shown to depend on the surface temperature of the star.}
   {We aim to extend the \teff$-$FIP bias relationship to a larger stellar sample and analyse the effect of other astrophysical parameters on the relation (e.g., surface gravity, age, activity indicators).}
   {We compiled FIP bias and other parameters for 59 stars for which coronal composition is available, now including evolved stars. Using principal component analysis and linear discriminant analysis, we searched for correlations with other astrophysical parameters within the sample which may influence the stellar FIP bias.}
   {Adding stars to the \teff$-$FIP bias diagram unveiled new features in its structure. In addition to the previously known relationship, there appears to be a second branch, a parallel sequence about 0.5\,dex above it. While the \teff\ remains the main determinant of the FIP bias, other parameters such as stellar activity indicators also have influence. We find three clusters in the FIP bias determinant parameter space. One distinct group is formed by the evolved stars. Two groups contain main sequence stars in continuation separated roughly by the sign change of the FIP-bias value.}
   {The new branch of the \teff$-$FIP bias diagram contains stars with higher activity level, in terms of X-ray flux and rotational velocity. The Rossby number also seems to be important, indicating possible dependence on the type of dynamo operating in these stars influencing their FIP bias. The two main sequence clusters run from the earliest spectral types of A-F with shallow convection zones through G-K-early M stars with gradually deeper convection zones, and end up with the fully convective M dwarf stars, depicting the change of the dynamo type with the internal differences of the main sequence stars in connection with the FIP-bias values.}


   \maketitle
%

\section{Introduction}

Differences in chemical composition between various regions of the solar corona and the photosphere were documented first by \citet{1963ApJ...137..945P}, who found significantly higher abundances for Si, Mg and Fe in the corona compared with the photospheric abundances. These elements all exhibit first ionization potentials (FIP) lower than 10\,eV, and hence called low-FIP elements, while elements with FIP values higher than 10\,eV (e.g., O, Ne, He) are called high-FIP elements. It has now become a well-known phenomenon that low-FIP element abundances are enhanced in the corona with respect to their photospheric values, while high-FIP elements show significantly lower enhancement, or even depletion \citep[see, e.g.][]{1985ApJS...57..151M,2000PhyS...61..222F,2003SSRv..107..665F}.

After high-energy space-borne observatories became available in the 1990s with the launch of the \textit{Extreme Ultraviolet Explorer} (\textit{EUVE}), similar abundance anomalies were discovered for several late-type stars \citep{1996ApJ...462..948L,1997ApJ...478..403D,1999ApJ...516..324L}, and since then, the \textit{XMM-Newton} and the \textit{Chandra} X-ray observatories have further increased the sample of stars with known FIP effect \citep[e.g.][etc.]{Wood_Linsky2010,2013ApJ...768..122W,Wood2018}.

Recently, \citet{Wood2018} have pointed out that stars with effective temperatures similar to or higher than that of the Sun show similar abundance enhancements, while cooler stars tend to exhibit lower abundance enhancements. At $T_{\mathrm{eff}}\approx 4500$\,K, coronal abundances become equal to the photospheric values, and furthermore, even cooler stars seem to exhibit an inverse FIP effect (IFIP) where low-FIP elements are depleted rather than enhanced in the corona compared to the photosphere. \citet{2015ApJ...808L...7D} and \citet{2016ApJ...825...36D} have also found signs of the IFIP effect on the Sun in flares, in small patches near sunspots, while \citet{2020ApJ...891..126K} observed IFIP effect in spatially unresolved segments of solar flare plasma.

The (I)FIP effect has been modeled by \citet{2004ApJ...614.1063L,2009ApJ...695..954L,2012ApJ...744..115L,2015ApJ...805..102L,2017ApJ...844..153L}, and lately, by \citet{2019ApJ...879..124L}, and it is explained with the effect of the ponderomotive force associated with Alfvén and fast mode waves, which separates ions from neutrals in the chromospheric regions. Most recently, it has been suggested that resonant Alfvén waves propagating along coronal loops can result in FIP effect \citep{2015ApJ...805..102L, laming2021}, while originally upward propagating p-modes or magneto-acoustic waves, which are converted to fast modes and propagate into the $\beta<1$ regions can be a plausible explanation for the IFIP effect \citep{laming2021,baker2019,2020ApJ...894...35B}. The competition between these two effects might be a deciding factor whether a star exhibits an overall FIP or IFIP effect.

To characterize the strength of the (I)FIP effect, the quantity known as the FIP bias is used, which is the mean coronal abundance of four high-FIP elements (C, N, O, Ne) relative to the Fe abundance (the most easily observable low-FIP element), normalized to the photospheric abundance ratios. The abundance ratios (and hence the FIP-bias values) are given in logarithmic units. In the case of solar-like FIP effect, the FIP bias is negative, while it is positive in the case of inverse-FIP effect. To measure the FIP bias, both the coronal abundances from X-ray spectra, and photospheric abundances from optical spectra are needed. While it is hard to acquire high quality X-ray observations even for nearby stars, there is no straightforward way to measure photospheric abundances of certain elements at all (mainly N and Ne). Thus, it is common to use solar photospheric composition or empirical formulae instead.

On the original \teff$-$FIP bias diagram of main sequence stars from \citet{2015LRSP...12....2L}, updated from \citet{Wood_Linsky2010}, the stars form a tight sequence with \teff, but lack diversity in other parameters. In this work we extend the diagram in two aspects: with a larger and more diverse stellar sample, and by involving other parameters, e.g., surface gravity, metallicity, age. We compile the parameters from existing literature, prioritizing homogeneity for stellar parameters, and transparency for the calculation of the FIP bias.

The structure of the paper is as follows: in Sect.\,\ref{sect:data} we describe the sample and the parameters of interest, in Sect.\,\ref{sect:methods} we describe the methods used to search for correlations within the dataset. In Sect.\,\ref{sect:results} we present the results, while their implications are discussed in Sect.\,\ref{sect:discussion}.

\section{Data}\label{sect:data}

In order to extend the \teff$-$FIP bias diagram, we collected all active stars regardless of spectral type and evolutionary stage for which coronal abundance values are available in the literature, thus where it is possible to calculate the FIP bias. In our final sample we selected a total of 59 stars, summarized in Table\,\ref{table:binarity}.
From the 59 stars, only 18 (31\%) are single. 41 are members of binaries or multiple systems, although 25 of them are
effectively single, i.e. gravitationally bound wide binaries or long period binaries which were formed together but do not interact, and the components evolve separately. Quite a few binaries are members of hierarchical multiple systems but no interaction between the system members is observed. 

The group of solar type stars with 15 members contains six single stars, two close and seven wide binaries. Among the nine flare stars there are two close binaries: YY\,Gem (M1Ve+M1Ve) and AT\,Mic (M4Ve+M4Ve) with two equal components, both stars of the binaries are active, the observed signals from the components are of common origin. We have altogether 13 K dwarfs, three of those are well-known fast rotating, very active and flaring stars. Most of the remaining ten K dwarfs are members of wide binaries as in the case of the solar-type group. The vast majority of the wide binaries are from these groups, as part of the observing strategy \citep[\textit{Chandra} can observe the two components in the same frame, see e.g.][]{2006ApJ...643..444W} beside the importance of having solar-like parameters and activity.

The sample of the evolved stars mostly consists of RS\,CVn type binaries. These are well-known magnetically active stars, and are bright enough for coronal observations.
Most of these 11 binaries are single-lined spectroscopic binaries (SB1), thus most of the observed features can be attributed to the subgiant or giant primary component. Double-lined binaries (SB2) are usually dominated by the primary as well, although the contribution of the secondary is not negligible. Spectral types of the binaries of evolved stars are from \citet{eker2008}.

The sample consists of dwarf stars from the main sequence (MS) and evolved stars from the red giant branch (RGB) of the Hertzsprung-Russell diagram. The MS stars of the sample start with the fully convective M dwarfs through K and G dwarfs to A-F stars, and in this sequence the depths of convective zones are continuously shrinking. Some stars are in an intermediate state: slightly before the MS or just turning off it. The  stars on the subgiant and giant branches are inflated. These stars have thick convective envelopes and low surface gravities. All of the sample stars have magnetic activity in common, which is the basis of the observed element fractionation, and through this, the FIP effect.
The magnetic fields of the stars are thought to be generated by different types of dynamos. Sun-like stars with $M\lesssim M_\odot$ and usually showing \mbox{(quasi-)cyclic} magnetic variability, are believed to maintain solar-type $\alpha\Omega$ dynamos, where differential rotation and large-scale convective flows play a crucial role. However, with vanishing differential rotation, e.g., in less massive stars with a fully convective interior, $\alpha^2$ dynamos are supposed to operate, driven rather by small-scale turbulent flows.
The rotational rates of the stars are from about 0.5 days to over 200 days which have  strong impact on the magnetic dynamo. Our sample is heterogeneous, and the task is to investigate how the FIP effect appears on stars of different masses and evolutionary status with different rotational velocities. 

For the 59 stars with known coronal composition, we compiled the following parameters from the literature, where available:
\begin{itemize}
    \item[-] \teff: effective temperature, mostly derived from optical spectra
    \item[-] $\log g$: surface gravity from optical spectra, in a few cases when no literature data existed it is calculated from mass and radius
    \item[-] [Fe/H]: metallicity
    \item[-] \logrhk: chromospheric contribution of the Ca H\&K lines (excluding the photospheric component). It measures the chromospheric activity, but since it varies strongly with activity level (activity cycle, e.g., for the Sun it varies between $-$4.88 and $-$5.02, more active stars could exhibit bigger change), rotational modulation of the activity features, and the incidental flares at the time of the observations. Thus it is only a rough index for the strength of activity in a given star \citep[see][and their references]{ramirez2014}. This parameter is missing for many objects, particularly for the evolved stars. Only 3 out of the 17 evolved stars have measured \logrhk.
    \item[-] $R$: stellar radius
    \item[-] $\log L_X$: X-ray luminosity
    \item[-] $\log F_X$: X-ray flux calculated from $\log L_X$ and radius
    \item[-] $t$: age
    \item[-] $Ro = P_\mathrm{rot} / \tau_c$: Rossby number, the ratio of rotational period and the convective turnover time. The latter is not an observable parameter. It was calculated using the empirical relation from \citet{convective_turnover_time}, with  $B-V$ values from the Simbad database. In the case of evolved stars, we used a different formula from \citet{rossby_evolved} to calculate $Ro$ directly from $\log{F_X}$ and $\log{g}$, without the use of $\tau_c$. We note that while this empirical formula has a large scatter, $Ro$ is itself strongly model dependent through $\tau_c$.
    \item[-] \prot: photometric rotational period from the rotational modulation of the light curve.
    It is only used to calculate $v_\mathrm{rot}$ and $Ro$.
    \item[-] $v_\mathrm{rot}$: rotational velocity at the surface, simply $v_{\mathrm{rot}}=2\pi R/P_\mathrm{rot}$
    \item[-] literature FIP bias: FIP-bias values compiled from various sources, and transformed to the same solar photospheric standard from \citet{asplund2009}, and Ne abundance from \citet{drake_testa_neon} ([Ne/O]=$\log_{10}{0.41}$). Thus we adopt the following abundance values with the normalization of A(H)=12: A(C)=8.43, A(N)=7.83, A(O)=8.69, A(Ne)=8.30. Where no photospheric abundance was available, solar composition was assumed. The FIP-bias values were corrected by $+0.084$ according to \citet{Wood2018} to deal with atomic data changes. The individual abundances used for the calculation are listed in Table\,\ref{table:lit_abund_table}.
    \item[-] KNN FIP bias: the same as the previous parameter, but recalculated with homogeneously predicted photospheric abundances with k-nearest neighbour (KNN) algorithm from LAMOST data \citep{lamost}, and with the $+0.084$ correction from \citet{Wood2018} to deal with atomic data changes. See Sect.\,\ref{sect:lamost} for details and Table\,\ref{table:knn_abund_table} for the individual abundances.
    \item[-] SME FIP bias: FIP bias calculated using photospheric composition determined by spectral synthesis with the \textit{Spectroscopy Made Easy} code (SME, \citealt{piskunovsme}), to check the consistency of the previous methods. Only available for a handful of stars, and only from C and O abundances. See Sect.\,\ref{sect:sme} for details and Table\,\ref{table:sme_results} for the abundances.
\end{itemize}

These parameters are summarized in Table\,\ref{table:sample_params}, with their sources listed in Appendix\,\ref{A2}. The uncertainty of the FIP bias is taken as the standard deviation of [C/Fe], [N/Fe], [O/Fe] and [Ne/Fe]. We chose this method over propagating the individual error bars of the abundances, as those are not available in some cases, and they are not homogeneous.

We note that in the case of GJ\,338\,AB the individual coronal abundance fractions are not available from emission measure analysis. \cite{Wood2012} estimated the FIP bias directly from Fe{\sc xvii}/O{\sc viii}. Thus we cannot use the same methods to reproduce the results for this binary, and adopt the FIP-bias value from \citet{Wood2018}.


\begin{table*}[t]
\small
\caption{The representation of different types of stars in our sample. The numbers of single stars are given in parentheses.}
\label{table:binarity}      
\centering          
\begin{tabular}{l l | l l}     
\hline\hline
 & \\
\multicolumn{2}{l |}{\bf flare stars, M-type:  9 (3)} &                        \multicolumn{2}{l}{\bf solar type: 15 (6)}  \\                           
AD\,Leo   &   single  M4V      &                                         Sun             & single G2V \\ 
CN\,Leo   &   single  M6V      &                                         $\alpha$\,Cen~A  & wide triple G2V \\
EQ\,Peg~A &  wide binary  M3.5V  &                                         $\alpha$\,Cen~B  & wide triple K0V \\
EQ\,Peg~B &  wide binary  M4.5V  &                                         $\pi^1$\,UMa   &  single + debris disk G1.5V \\
Proxima~Cen & wide triple M5.5V &                                         EK\,Dra        &  long period binary (45 yr) G5V \\
EV\,Lac  &   single M3.5V   &                                              $\xi$\,Boo A   &  wide binary G8V \\
YY\,Gem  &   ecl. binary Castor sys. M1Ve+M1Ve SB2  &                 $\chi^1$\,Ori  &  wide binary G0V \\
AU\,Mic  &  triple with AT\,Mic M0.5V    &                                  $\kappa$\,Cet  &  single G5V \\
AT\,Mic  &  spectroscopic binary M4Ve+M4Ve SB2 &                                  $\beta$\,Com   &  single G0V\\                                         
        &                        &                                  47\,Cas B      &  wide binary G0-2V \\
\multicolumn{2}{l |}{\bf fast rotating K dwarfs, flares: 3 (2)}  &      $\iota$\,Hor   &  single + planet F8V\\
AB\,Dor  &   quadruple  K0V    &                                        11\,LMi        &  long period binary (201 yr) G8V \\                                
LO\,Peg  &   single  K3V       &                                        HR 7291       &  single + planet F8.5V \\
V471\,Tau & binary (triple?)  K2V    &                                        $\sigma^2$\,CrB  & spec. binary in a quintuple system F9V+G0V SB2 \\
        &                  &                                    $\xi$\,UMa B    &    spec. binary in a quintuple  system G5V SB1 \\ 
\multicolumn{2}{l |}{\bf K(-M) dwarfs: 10 (1)}    &   \\                       
GJ\,338 A &    wide binary M0V     &                                 \multicolumn{2}{l}{\bf evolved stars: 17 (4)} \\
GJ\,338 B &    wide binary M0V     &                                      HR\,1099     &  RS\,CVn binary  K1IV+G5IV SB2     \\  
36\,Oph A &    wide binary K1V    &                                       UX\,Ari      & RS\,CVn binary  K0IV+G5V  SB2     \\ 
36\,Oph~B &   wide binary  K1V     &                                     $\lambda$\,And & RS\,CVn binary  G8III-IV  SB1   \\
$\xi$\,Boo B & wide binary  K4V    &                                     VY\,Ari       & RS\,CVn binary  K3-4V-IV  SB1   \\
61\,Cyg A  &   wide binary  K5V   &                                      Capella      &    spec. binary in a quintuple system K0III \\
61\,Cyg B  &   wide binary K7V    &                                      $\sigma$\,Gem  &   RS\,CVn binary  K1III     SB1    \\
70\,Oph A  &   wide binary K0V    &                                       31\,Com       &    single G0III \\                            
70\,Oph~B  &   wide binary K5V    &                                       $\mu$\,Vel    &    long period binary (138 yr) G6III \\
$\epsilon$\,Eri  & single with planet  K2V  &                            $\beta$\,Cet  &    single G9II-III\\  
  & &                                                                      FK\,Com       &    single G2III \\ 
\multicolumn{2}{l |}{\bf A-F stars, hotter than 6300K: 5 (3)} &          YY\,Men       &    single K1III  \\
$\eta$ Lep    &    single  F2V  &                                       EI\,Eri       & RS\,CVn binary  G5IV+G0V	 SB1   \\
$\pi^3$\,Ori   &   single  A3V   &                                       V851\,Cen     & RS\,CVn binary  K3V-IV	 SB1   \\
$\tau$\,Boo A  &  wide binary    &                                    AR\,Psc       &  RS\,CVn binary K1IV-V+G5-6V SB2    \\
Procyon       & long period binary (40.8 yr) F5IV-V  &                     AY\,Cet       &  RS\,CVn binary WD+G5III SB1   \\
Altair        &  single, A type  (A7V)  &                                 II\,Peg       &  RS\,CVn binary K2IV+M0-3V SB1   \\
  & &                                                                AR\,Lac       &   RS\,CVn binary G2IV+K0IV SB2    \\                            
\hline       
\end{tabular}
\end{table*}


\section{Methods}\label{sect:methods}

\begin{table*}[ht!]
\caption{Compiled astrophysical parameters for the sample. See Sect.\,\ref{sect:data} for the description of each parameter, and Appendix\,\ref{A2} for the sources.}
\label{table:sample_params}
\centering
\tiny
\begin{tabular}{c|r|r|l|r|r|r|r|r|r|r|r|r|r}     
\hline
\hline
\multicolumn{1}{c|}{Star} & \multicolumn{1}{c|}{FIP bias} & \multicolumn{1}{c|}{FIP bias} & \multicolumn{1}{c|}{\teff $^{(a)}$} & \multicolumn{1}{c|}{$\log g$} & \multicolumn{1}{c|}{[Fe/H]} & \multicolumn{1}{c|}{$t^{(b)}$} & \multicolumn{1}{c|}{$R$} & \multicolumn{1}{c|}{\prot} & \multicolumn{1}{c|}{$v_\mathrm{rot}$} & \multicolumn{1}{c|}{$Ro$} & \multicolumn{1}{c|}{\logrhk} & \multicolumn{1}{c|}{$\log L_X$} & \multicolumn{1}{c}{$\log F_X$} \\

\multicolumn{1}{c|}{} & \multicolumn{1}{c|}{literature} & \multicolumn{1}{c|}{KNN} & \multicolumn{1}{c|}{K} & \multicolumn{1}{c|}{(cgs)} & \multicolumn{1}{c|}{} & \multicolumn{1}{c|}{Gyr} & \multicolumn{1}{c|}{$R_\odot$} & \multicolumn{1}{c|}{days} & \multicolumn{1}{c|}{km\,s$^{-1}$} & \multicolumn{1}{c|}{} & \multicolumn{1}{c|}{} & \multicolumn{1}{c|}{$^{(c)}$} & \multicolumn{1}{c}{$^{(d)}$} \\

\hline
AD\,Leo & $0.49 \pm 0.11$ & $0.74 \pm 0.10$ & $3390 \pm 19$ & $4.81$ & $0.28$ & $0.16$ & $0.42$ & $2.24$ & $9.53$ & $0.09$ & $-4.19$ & $28.70$ & $6.66$ \\
CN\,Leo & $0.27 \pm 0.17$ & $0.46 \pm 0.32$ & $3100 \pm 100$ & $5.21$ & $0.18$ & $0.23$ & $0.14$ & $2.71$ & $2.52$ & $0.08$ & -- & $27.01$ & $5.96$ \\
EQ\,Peg\,A & $0.53 \pm 0.07$ & $0.65 \pm 0.25$ & $3353 \pm 60$ & $4.91$ & $0.03$ & $0.10$ & $0.35$ & $1.06$ & $16.71$ & $0.04$ & $-5.16$ & $28.71$ & $6.84$ \\
EQ\,Peg\,B & $0.49 \pm 0.08$ & $0.61 \pm 0.22$ & $3152 \pm 2$ & $4.92$ & $0.03$ & $0.10$ & $0.25$ & $0.40$ & $31.62$ & $0.01$ & -- & $27.89$ & $6.31$ \\
Prox\,Cen & $0.55 \pm 0.30$ & $0.67 \pm 0.31$ & $2879 \pm 50$ & $5.23$ & $0.10$ & $5.03$ & $0.13$ & $89.80$ & $0.07$ & $3.00$ & $-5.00$ & $27.22$ & $6.20$ \\
EV\,Lac & $0.55 \pm 0.15$ & $0.65 \pm 0.24$ & $3291 \pm 51$ & $5.11$ & $-0.19$ & $0.16$ & $0.34$ & $4.38$ & $3.98$ & $0.16$ & $-3.97$ & $28.99$ & $7.13$ \\
YY\,Gem & $0.69 \pm 0.17$ & $0.80 \pm 0.23$ & $3775 \pm 110$ & $4.63$ & $0.00$ & $0.37$ & $0.62$ & $0.81$ & $38.47$ & $0.03$ & -- & $29.27$ & $6.90$ \\
AU\,Mic & $0.68 \pm 0.13$ & $0.93 \pm 0.09$ & $3679 \pm 6$ & $5.00$ & $0.15$ & $0.02$ & $0.69$ & $4.86$ & $7.18$ & $0.18$ & $-3.99$ & $29.36$ & $6.90$ \\
AT\,Mic & $0.51 \pm 0.07$ & $0.70 \pm 0.23$ & $3123 \pm 12$ & $4.67$ & $0.15$ & $0.02$ & $0.60$ & $1.00$ & $30.36$ & $0.04$ & -- & $29.47$ & $7.67$ \\
AB\,Dor & $0.60 \pm 0.09$ & $0.49 \pm 0.20$ & $5081 \pm 50$ & $4.55$ & $0.18$ & $0.15$ & $0.96$ & $0.51$ & $94.35$ & $0.02$ & $-3.88$ & $30.06$ & $7.31$ \\
LO\,Peg & $0.59 \pm 0.07$ & $0.46 \pm 0.31$ & $4739 \pm 138$ & $4.36$ & $0.00$ & $0.15$ & $0.66$ & $0.42$ & $78.79$ & $0.02$ & $-3.91$ & $29.70$ & $7.28$ \\
V471\,Tau & $0.39 \pm 0.10$ & $0.53 \pm 0.17$ & $4980 \pm 10$ & $4.50$ & $0.12$ & $0.75$ & $0.91$ & $0.52$ & $88.33$ & $0.02$ & -- & $30.00$ & $7.34$ \\
GJ\,338\,A & $0.39 \pm 0.25$ & $0.39 \pm 0.25$ & $4024 \pm 51$ & $4.68$ & $-0.05$ & $0.14$ & $0.58$ & $16.30$ & $1.79$ & -- & $-4.65$ & $27.92$ & $5.61$ \\
GJ\,338\,B & $0.39 \pm 0.25$ & $0.39 \pm 0.25$ & $4005 \pm 51$ & $4.68$ & $-0.03$ & $0.14$ & $0.58$ & $16.61$ & $1.78$ & -- & $-4.42$ & $27.92$ & $5.60$ \\
36\,Oph\,A & $-0.14 \pm 0.09$ & $-0.18 \pm 0.18$ & $5103 \pm 29$ & $4.64$ & $-0.29$ & $1.39$ & $0.69$ & $20.69$ & $1.69$ & $0.99$ & $-4.57$ & $28.02$ & $5.56$ \\
36\,Oph\,B & $-0.27 \pm 0.10$ & $-0.28 \pm 0.17$ & $5199 \pm 63$ & $4.62$ & $-0.30$ & $1.44$ & $0.59$ & $21.11$ & $1.41$ & $1.01$ & $-4.56$ & $27.89$ & $5.56$ \\
$\xi$ Boo\,B & $-0.16 \pm 0.15$ & $0.16 \pm 0.10$ & $4359 \pm 38$ & $4.69$ & $-0.08$ & $0.27$ & $0.61$ & $11.94$ & $2.58$ & $0.49$ & $-4.42$ & $28.08$ & $5.72$ \\
61\,Cyg\,A & $-0.01 \pm 0.04$ & $0.08 \pm 0.12$ & $4374 \pm 22$ & $4.63$ & $-0.20$ & $2.12$ & $0.67$ & $35.37$ & $0.96$ & $1.45$ & $-4.76$ & $27.03$ & $4.59$ \\
61\,Cyg\,B & $0.33 \pm 0.08$ & $0.54 \pm 0.12$ & $4044 \pm 32$ & $4.67$ & $-0.27$ & $1.87$ & $0.60$ & $37.84$ & $0.80$ & $1.46$ & $-4.89$ & $26.97$ & $4.63$ \\
70\,Oph\,A & $-0.29 \pm 0.06$ & $-0.28 \pm 0.11$ & $5320 \pm 40$ & $4.52$ & $-0.02$ & $1.30$ & $0.83$ & $20.00$ & $2.10$ & -- & $-4.55$ & $28.09$ & $5.47$ \\
70\,Oph\,B & $0.15 \pm 0.10$ & $0.31 \pm 0.04$ & $4400 \pm 100$ & $4.47$ & $-0.02$ & $1.30$ & $0.67$ & -- & -- & -- & $-3.61$ & $27.97$ & $5.53$ \\
$\epsilon$ Eri & $0.06 \pm 0.07$ & $-0.04 \pm 0.11$ & $5050 \pm 10$ & $4.48$ & $-0.11$ & $0.44$ & $0.74$ & $11.68$ & $3.21$ & $0.54$ & $-4.46$ & $28.31$ & $5.79$ \\
Sun & $-0.60 \pm 0.10$ & $-0.37 \pm 0.10$ & $5777 \pm 10$ & $4.44$ & $0.00$ & $4.57$ & $1.00$ & $26.09$ & $1.94$ & -- & $-4.94$ & $27.35$ & $4.57$ \\
$\alpha$ Cen\,A & $-0.54 \pm 0.30$ & $-0.47 \pm 0.31$ & $5829 \pm 6$ & $4.35$ & $0.23$ & $5.03$ & $1.23$ & $28.80$ & $2.16$ & $1.88$ & $-5.00$ & $26.99$ & $4.03$ \\
$\alpha$ Cen\,B & $-0.38 \pm 0.15$ & $-0.33 \pm 0.17$ & $5189 \pm 18$ & $4.30$ & $0.22$ & $5.03$ & $0.87$ & $38.70$ & $1.14$ & $1.80$ & $-4.92$ & $27.32$ & $4.66$ \\
$\pi^1$ UMa & $-0.45 \pm 0.18$ & $-0.32 \pm 0.21$ & $5950 \pm 70$ & $4.53$ & $-0.12$ & $0.19$ & $0.91$ & $4.69$ & $9.82$ & $0.46$ & $-4.38$ & $28.99$ & $6.29$ \\
EK\,Dra & $-0.00 \pm 0.13$ & $-0.04 \pm 0.17$ & $5840 \pm 100$ & $4.57$ & $-0.01$ & $0.15$ & $0.93$ & $2.68$ & $17.54$ & $0.24$ & $-4.15$ & $30.06$ & $7.34$ \\
$\xi$ Boo\,A & $-0.29 \pm 0.04$ & $-0.19 \pm 0.07$ & $5550 \pm 100$ & $4.66$ & $-0.19$ & $0.19$ & $0.86$ & $6.31$ & $6.90$ & $0.39$ & $-4.36$ & $28.91$ & $6.26$ \\
$\chi^1$ Ori & $-0.47 \pm 0.33$ & $-0.39 \pm 0.31$ & $6020 \pm 10$ & $4.45$ & $-0.06$ & $0.29$ & $0.98$ & $5.36$ & $9.25$ & $0.59$ & $-4.43$ & $28.99$ & $6.22$ \\
$\kappa$ Cet & $-0.38 \pm 0.34$ & $-0.28 \pm 0.32$ & $5740 \pm 20$ & $4.46$ & $-0.02$ & $0.52$ & $0.92$ & $9.24$ & $5.04$ & $0.70$ & $-4.42$ & $28.79$ & $6.08$ \\
$\beta$ Com & $-0.58 \pm 0.25$ & $-0.52 \pm 0.23$ & $6090 \pm 60$ & $4.41$ & $-0.03$ & $1.63$ & $1.11$ & $12.35$ & $4.55$ & $1.45$ & $-4.75$ & $28.21$ & $5.33$ \\
47\,Cas\,B & $0.19 \pm 0.36$ & $0.14 \pm 0.33$ & $5780 \pm 100$ & $4.50$ & $-0.11$ & $0.10$ & $1.00$ & $1.00$ & $50.59$ & $2.50$ & -- & $30.39$ & $7.61$ \\
$\iota$ Hor & $0.21 \pm 0.25$ & $0.15 \pm 0.19$ & $6057 \pm 60$ & $4.37$ & $0.15$ & $0.75$ & $1.17$ & $8.50$ & $6.96$ & $1.14$ & $-4.56$ & $28.20$ & $5.28$ \\
11\,LMi & $-0.10 \pm 0.22$ & $-0.12 \pm 0.32$ & $5376 \pm 43$ & $4.48$ & $0.33$ & $2.39$ & $1.00$ & $17.88$ & $2.83$ & $0.98$ & $-4.69$ & $28.51$ & $5.73$ \\
HR\,7291 & $-0.51 \pm 0.12$ & $-0.53 \pm 0.08$ & $6131 \pm 32$ & $4.34$ & $0.14$ & $2.56$ & $1.19$ & $7.60$ & $7.92$ & $1.34$ & $-4.79$ & $28.57$ & $5.63$ \\
$\sigma^2$ CrB & $0.06 \pm 0.04$ & $0.01 \pm 0.03$ & $5950 \pm 50$ & $4.12$ & $-0.06$ & $1.00$ & $1.24$ & $1.16$ & $54.40$ & $0.14$ & $-3.83$ & $30.68$ & $7.80$ \\
$\xi$ UMa & $0.35 \pm 0.06$ & $0.39 \pm 0.33$ & $5796 \pm 100$ & $4.43$ & $-0.29$ & -- & $0.95$ & $3.98$ & $12.07$ & $0.47$ & -- & $29.43$ & $6.69$ \\
HR\,1099 & $0.76 \pm 0.12$ & $0.76 \pm 0.05$ & $4833 \pm 100$ & $2.84$ & $-0.16$ & -- & $3.90$ & $2.84$ & $69.55$ & $0.00$ & $-3.84$ & $31.28$ & $7.31$ \\
UX\,Ari & $1.13 \pm 0.25$ & $1.12 \pm 0.21$ & $4560 \pm 100$ & $3.06$ & $0.30$ & $5.60$ & $5.60$ & $6.44$ & $44.01$ & $0.03$ & -- & $31.15$ & $6.87$ \\
$\lambda$ And & $0.52 \pm 0.07$ & $0.48 \pm 0.09$ & $4630 \pm 100$ & $2.57$ & $-0.56$ & $5.62$ & $6.41$ & $53.95$ & $6.01$ & $0.07$ & $-4.48$ & $30.57$ & $6.17$ \\
VY\,Ari & $0.74 \pm 0.09$ & $0.75 \pm 0.09$ & $4800 \pm 100$ & $3.10$ & $-0.09$ & -- & $2.66$ & $16.20$ & $8.31$ & $0.01$ & -- & $31.09$ & $7.46$ \\
Capella & $-0.13 \pm 0.20$ & $-0.11 \pm 0.18$ & $4970 \pm 50$ & $2.69$ & $-0.04$ & $0.62$ & $11.98$ & $104.00$ & $5.83$ & $0.34$ & -- & $30.62$ & $5.68$ \\
$\sigma$ Gem & $0.66 \pm 0.26$ & $0.60 \pm 0.19$ & $4630 \pm 100$ & $2.79$ & $-0.10$ & $5.00$ & $10.10$ & $19.60$ & $26.07$ & $0.03$ & -- & $31.48$ & $6.69$ \\
31\,Com & $0.07 \pm 0.19$ & $0.05 \pm 0.15$ & $5660 \pm 42$ & $3.51$ & $-0.15$ & $0.54$ & $8.74$ & $6.76$ & $65.41$ & $0.40$ & -- & $30.90$ & $6.23$ \\
$\mu$ Vel & $-0.16 \pm 0.17$ & $-0.15 \pm 0.18$ & $5030 \pm 40$ & $2.73$ & -- & $0.36$ & $13.00$ & -- & -- & $0.61$ & -- & $30.51$ & $5.50$ \\
$\beta$ Cet & $-0.08 \pm 0.14$ & $-0.03 \pm 0.21$ & $4720 \pm 100$ & $2.65$ & $-0.15$ & $0.46$ & $16.78$ & $215.00$ & $3.95$ & $1.19$ & $-4.79$ & $30.43$ & $5.20$ \\
FK\,Com & $0.29 \pm 0.07$ & $0.30 \pm 0.06$ & $5000 \pm 100$ & $3.50$ & $-0.89$ & -- & $6.99$ & $2.40$ & $147.35$ & $0.17$ & -- & $31.00$ & $6.53$ \\
YY\,Men & $0.49 \pm 0.31$ & $0.47 \pm 0.25$ & $4700 \pm 100$ & $2.40$ & $-0.37$ & -- & $12.70$ & $9.55$ & $67.28$ & $0.00$ & -- & $32.50$ & $7.51$ \\
EI\,Eri & $0.23 \pm 0.08$ & $0.32 \pm 0.15$ & $5500 \pm 100$ & $3.73$ & $-0.03$ & $6.15$ & $2.37$ & $1.95$ & $61.58$ & $0.01$ & -- & $31.13$ & $7.60$ \\
V851\,Cen & $0.67 \pm 0.22$ & $0.74 \pm 0.27$ & $4700 \pm 80$ & $3.00$ & $-0.13$ & -- & $3.50$ & $12.27$ & $14.43$ & $0.03$ & -- & $30.61$ & $6.74$ \\
AR\,Psc & $0.68 \pm 0.19$ & $0.83 \pm 0.27$ & $4880 \pm 100$ & $2.89$ & $-0.91$ & $7.50$ & $2.70$ & $12.25$ & $11.15$ & $0.00$ & -- & $31.24$ & $7.59$ \\
AY\,Cet & $0.53 \pm 0.32$ & $0.59 \pm 0.34$ & $5044 \pm 17$ & $3.05$ & $-0.38$ & $1.32$ & $6.80$ & $75.12$ & $4.58$ & $0.06$ & -- & $31.00$ & $6.55$ \\
II\,Peg & $1.22 \pm 0.14$ & $1.22 \pm 0.17$ & $4600 \pm 100$ & $3.20$ & $-0.40$ & -- & $3.40$ & $6.72$ & $25.61$ & $0.03$ & -- & $30.78$ & $6.93$ \\
AR\,Lac$^{(e)}$ & $0.49 \pm 0.17$ & $0.49 \pm 0.25$ & $4820 \pm 100$ & $3.68$ & $-0.70$ & -- & $2.68$ & $1.98$ & $68.48$ & $0.01$ & -- & $31.02$ & $7.87$ \\
$\eta$ Lep & $-0.31 \pm 0.09$ & $-0.30 \pm 0.09$ & $6920 \pm 70$ & $4.19$ & $-0.09$ & $1.80$ & $1.56$ & -- & -- & -- & $-3.94$ & $28.50$ & $5.33$ \\
$\pi^3$ Ori & $-0.41 \pm 0.07$ & $-0.31 \pm 0.12$ & $6430 \pm 40$ & $4.25$ & $-0.04$ & $1.55$ & $1.32$ & $4.00$ & $16.70$ & $1.82$ & $-4.65$ & $28.99$ & $5.96$ \\
$\tau$ Boo\,A & $-0.21 \pm 0.09$ & $-0.27 \pm 0.13$ & $6370 \pm 30$ & $4.14$ & $0.16$ & $2.30$ & $1.47$ & $3.31$ & $22.47$ & $0.88$ & $-4.73$ & $28.76$ & $5.64$ \\
Procyon & $0.29 \pm 0.17$ & $0.25 \pm 0.15$ & $6530 \pm 50$ & $3.96$ & $-0.05$ & $1.87$ & $2.03$ & $23.00$ & $4.47$ & $13.37$ & $-4.63$ & $28.39$ & $4.99$ \\
Altair & $-0.56 \pm 0.42$ & $-0.25 \pm 0.36$ & $7655 \pm 150$ & $4.29$ & $-0.20$ & $1.20$ & $1.84$ & $0.37$ & $252.28$ & $7.00$ & $-3.97$ & $27.15$ & $3.84$ \\
\hline       
\end{tabular}
\tablefoot{${(a)}$ Where no error for \teff\ was given in the literature source, we assume $\pm$100\,K. 
${(b)}$ Estimated age in $10^9$\,yr.
${(c)}$ Logarithmic values measured in erg\,s$^{-1}$.
${(d)}$ Logarithmic values measured in erg\,s$^{-1}$cm$^{-2}$.
${(e)}$ The activity of AR\,Lac originates from both components, however, the K1V secondary is more luminous and its activity is higher according to \citet{1998A&A...332..541L}, while the X-ray fluxes of the components and X-ray flares are discussed in \citet{2014ApJ...783....2D}; we consider the secondary component as the source of the FIP effect.}

\end{table*}

\subsection{Photospheric abundances from LAMOST}\label{sect:lamost}

Most of the time when the stellar FIP bias was calculated, there were no stellar photospheric abundances available, so the coronal abundances had to be compared to the solar photosphere. However, different stars can have significantly different chemical composition, which may affect the derived FIP-bias values. To take this into account, individual photospheric abundances are needed. Ideally, one would need a homogeneous survey of high resolution optical/near-infrared spectra with high signal-to-noise ratio, and observable transitions for each of the required elements. Given the absence of such data, to have an independent set of stellar photospheric abundances, as homogeneous as possible, we used the results of \citet{lamost}, where a large set of abundances were predicted from medium resolution LAMOST spectra. They used  deep learning with a training set compiled from APOGEE results for the common stars in the LAMOST and APOGEE sample. Instead of using individually measured equivalent widths, the full spectra were used in the prediction. Since the lines of some elements are weak or missing in the optical, the method inherently involves hidden correlations between abundances. Their cited precision is 0.06--0.12\,dex for 11 elements, also including C, N and O.

\begin{figure}[thb]
\includegraphics[width=\columnwidth]{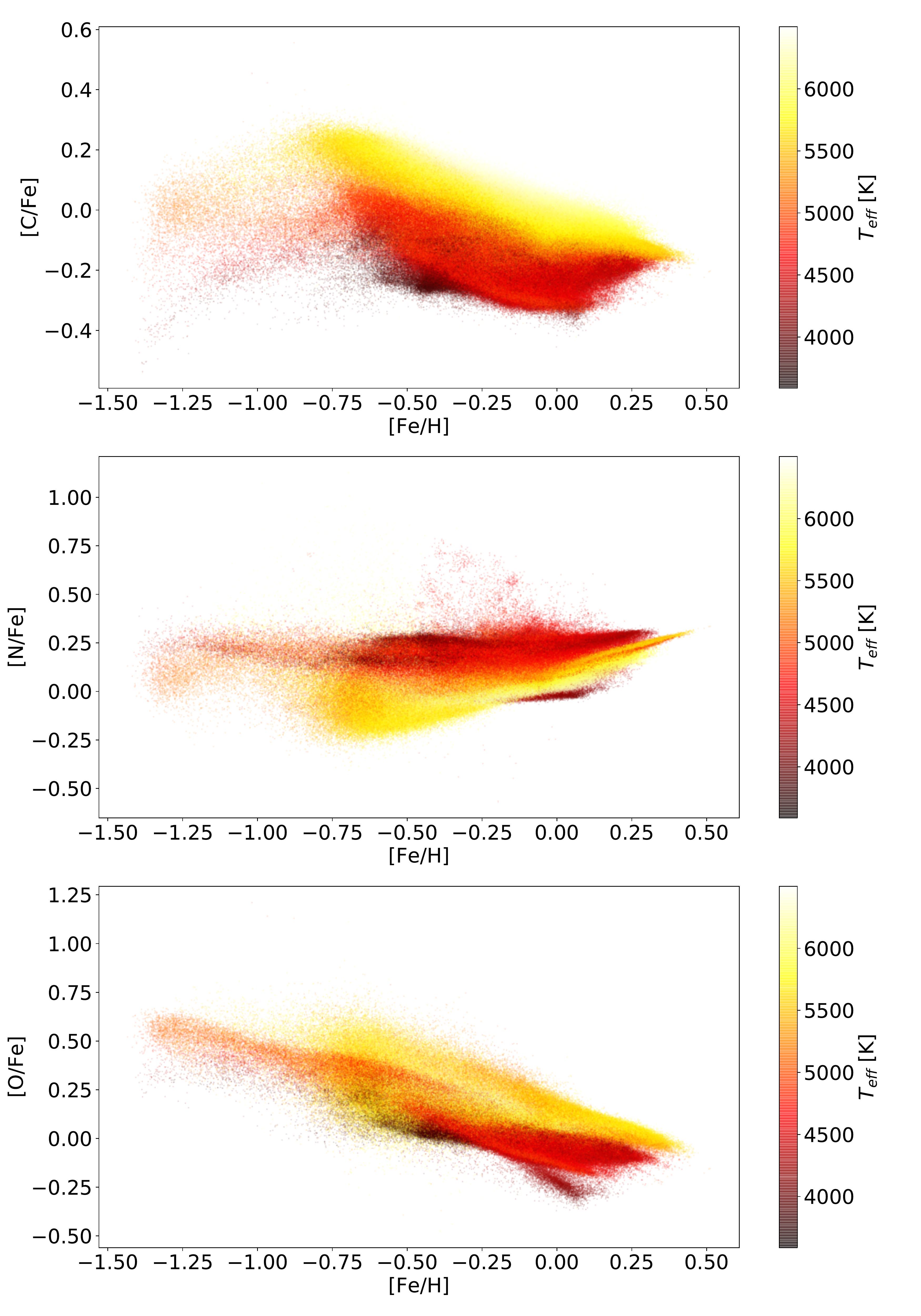}
\caption{Abundance trends based on data from \citet{lamost} for the LAMOST sample, for the three relevant high-FIP elements (there are no available Ne abundances).}
\label{fig:lamost}
\end{figure}

Since the LAMOST survey only extends to the northern hemisphere, roughly half of our sample is not covered by it. To circumvent this problem, we used a k-nearest neighbours regressor (KNN, \texttt{KNeighborsRegressor} from \texttt{sklearn.neighbors}\footnote{\url{https://scikit-learn.org}}) to predict the abundances of interest from \teff, $\log g$ and [Fe/H]. This essentially means finding the nearest few stars in this three-dimensional parameter space, and getting their mean abundances as the result. For the prediction, we used $k$=6 neighbours. The root-mean-square error of the method is 0.05\,dex, using 10\% of the whole LAMOST sample as test set. The method gives the following abundances for the Sun: [C/Fe]=$-$0.09$\pm$0.03, [N/Fe]=0.06$\pm$0.08 and [O/Fe]=0.10$\pm$0.03. A lower dimensional cut of this parameter space can be seen in Fig.\,\ref{fig:lamost}. The tempting aspects of this method are the following: it is really accessible (only needs the three most fundamental parameters), it gives even N and O abundances, and since it practically involves interpolation, it cannot give too extreme results (which is a problem with spectral synthesis and NLTE correction, giving occasionally clearly unrealistic O abundances). But since the predictions are quite indirect, we only treat the results as a little better first approximation, after the initial zeroth approximation of solar composition.

\subsection{Photospheric abundances from spectral synthesis}\label{sect:sme}

To derive homogeneous photospheric abundances for as many stars as possible from the sample, we used spectra from CFHT-ESpaDoNS \citep{2003ASPC..307...41D} and TBL-NARVAL \citep{2003EAS.....9..105A}. The high spectral resolution enables the use of spectral synthesis, where the whole line profiles are fitted with model spectra calculated from a model atmosphere with varying physical parameters. The results are used to validate the process described in the previous section.

After acquiring the optical echelle spectra, we begin with a 99\% percentile smoothing to locate the continuum, and renormalize it where necessary. We used SME \citep{piskunovsme} for the spectral synthesis. During the synthesis, MARCS models were used \citep{gustafsson_marcs} with atomic parameters from VALD \citep{kupka_vald} using a transition list typical for the approximate effective temperature of the star in question. Macroturbulence was computed with the following equation from \citet{valenti_macro}:
\begin{equation}
    v_{\mathrm{mac}}=\Bigg(3.98-\frac{T_{\mathrm{eff}}-5770 \mathrm{\,K}}{650 \mathrm{\,K}}\Bigg)\,\mathrm{km\,s}^{-1}
\end{equation}

Astrophysical parameters were determined by the following procedure:
\begin{enumerate}
    \item Fitting $v\sin i$ and microturbulence using rough astrophysical parameters and assuming solar abundances.
    \item Fitting \teff.
    \item Fitting metallicity and microturbulence (microturbulence values can be unreliable sometimes, and can alter the other astrophysical parameters slightly, thus we fit it simultaneously with another parameter).
    \item Fitting $\log g$ for the Na doublet and the Mg triplet, or in some cases, for a special line list only containing transitions with $\log gf > 0$ (stronger lines tend to be more sensitive to changes in $\log g$).
    \item Refitting \teff\ and metallicity.
    \item Fitting individual abundances (C, O, Ca, Mg, Si, Ti, Fe, Ni).
\end{enumerate}
From the last step, only C, O and Fe are of interest, the others were only added to help with line blends. Determining the Fe abundance is mostly straightforward, thanks to the multitude of Fe lines visible in practically any stellar spectra. The carbon abundance was fitted using several atomic and molecular lines, but most of them are really weak and blended. The oxygen abundance was fitted using only the O triplet at 7774\,\AA, since the 6300\,\AA\ line was not visible in any of the spectra. This is a highly NLTE-sensitive triplet, so we used an NLTE correction from \citet[][]{NLTE_MPIA} with parameters from \citet[][]{NLTE_oxygen} that resulted in slightly different values. This correction uses \teff, $\log{g}$, [Fe/H] and microturbulence, and outputs separate correction factors for each line in the O triplet, which we averaged. As there are no N lines in the optical, the N abundance was estimated through proxies (i.e., we assumed the same relative N/O abundance as in the Sun). For Ne we used the empirical relation from \citet{drake_testa_neon} ([Ne/O]=$\log_{10}{0.41}$). We note however, that the abundance of 3 out of 4 high-FIP elements depend on the NLTE-sensitive O triplet. Thus we argue that in the case of the high-FIP elements, the photospheric abundances calculated with spectral synthesis are not much more reliable than the values from broader abundance trends or empirical correlations. The results of the spectral synthesis are listed in Table\,\ref{table:sme_results}.

\subsection{Principal component analysis}

In the space of physical parameters of the stars in the sample (\teff, $\log g$, [Fe/H], age, radius $R$, $v_\mathrm{rot}$, Rossby number $Ro$, \logrhk, $\log L_X$, $\log F_X$, literature and KNN FIP bias) there are several clear and not so clear correlations. To find these, we used principal component analysis (PCA, \citealt{pca_pearson}), with the \texttt{PCA} implementation of \texttt{sklearn.decomposition}. PCA is an algorithm that eliminates linear correlations by selecting directions in the parameter space with the highest variance, and defining new orthonormal basis vectors in those directions, thus creating a basis where a large portion of the variance can be described by only a few components. It is a powerful technique for dimensionality reduction and data compression, and it is often the first step when dealing with clustering problems (for a similar treatment, see e.g. \citealt{pca_geza}). We use PCA to identify hidden correlations between parameters, and to find the parameters that are not really useful. Before running PCA on the full parameter set, we scaled the features to have zero mean and unit variance. In the case of missing parameters, we refilled them using the average value of the given parameter for the 2 nearest neighbours in the 12 dimensional parameter space. For parameters that change orders of magnitude (age, radius, $v_\mathrm{rot}$ and Rossby number) we used their base 10 logarithm.

\subsection{Unsupervised clustering}

To separate different groups of stars based on the available parameters, we used unsupervised k-means clustering \citep{kmeans_macqueen}, with the \texttt{KMeans} implementation of \texttt{sklearn.cluster}. It identifies $k$ different clusters, based on proximity in the parameter space, by minimizing the variance within each cluster. The resulting clusters are basically Voronoi cells around $k$ number of centers (cluster means), where $k$ is defined a priori. Since some of the parameters in the original 12 dimensional dataset are correlated, they do not form an orthogonal basis, which can bias the clustering. Therefore we run the clustering after dimensionality reduction with PCA, only keeping 4 PCs.

\subsection{Linear discriminant analysis}

To find the main physical difference between the two emerging clusters on the \teff$-$FIP bias diagram, we carried out linear discriminant analysis (hereafter LDA, see \citealt{lda_fisher} for the original paper, \citealt{lda_book} for a review of the method, or \citealt{lda_example} for an application), using the \texttt{LinearDiscriminantAnalysis} implementation of \texttt{sklearn.discriminant\_analysis}. It finds a linear combination of input features that maximizes the separation between some pre-defined clusters, then uses linear decision boundaries for classification. Interpreting the model given by LDA is relatively straightforward, as it is a simple linear combination of the scaled input parameters. The LDA method assumes that the input features follow a multivariate normal distribution, and are independent with no correlations between them. We note that in our case, none of these assumptions hold strictly. From the 12 input parameters 9 are independent measurements, while $\log F_X$ follows from $\log L_X$ and radius, $v_\mathrm{rot}$ contains radius and rotational period, and the Rossby number for giants is derived from $v_\mathrm{rot}$ and $\log g$ for the evolved sample. Also, there are deviations from normality. For the cluster labels, we used a simple linear boundary on the \teff$-$FIP bias diagram. Since the FIP bias was removed from the input features, LDA had to find a combination of the remaining ten parameters that separate the two clusters.

\section{Results}\label{sect:results}

\begin{figure*}
\includegraphics[width=\textwidth]{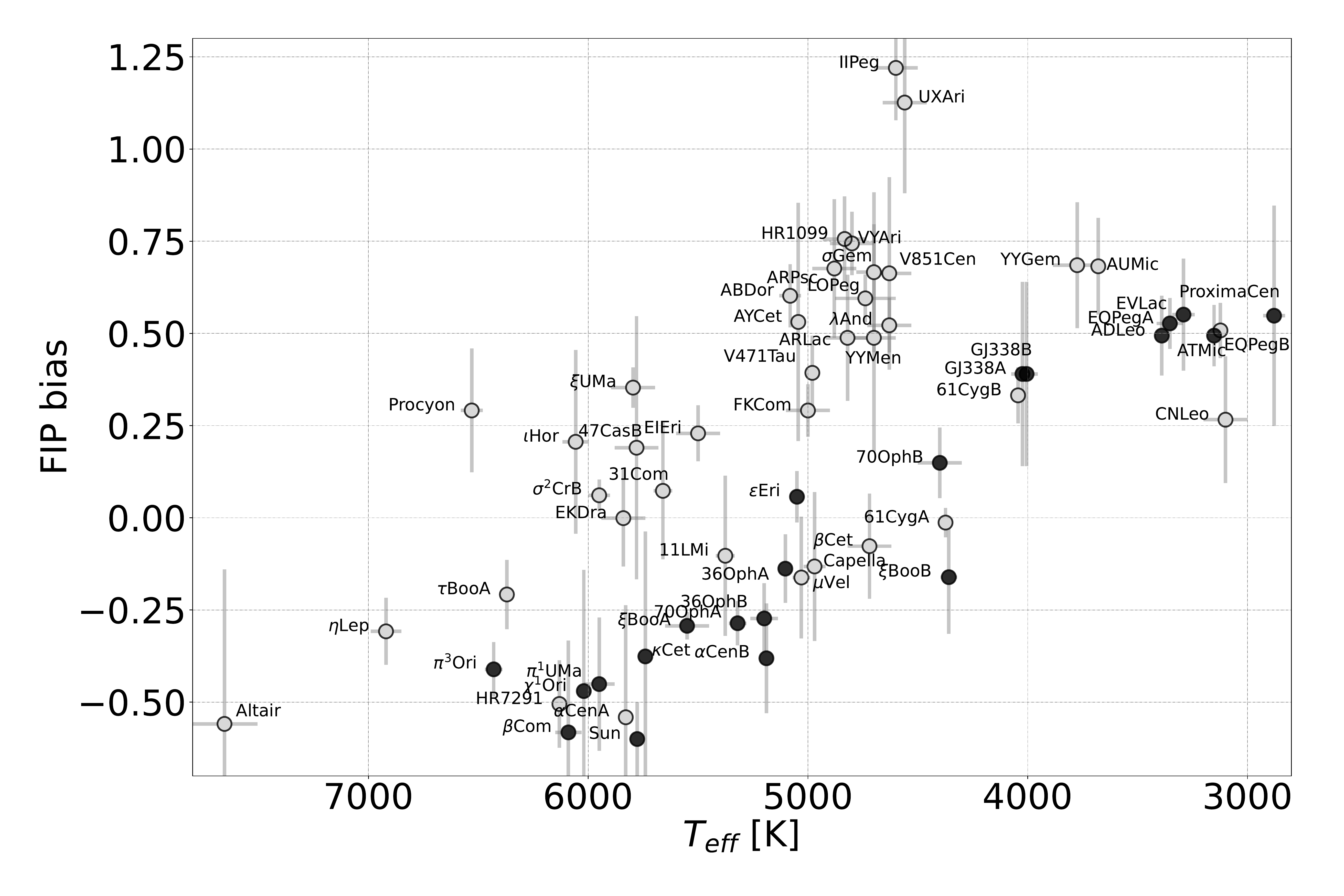}
\caption{Literature FIP bias vs. \teff. Black dots indicate the stars from the original FIP-bias diagram of \citet{2015LRSP...12....2L}, updated from \citet{Wood_Linsky2010}, forming a tight sequence with \teff.}
\label{fig:fip_names}
\end{figure*}

\begin{figure}
\includegraphics[width=\columnwidth]{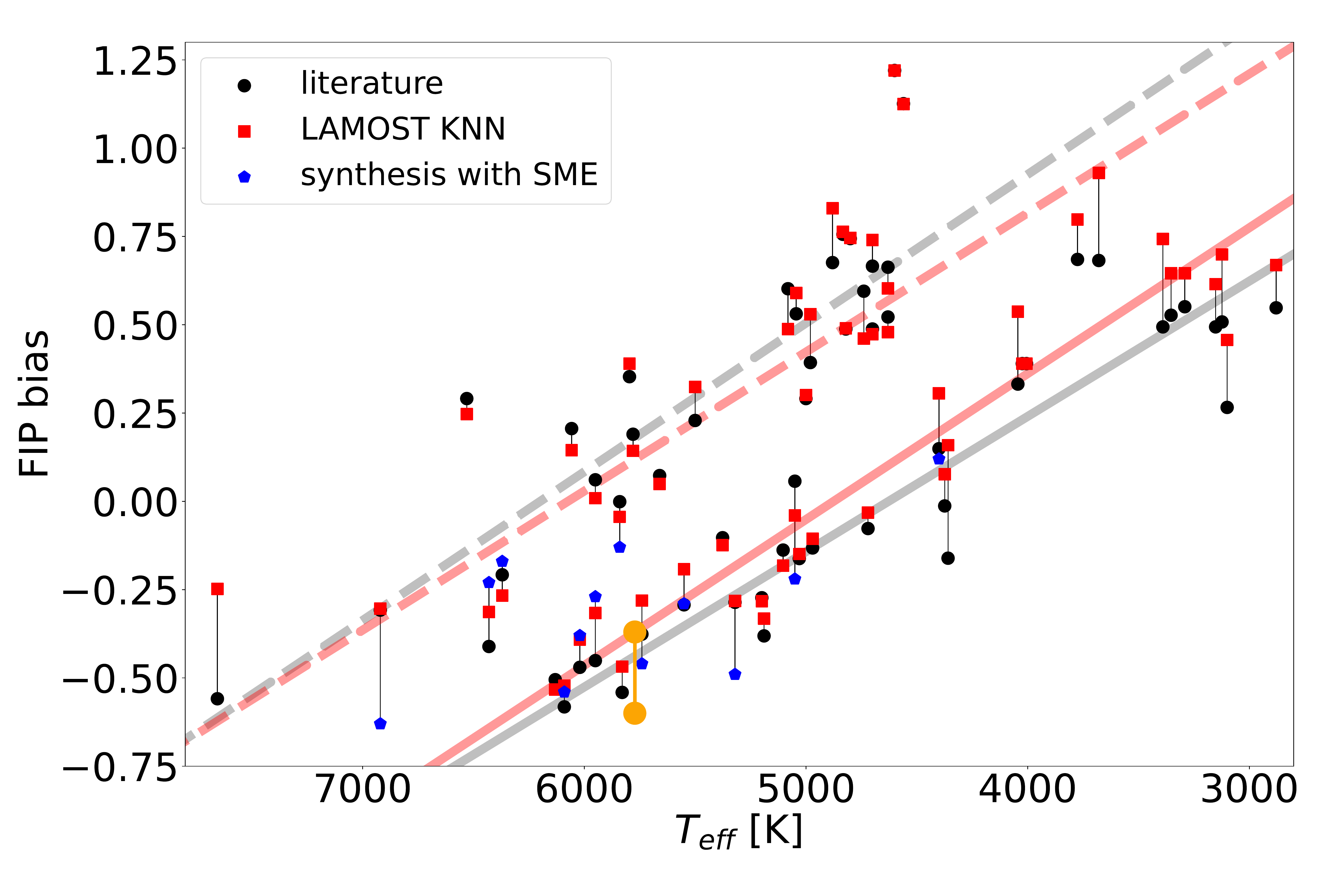}
\caption{The \teff$-$FIP bias diagram with recalculated FIP-bias values. For the Sun, orange points mark its activity minimum and maximum. Solid and dashed lines show two component linear fits with a RANSAC regressor, with gray and red colours corresponding to literature and KNN FIP bias, respectively. } 
\label{fig:fip_3_values}
\end{figure}

Our first step was to plot a new \teff$-$FIP bias diagram with all stars in the extended sample using the already published and homogenized FIP-bias values and their error bars (see Sect.\,\ref{sect:methods}) in Fig.\,\ref{fig:fip_names}. The figure shows the names of the stars for easier identification, black points mark the stars from the original FIP-bias diagram of \citet{2015LRSP...12....2L}, updated from \citet{Wood_Linsky2010}. Our important result, that the relation has now nearly parallel branches separated by about 0.5 in FIP bias, is well seen and will be discussed later. The low mass M dwarf stars (\teff$<4000$\,K) show a separate blob in the continuation of the original relation. A hint of a bimodal distribution of the FIP-bias values is found in \citet[][see their Fig.\,7a]{Wood2018}: all four stars lying above their diagram ($\tau$\,Boo~A, EK\,Dra, AB\,Dor~A and AU\,Mic) are included now in the parallel, second branch of the relation, see Fig.\,\ref{fig:fip_names}.

\subsection{Comparison with FIP-bias values derived with different methods}

All stars from the sample have both literature and KNN FIP-bias values. No systematic difference is found between them except for the M dwarfs below \teff$\sim$4200\,K, as those stars are at the edge of the LAMOST \teff$-\log{g}-$[Fe/H] parameter grid, thus their predicted photospheric abundances are similar. There is generally a good agreement between the FIP-bias values calculated from the original photospheric composition and the newly derived one from spectral synthesis. Fig.\,\ref{fig:fip_3_values} shows a comparison between the literature and KNN FIP-bias values. The homogeneously predicted photospheric abundances from LAMOST slightly decrease the scatter, the Spearman correlation coefficient changes from $-$0.67$\pm$0.07 for the literature values to $-$0.74$\pm$0.05 for LAMOST KNN, with error bars calculated from bootstrap resampling.

The extended \teff$-$FIP bias diagram includes 59 stars, which seem to lie on two separate lines, shifted by a FIP-bias value of $\sim$0.5. To show these trends in Fig.\,\ref{fig:fip_3_values}, we used a RANSAC (RANdom SAmple Consensus)\footnote{The RANSAC algorithm selects a subset of the original data points, and splits it to inlier and outlier points after fitting them by a model. By iterating these steps a final set of outliers is created where the outliers will have no influence on the final fit. We used \texttt{RANSACRegressor} from \texttt{sklearn.linear\_model}.} regressor to fit two lines.  We use this robust regression method to first identify the most prominent linear correlation (the \teff$-$FIP bias diagram of MS stars from \citet{2015LRSP...12....2L} shown in Fig.\,\ref{fig:fip_names} with black points), and then fit the outlier points with another RANSAC line. To match visual inspection, the residual threshold parameter was set to 0.2\,dex in FIP bias, it is the maximum residual to be classified as an inlier. The fitted lines seem to be similar for both literature and KNN FIP bias, see the red and gray lines in Fig.\,\ref{fig:fip_3_values}. For the lower branch, the fitted parameters are as follows:
\begin{equation} 
    \mathrm{FIP\,bias_{lit}} = (-0.00038 \pm 0.00002)  T_\mathrm{eff} + (1.77 \pm 0.08)
\end{equation}
\begin{equation} 
    \mathrm{FIP\,bias_{KNN}} = (-0.00041 \pm 0.00002)  T_\mathrm{eff} + (2.01 \pm 0.09)
\end{equation}
And for the upper branch:
\begin{equation} 
    \mathrm{FIP\,bias_{lit}} = (-0.00042 \pm 0.00003)  T_\mathrm{eff} + (2.61 \pm 0.18)
\end{equation}
\begin{equation} 
    \mathrm{FIP\,bias_{KNN}} = (-0.00040 \pm 0.00003)  T_\mathrm{eff} + (2.39 \pm 0.16)
\end{equation}

\begin{figure*}
\includegraphics[height=5.9cm, width=9.0cm]{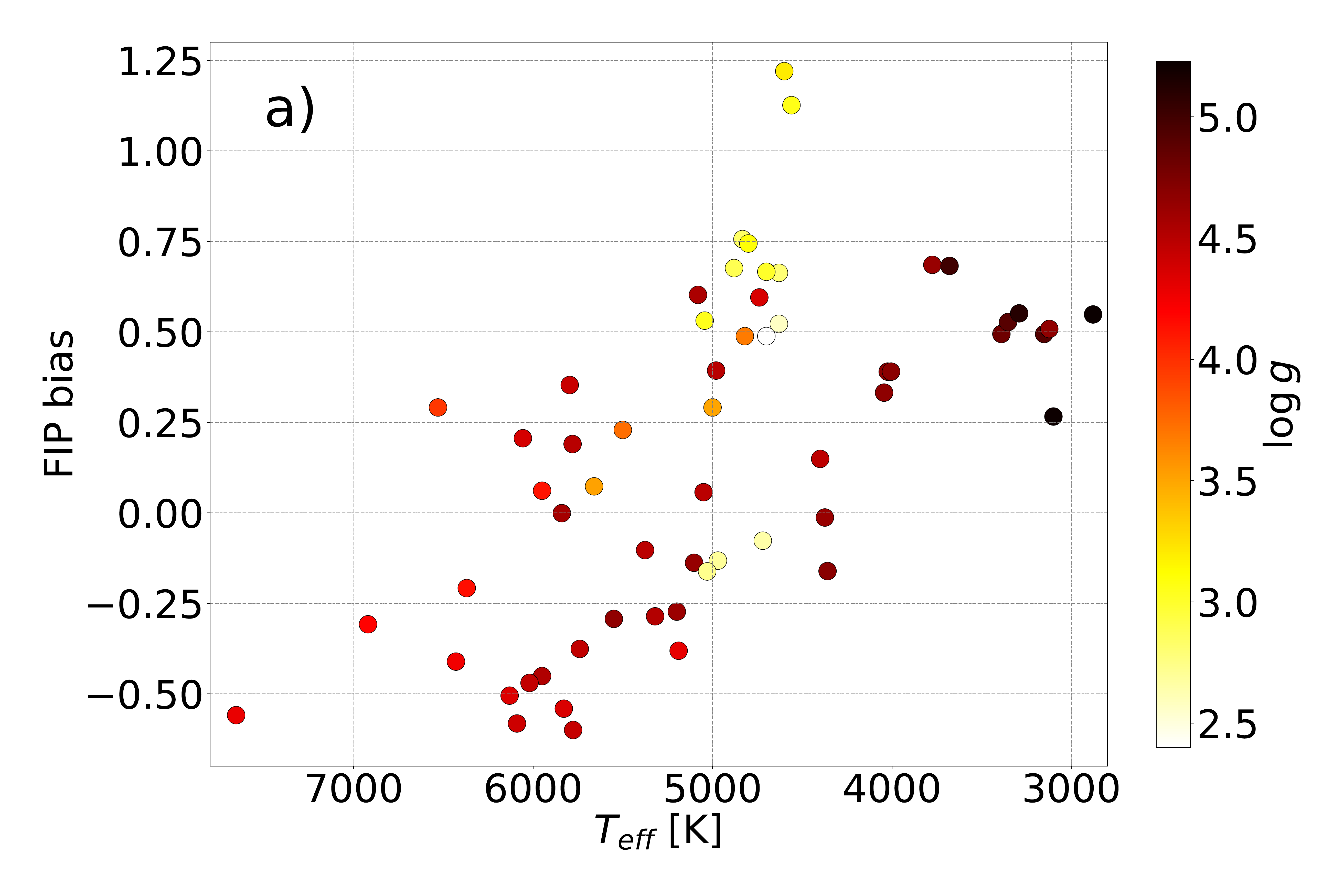}%
\hspace{0.15cm}\includegraphics[height=5.9cm, width=9.40cm]{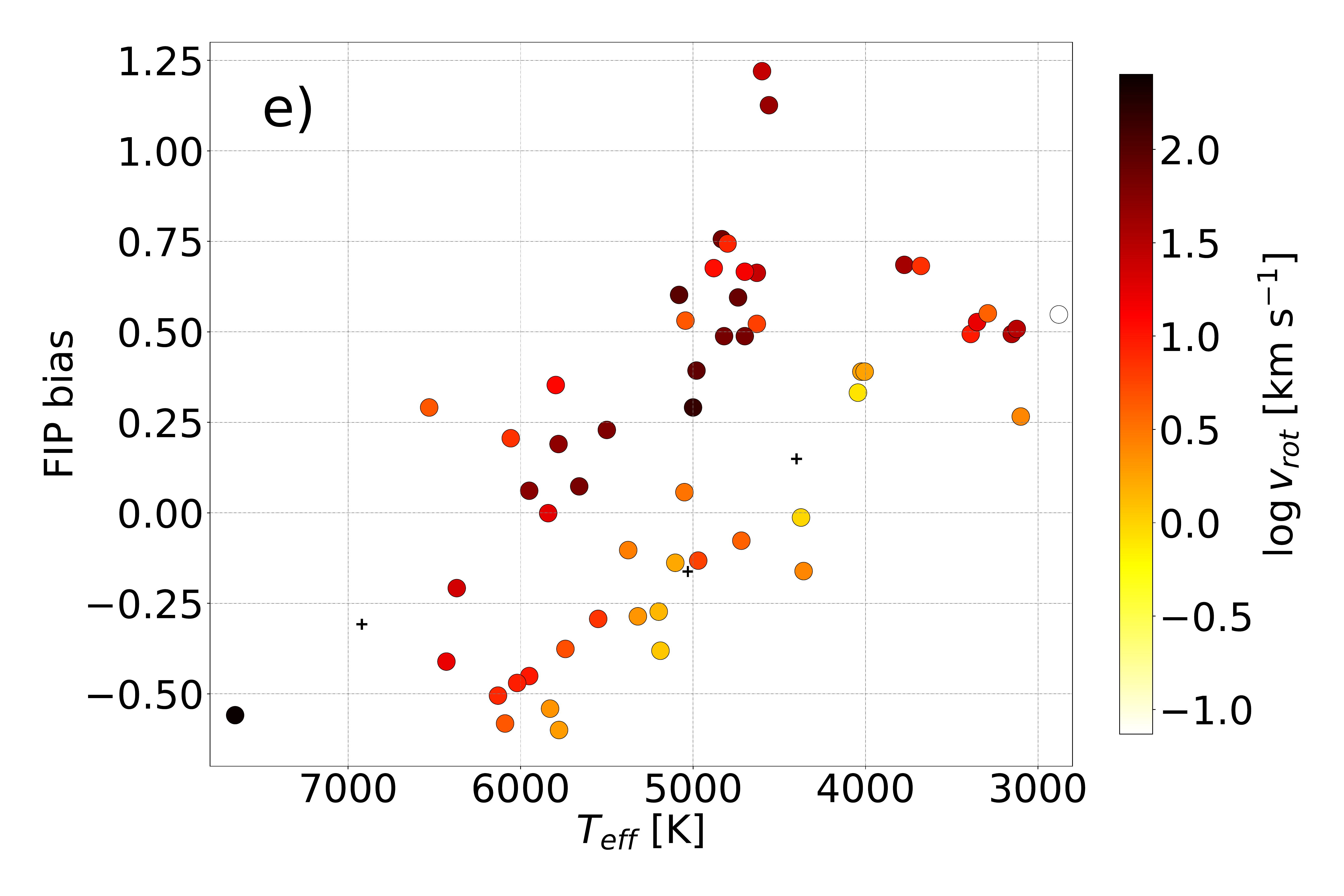}

\includegraphics[height=5.9cm, width=8.95cm]{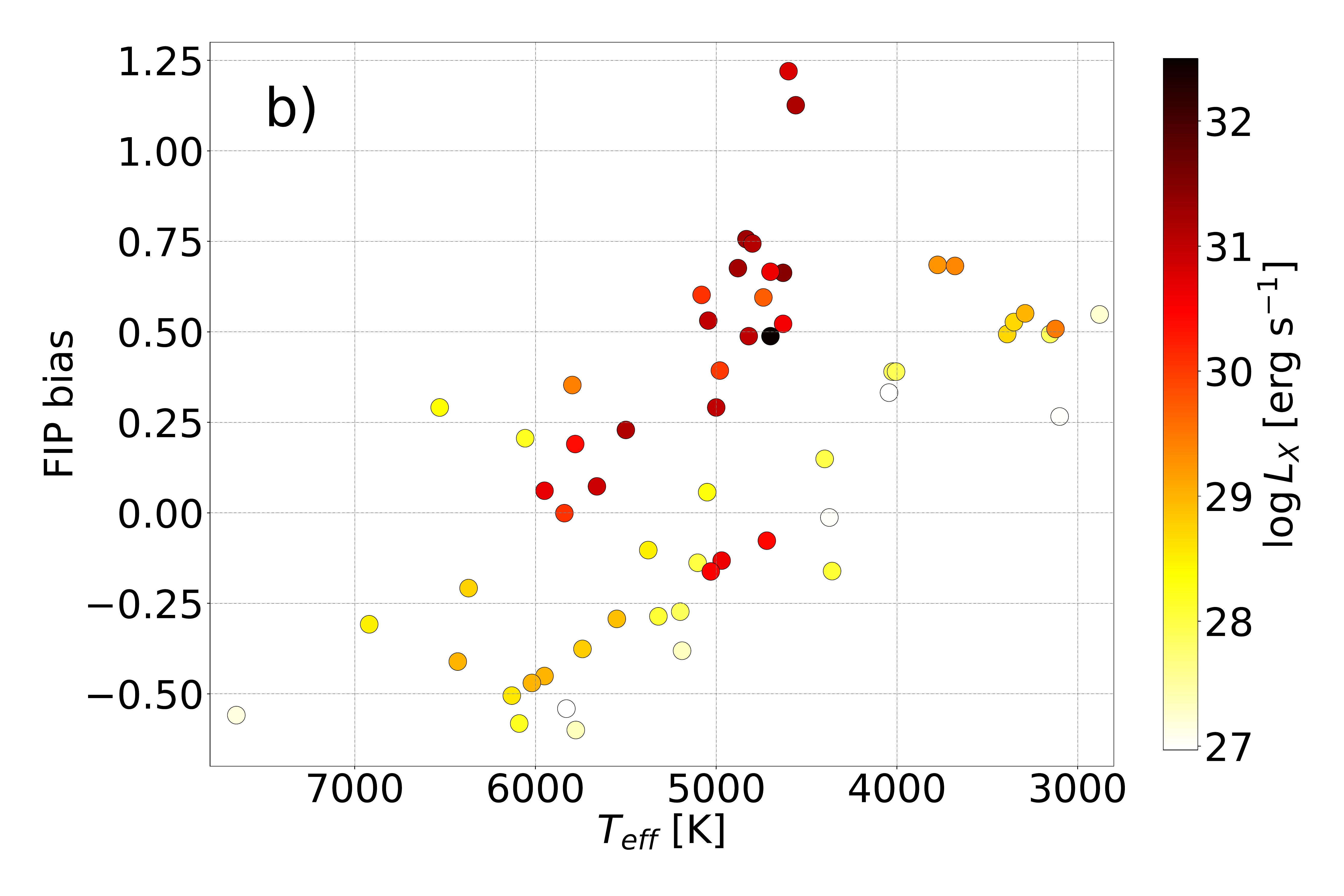}%
\hspace{0.30cm}\includegraphics[height=5.9cm, width=8.77cm]{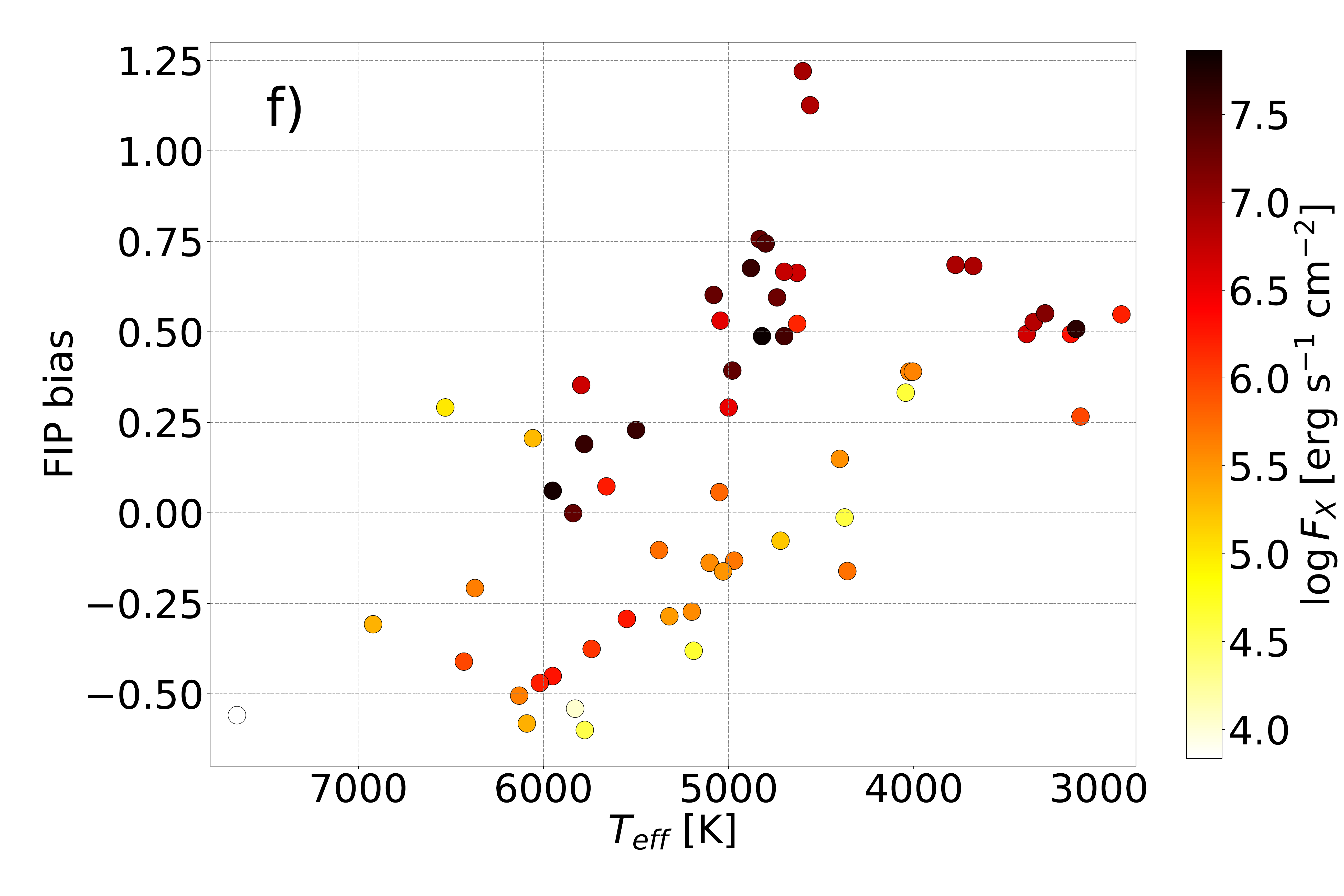}

\includegraphics[height=5.9cm, width=9.22cm]{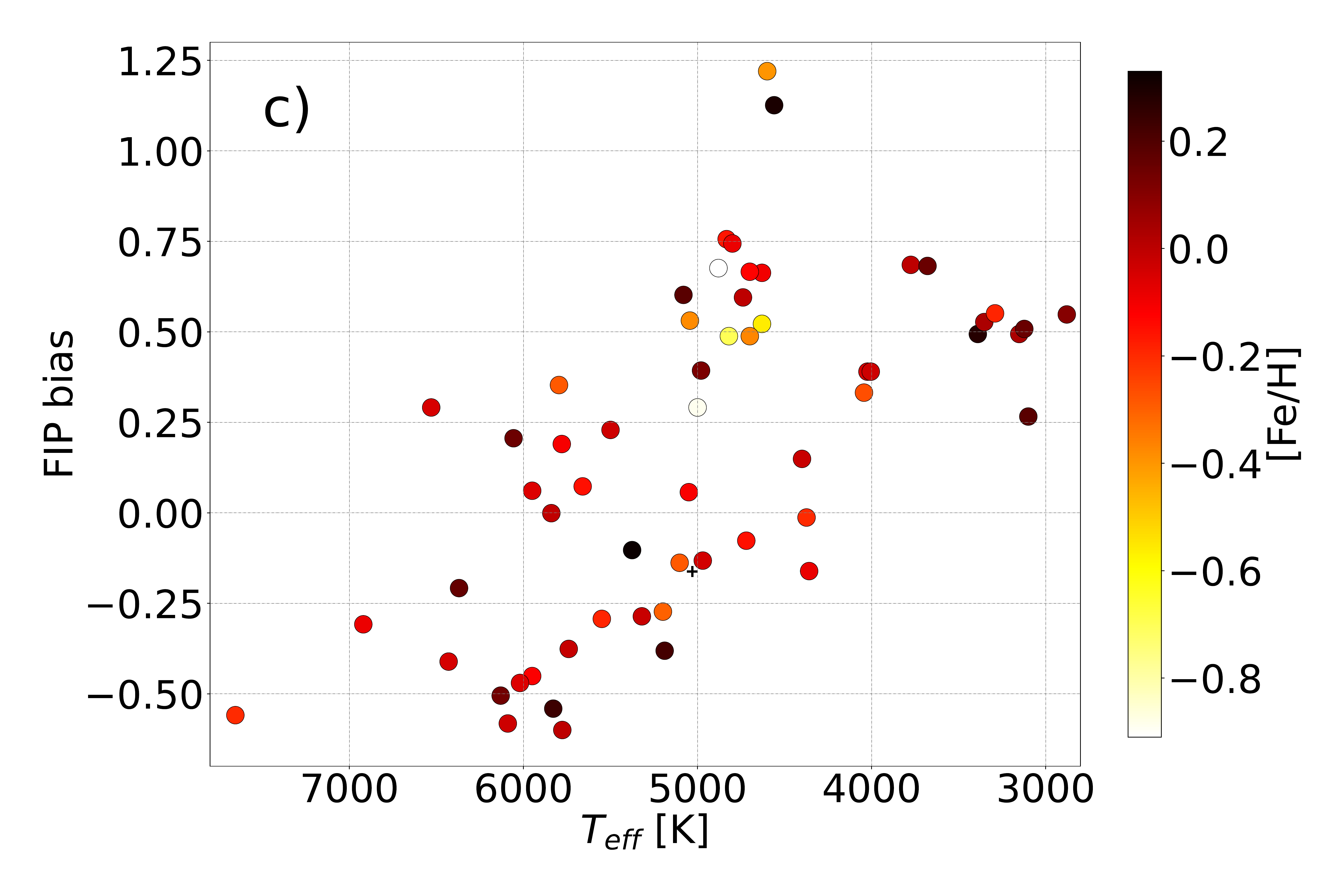}%
\hspace{-0.06cm}\includegraphics[height=5.9cm, width=9.35cm]{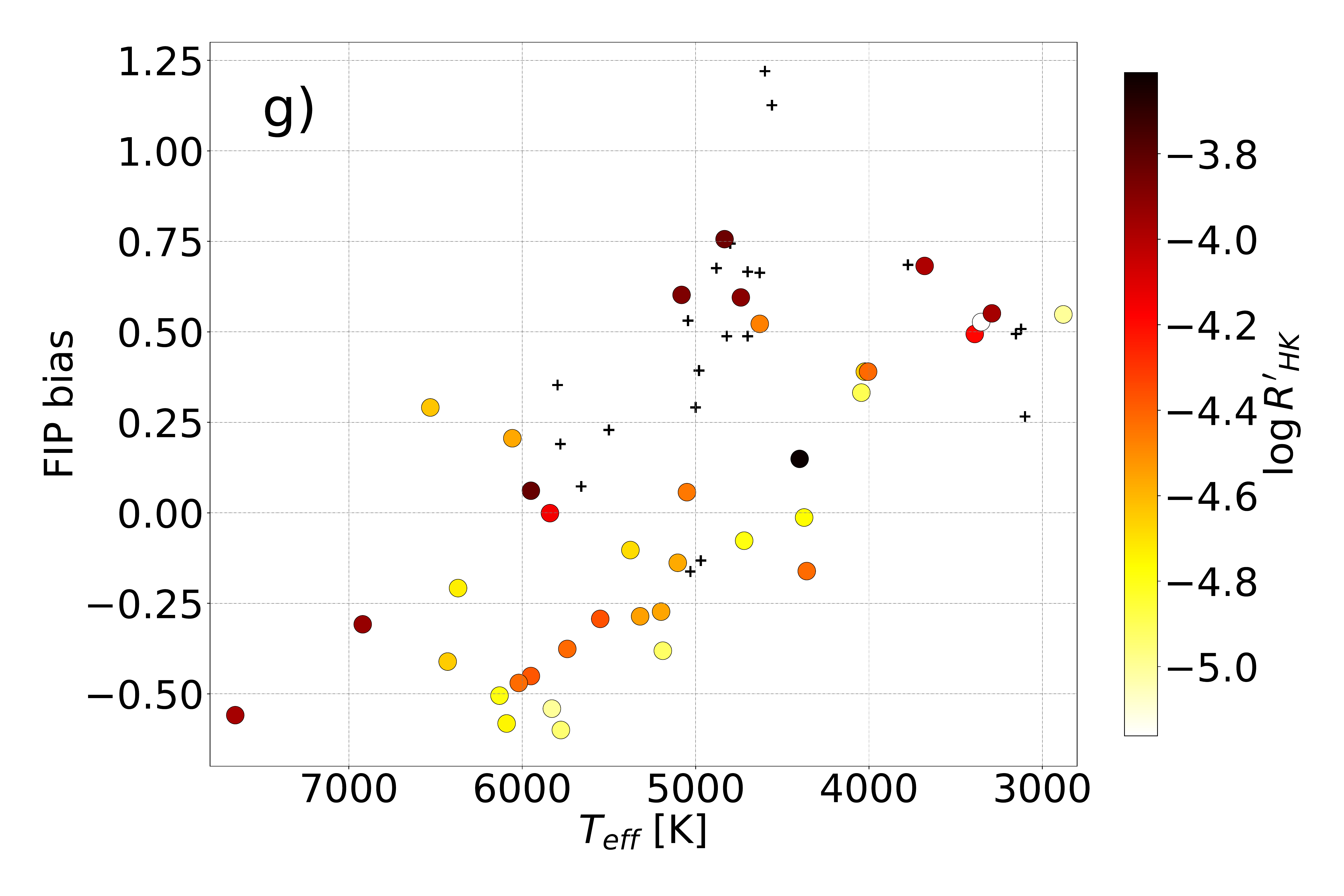}

\includegraphics[height=5.9cm, width=9.25cm]{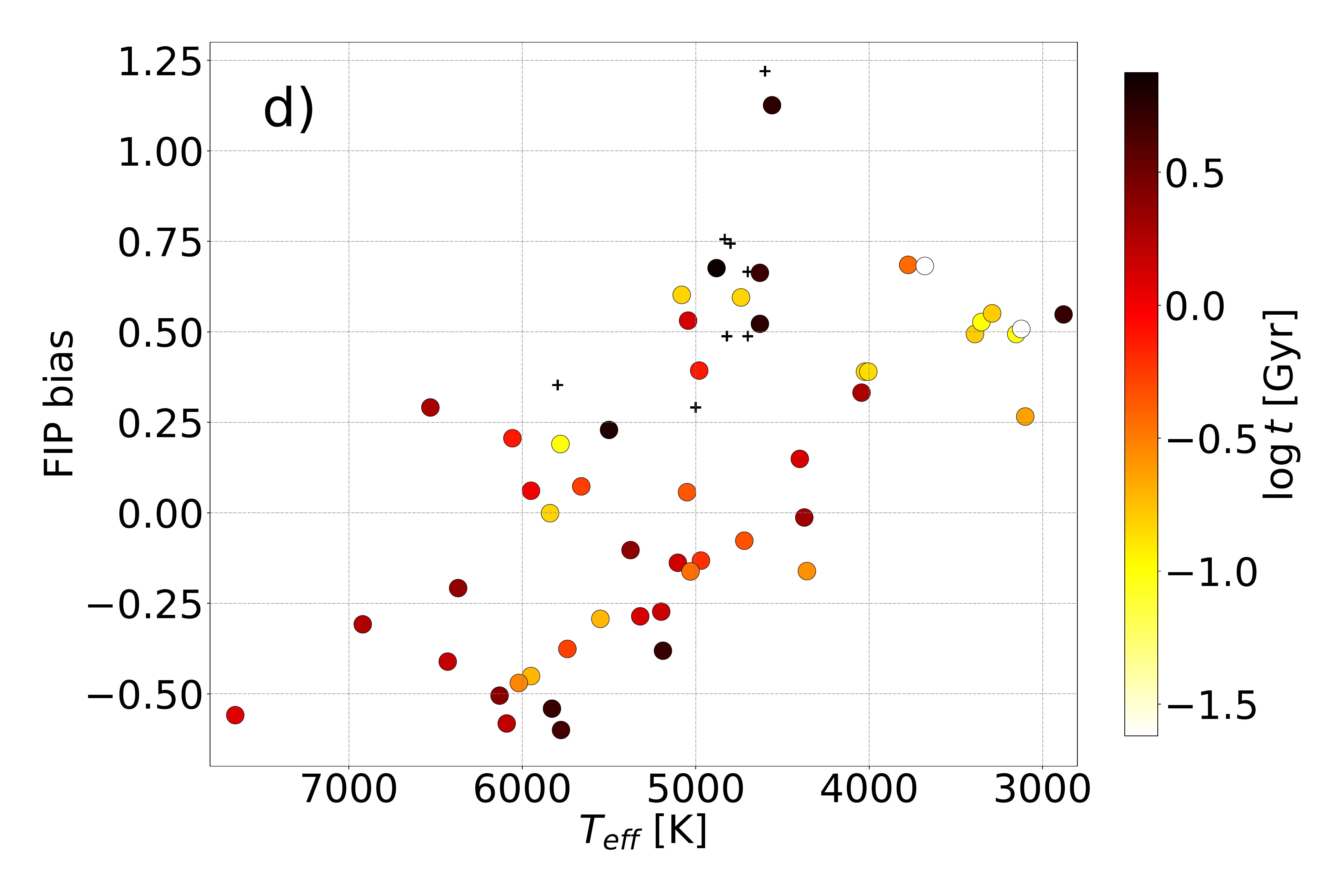}%
\hspace{-0.04cm}\includegraphics[height=5.9cm, width=9.00cm]{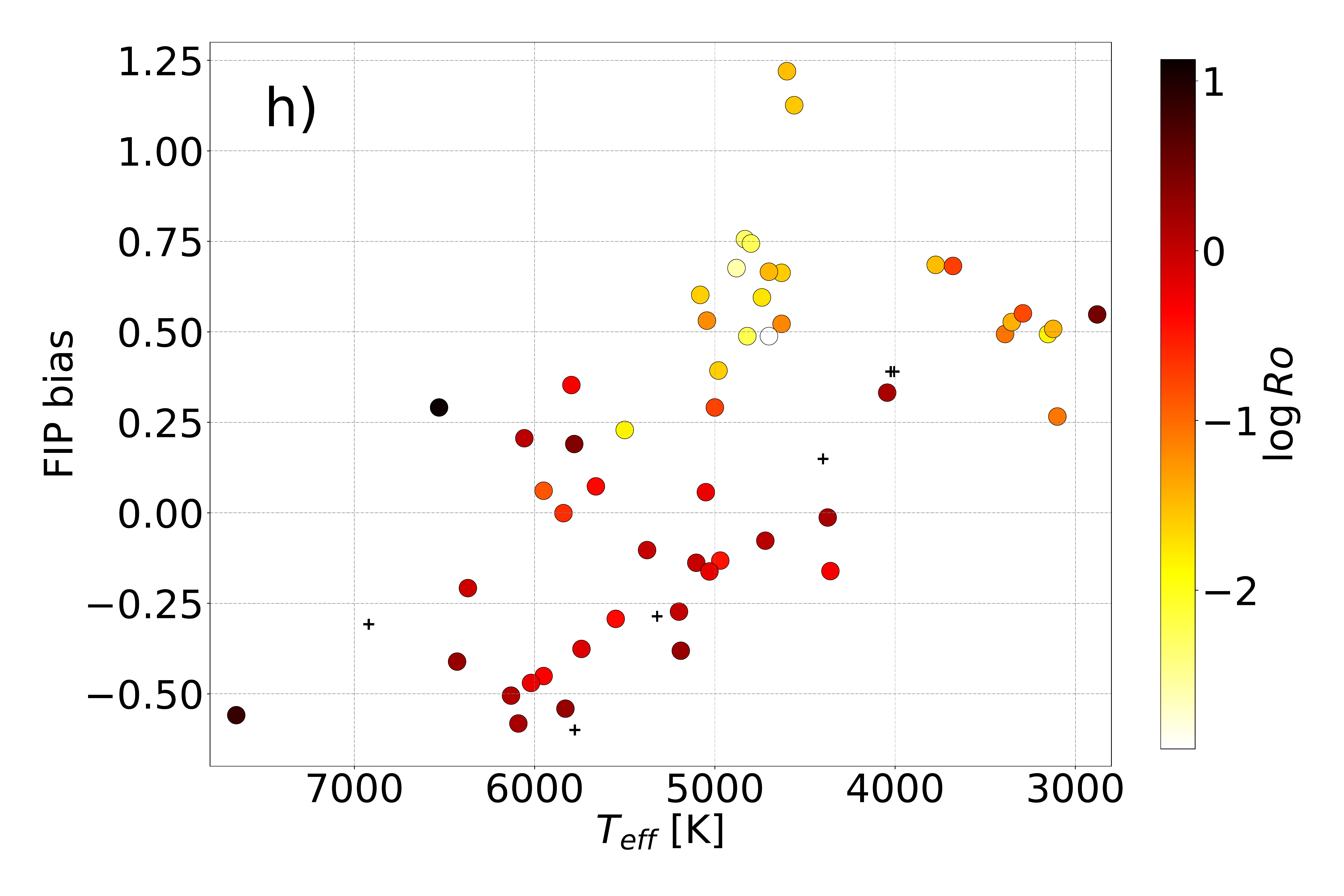}
\caption{\teff$-$FIP bias diagrams with data points coloured according to various parameters. Missing parameters are indicated with black crosses. See the text for the details.}
\label{fig:correlations_8}
\end{figure*}

\subsection{Contributions of the different stellar parameters to the structure of the \teff$-$FIP bias diagram}

To check the contribution of the different stellar parameters to the structure of the \teff$-$FIP bias diagram, we plotted several relations with the data points weighted by a given parameter. In the following we discuss which parameters have strong influence on the FIP-bias values of the sample, thus ultimately shape the diagram, whereas other parameters have seemingly no direct relation to the FIP bias and its bimodal distribution.

In the left panels of Fig.\,\ref{fig:correlations_8} from top to bottom, data points in the \teff$-$FIP bias diagram are coloured increasingly dark with increasing values of four stellar parameters: {\it a)} the $\log g$ surface gravity, {\it b)} the $\log L_X$ X-ray luminosity, {\it c)} the [Fe/H] metallicity, and {\it d)} the $\log t$ age. The evolved stars seem to form a separate, steep branch around \teff$\sim4500-5000$\,K, as seen from the points with light yellow hue accounting for their low $\log g$ values in Fig.\,\ref{fig:correlations_8}\,{\it a}. The giant stars $\mu$\,Vel, Capella and $\beta$\,Cet, all around \teff$\sim5000$\,K with FIP bias slightly below zero, have low gravities and high X-ray luminosities as seen in Fig.\,\ref{fig:correlations_8}\,{\it b}. 

Metallicity is close to solar for almost all stars; see Fig.\,\ref{fig:correlations_8}\,{\it c}. This parameter is uncertain anyway, and since all stars are close to the Sun in space, we cannot expect great differences. It looks that the observed age does not have an influence on FIP bias at all; see Fig.\,\ref{fig:correlations_8}\,{\it d}. However, we call attention to the uncertainty of the age determinations. The measured stellar parameters which are used for finding the age (like effective temperature or rotation) are modified in different extents by stellar activity originating from magnetic fields of different strengths. Additionally, gyrochronology is valid on the MS only, since off the MS the internal structure of the stars is changing fast \citep{Barnes2007}, therefore it is not possible to use it for the whole sample.

The MS sample consists of mostly young stars with only a few exceptions. Among the solar-type stars, only the two components of the $\alpha$\,Cen system are of similar age than the Sun, and the third component Proxima~Cen is among the M-type flare stars. From the 17 evolved stars five are older than 5\,Gyr and also five are younger than 1.4\,Gyr, the rest do not have derived ages. To find the age of the 10 close RS\,CVn systems in this group is problematic due to the non-trivial evolution of the close binaries with strong magnetic activity. If there exists any age effect on the FIP-bias values, we cannot explore it on the basis of the available data. 

In right panels of Fig.\,\ref{fig:correlations_8} from top top bottom, data points are coloured increasingly dark with increasing values of: {\it e)} the $\log v_\mathrm{rot}$ stellar rotational velocity, {\it f)} the $\log F_X$ X-ray flux, {\it g)} the \logrhk\ chromopheric activity index, and {\it h)} the $\log Ro$ Rossby number. Fast rotating stars (with higher rotational velocity), regardless of spectral type and evolutionary state, have higher FIP-bias values by about 0.5 at the same \teff. The evolved stars with their high rotational velocity make up the majority of the stars on the upper branch of the diagram. Apart from the evolved stars the three fast rotating K dwarfs (AB\,Dor, LO\,Peg and V471\,Tau), all F-stars except Procyon, and a few fast rotating G dwarfs are situated in the upper branch as well. From the evolved stars Capella and $\beta$\,Cet are the slowest rotators with 104 and 205 days periods, respectively, and for $\mu$\,Vel no rotational period is known; these stars lay on the original lower branch of the \teff$-$FIP bias diagram (see the black dots in Fig.\,\ref{fig:fip_names}). 

Together with the higher rotational velocities, the X-ray fluxes are also higher in the parallel, upper relation (Fig.\,\ref{fig:correlations_8}\,{\it f}) as follows from the rotation-activity relation. We note that fast rotating stars did not simply increase the scatter of the relation towards the higher FIP-bias values, but appear on the new branch with a clear separation from the original one, which was implicitly suggested by \citet[][see their Fig. 7a]{Wood2018}, depicting the Vaughan-Preston gap \citep{VPG1980}. For the fast rotating M dwarf stars the X-ray activity saturates \citep[see, e.g.][]{Magaudda2020}. $\mu$\,Vel, Capella and $\beta$\,Cet have large radii, these are three stars out of the four having larger that 10\,$R_{\odot}$. The high X-ray luminosity of $\mu$\,Vel, Capella and $\beta$\,Cet (Fig.\,\ref{fig:correlations_8}\,{\it b}) combined with their large radii result in lower X-ray fluxes which fit well in the lower, original branch of the relation. 

The important chromospheric activity parameter \logrhk\ is missing for many stars, especially for the evolved stars. However, as a direct activity index, \logrhk\ affects the FIP bias at least to some extent; see Fig.\,\ref{fig:correlations_8}\,{\it g}, and see the lower panel of Fig.\,\ref{fig:pca_explained_variance} for more details.
The Rossby numbers for the sample were calculated separately for MS \citep[][]{convective_turnover_time} and evolved stars \citep[][]{rossby_evolved} with different empirical methods, see Sect.\,\ref{sect:data}. From  Fig.\,\ref{fig:correlations_8}\,{\it h} it seems that stars with lower Rossby numbers have higher FIP-bias values (M-type MS and evolved stars). However, due to the non-uniform calculation of the Rossby number this should be taken with caution. The visible trends would be the same with the dynamo number $N_D$, if we assume a power law $N_D(Ro)$$\sim$$Ro^\alpha$ with constant exponent. The role of the Rossby number in interpreting the magnetic dynamo inside the stars of the sample is further discussed in Sect.\,\ref{sect:discussion}.

\begin{figure}
\includegraphics[width=\columnwidth]{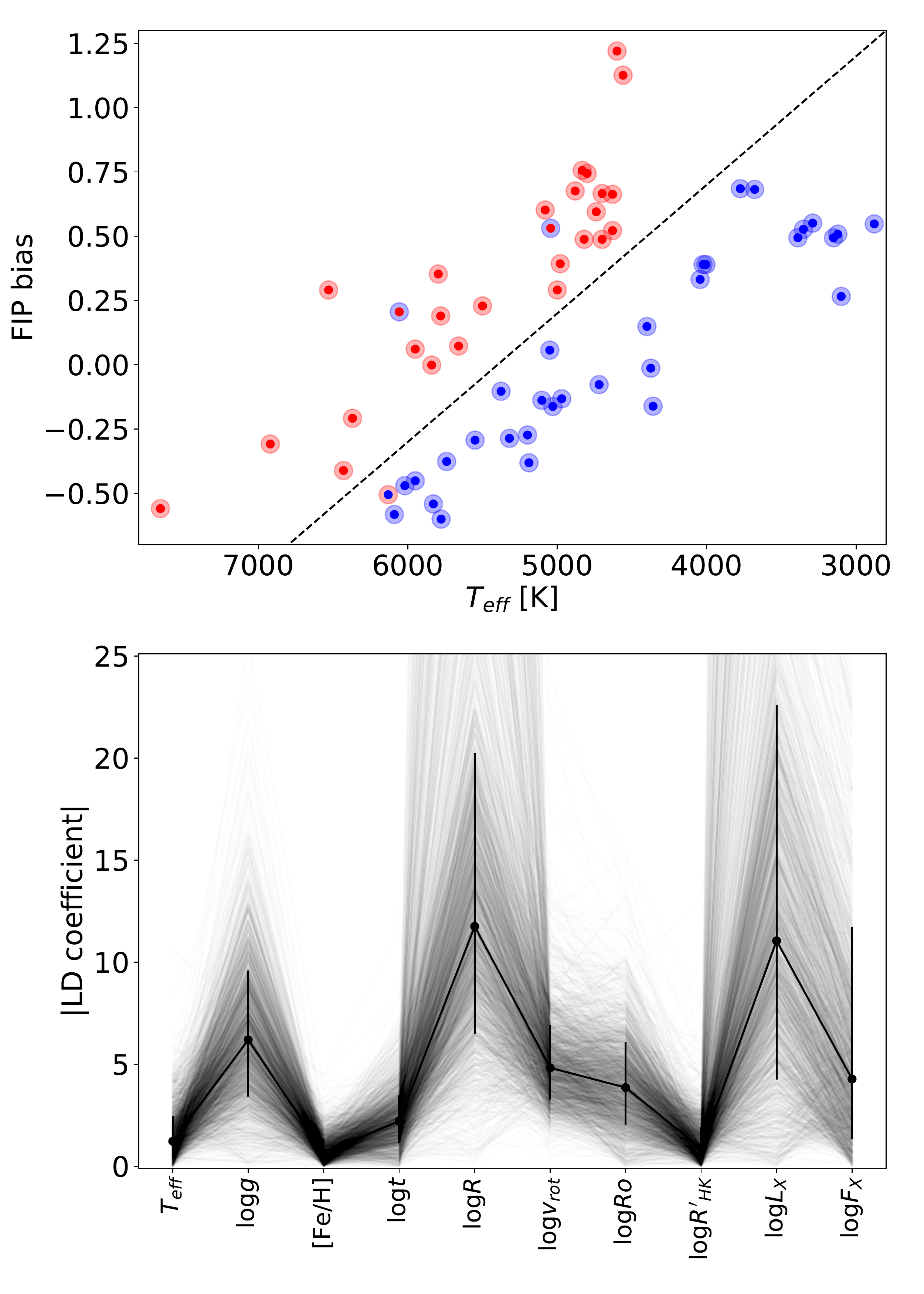}
\caption{Results of the linear discriminant analysis (LDA). \textit{Upper panel:} the pre-defined boundary between the two clusters (red and blue points) is shown with dashed lines on the \teff$-$FIP bias diagram. The larger circles indicate the clusters assigned by the LDA. The three misclassified stars are $\iota$\,Hor, HR\,7291 and AY\,Cet. \textit{Lower panel:} the absolute value of the LD components corresponding to the given parameters. Thin lines come from bootstrap resampling, error bars indicate the region between the 16th and 84th percentiles.}
\label{fig:lda}
\end{figure}

As a further insight on the importance of each parameter in the formation of the bimodality, the result of the LDA can be used. The upper panel of Fig.\,\ref{fig:lda} shows the linear boundary we used to a priori separate the two parallel sequences. The FIP bias was removed from the input parameters, so the LDA only used ten scaled features to predict the membership to the lower or upper sequence. The absolute values of the LDA coefficients are plotted in the lower panel of Fig.\,\ref{fig:lda}. Since the input parameters were scaled to have zero mean and unit variance, these coefficients can be treated as a sort of feature importance. The thin lines show the result of bootstrap analysis, where all the input points were resampled with replacement, in order to quantify the uncertainty of the method. It appears that the most important parameters separating the two sequences are the stellar radius and $\log{L_X}$. Naturally, these are correlated with $\log{g}$, $\log{F_X}$ and $\log{v_\mathrm{rot}}$. An interesting result is that the ratio of the coefficients corresponding to $\log{R}$ and $\log{g}$ is $\sim$2, indicating that $2\log{R}$+$\log{g}$$\sim$$\log{M}$ is a variable that separates the two branches, where $M$ is the stellar mass. The dependence on metallicity, Rossby number and age is negligible. The small contribution of \logrhk\ is due to the numerous missing values.

\subsection{Correlations between the parameters of the stars in the \teff$-$FIP bias diagram}

\begin{figure}
\includegraphics[width=\columnwidth]{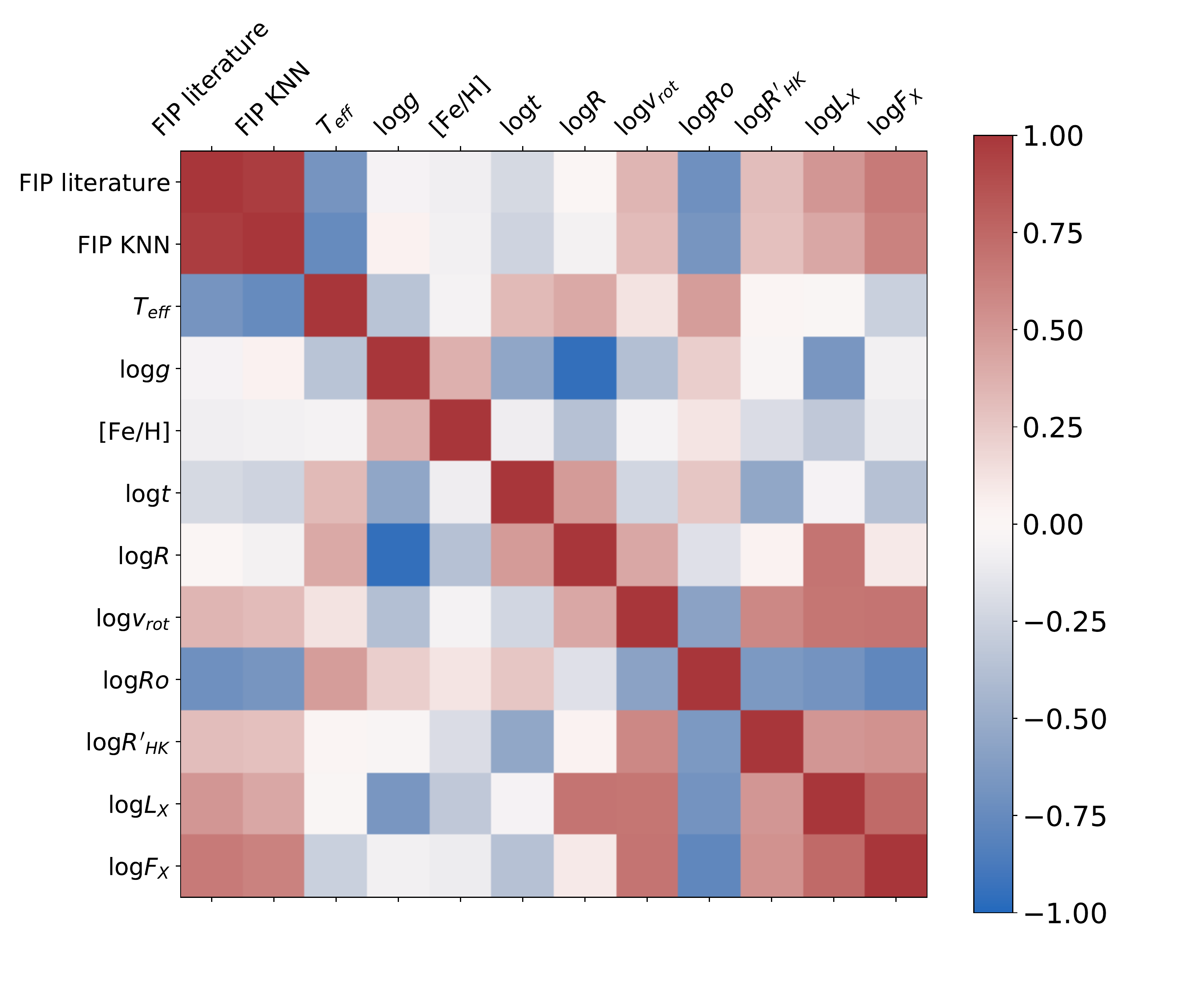}
\caption{Correlation matrix using Spearman correlation coefficients.}
\label{fig:matrix}
\end{figure}

As a start, we plotted the correlation matrix between the stellar parameters with Spearman correlation coefficients in Fig.\,\ref{fig:matrix}, showing the most direct linear correlations. One can see for example that $\log L_X$ is correlated with $\log v_\mathrm{rot}$. The FIP bias shows the strongest correlation with \teff, but $\log F_X$ also seems to be important.

\begin{figure}
\includegraphics[width=\columnwidth]{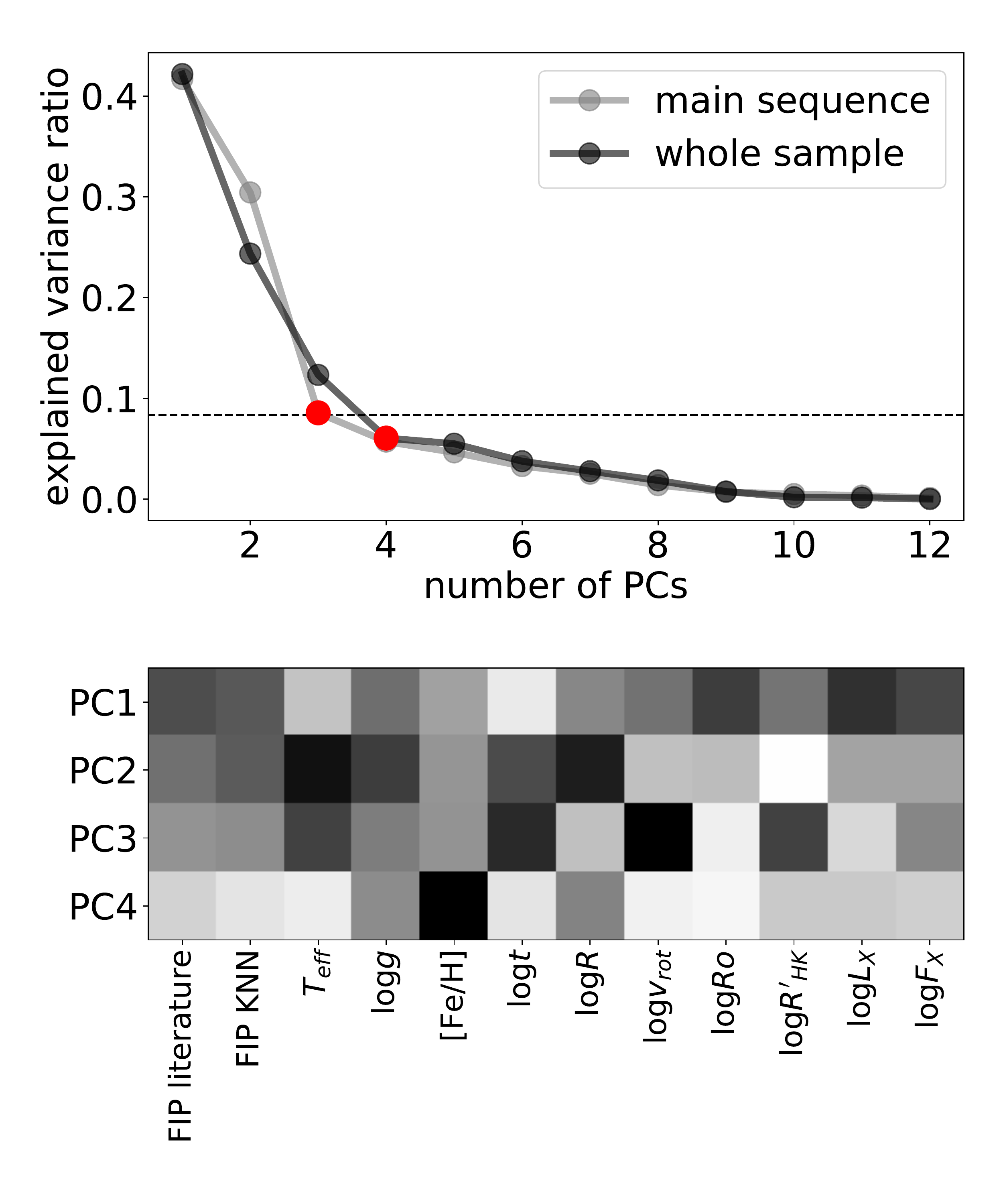}
\caption{\textit{Upper panel:} Explained variance ratio of each principal component (PC). The dashed line denotes the Kaiser criterion \citep{kaiser_criterion}, and the red points show the elbow points.
\textit{Lower panel:}  The importance of each parameter to a given principal component, with darker colours indicating stronger contribution (in absolute values).}
\label{fig:pca_explained_variance}
\end{figure}

Using all 12 parameters, we run PCA to represent the sample in a lower dimensional space, making it easier to look for emerging clusters. To find the optimal number of principal components (PCs) to keep, several heuristic methods can be used. Figure\,\ref{fig:pca_explained_variance} shows the explained variance ratio for different number of PCs, i.e. the fraction of the sample variance that a given component can explain. An elbow point (abrupt change in the slope) appears around three or four components, hinting that the optimal number should be no greater, as adding more parameters does not cause such a steep increase in information content. Indeed, using the first four PCs can explain $\sim$80\% of the sample variance. It is interesting (although not surprising) that the elbow point is clearly at $N$=3 for the MS stars, since there are parameters that do not change much if one excludes giants (e.g. $\log g$). The Kaiser criterion \citep{kaiser_criterion} states that one has to retain at least the PCs with eigenvalues larger than 1, which carry more information than an average single parameter from the original dataset (see the dashed line in Fig.~\ref{fig:pca_explained_variance}). In our case the eigenvalues drops below 1 after PC3. In the lower panel of Fig.\,\ref{fig:pca_explained_variance}, one can see how each of the PCs are made up. The following rough trends can be observed:
\begin{itemize}
    \item[-] PC1 contains parameters related to the corona, namely the FIP bias, $\log L_{\rm X}$, $\log F_{\rm X}$ and the Rossby number;
    \item[-] PC2 contains \teff\ and parameters related to the ,,size'' of the star, $\log g$ and radius;
    \item[-] PC3 contains \teff\, and activity-related parameters $v_\mathrm{rot}$ and \logrhk;
    \item[-] PC4 contains the otherwise unimportant metallicity
\end{itemize}

\begin{figure}
\includegraphics[width=\columnwidth]{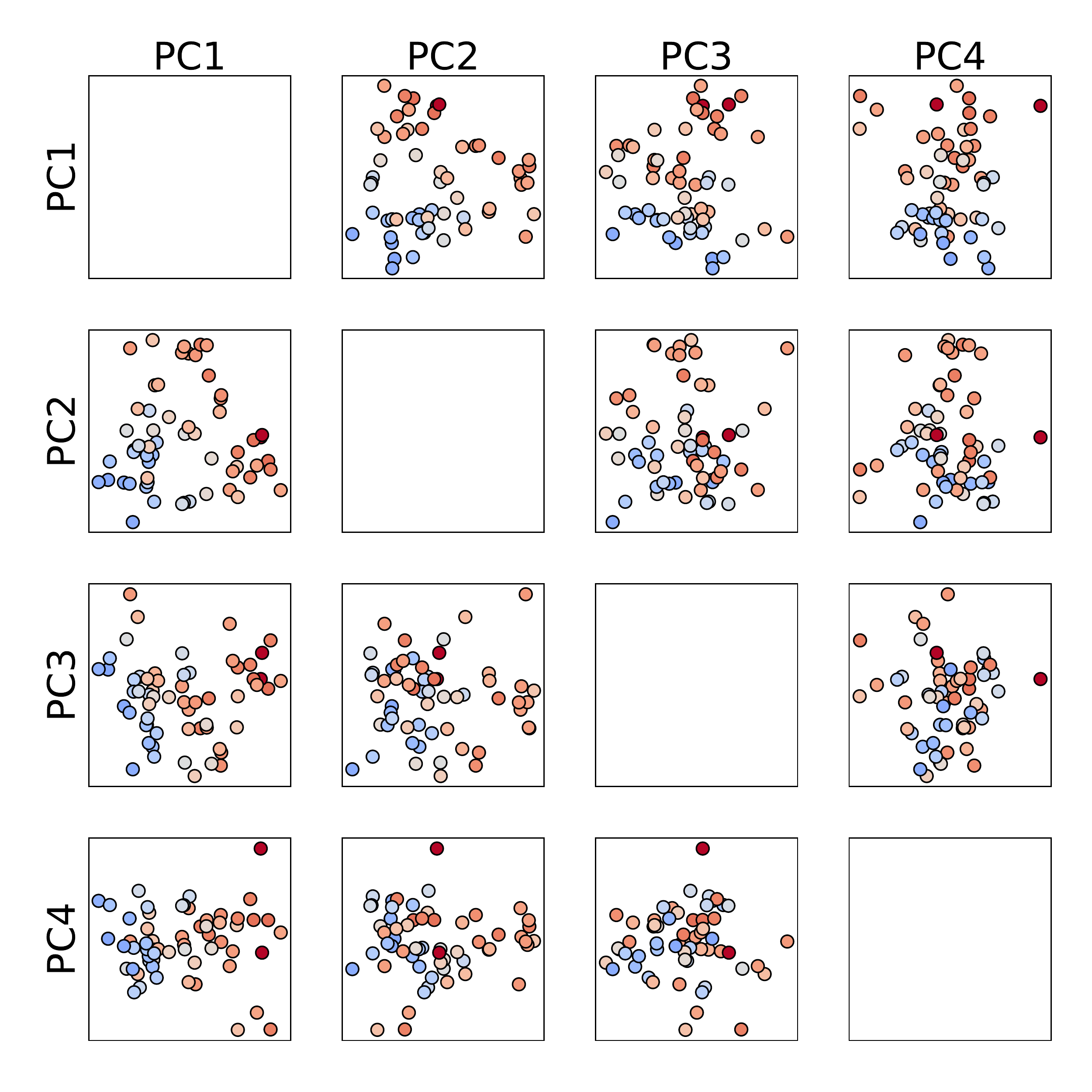}
\caption{The first four PCs, coloured with the FIP bias (see Fig.~\ref{fig:clustering} for the colour scale). The clear colour gradient shows the importance of the FIP bias parameter.}
\label{fig:pca_components}
\end{figure}

Figure\,\ref{fig:pca_components} shows two-dimensional projections in PC space. It is clearly seen from the colour gradient that the FIP bias plays an important role (red colour indicates positive, blue negative FIP bias).

\section{Discussion}\label{sect:discussion}
\subsection{On the bimodality of the \teff$-$FIP bias diagram}\label{sect:disc_bimod}
\begin{figure*}
\includegraphics[width=\textwidth]{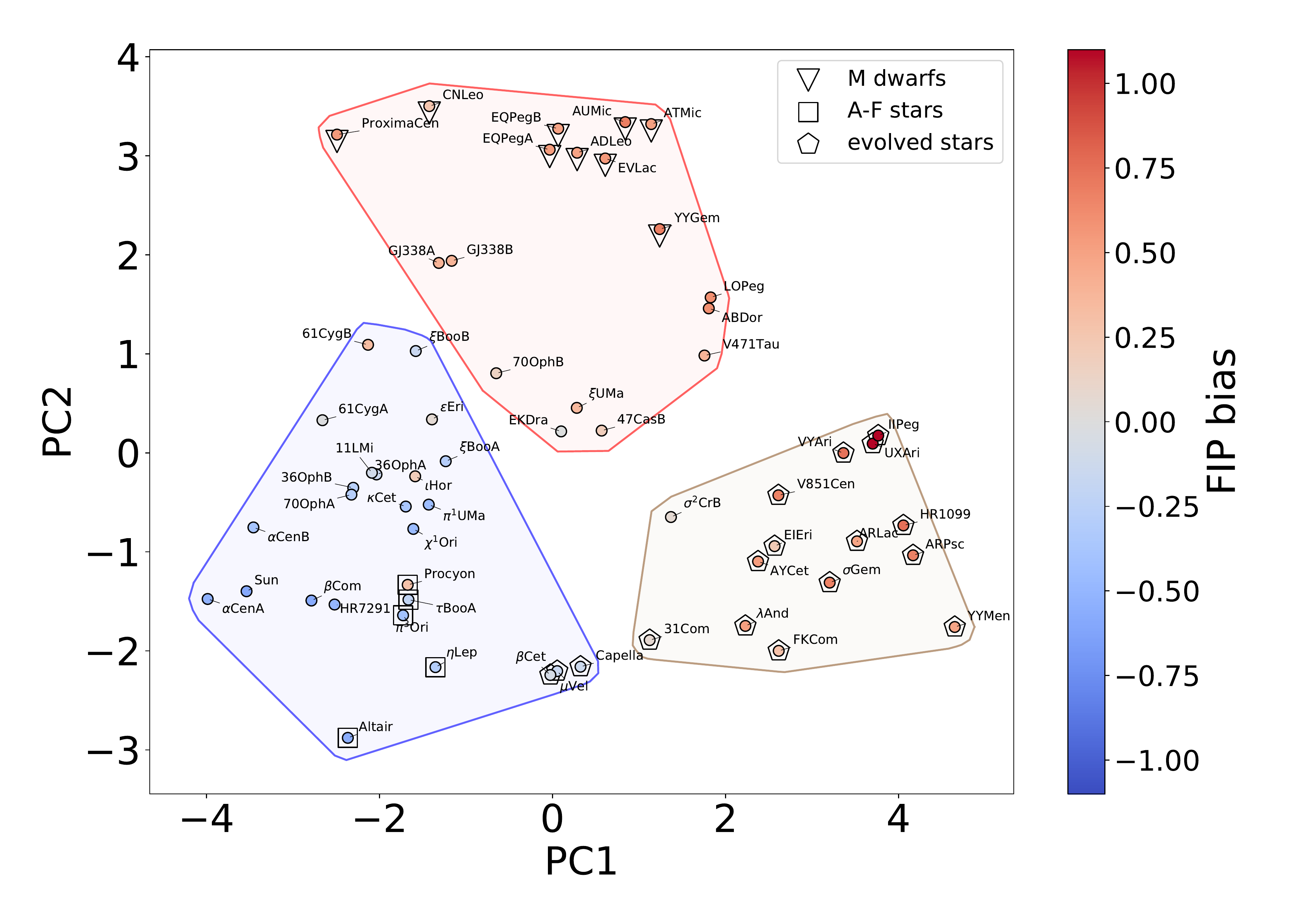}
\caption{The first two principal components, coloured (in the symbols' centre) with FIP bias. Special stars (evolved, M dwarfs and A--F stars) are indicated separately with different symbols. The clusters found with k-means are shown with red, blue and brown polygons.}
\label{fig:clustering}
\end{figure*}


\begin{figure}
\includegraphics[width=\columnwidth]{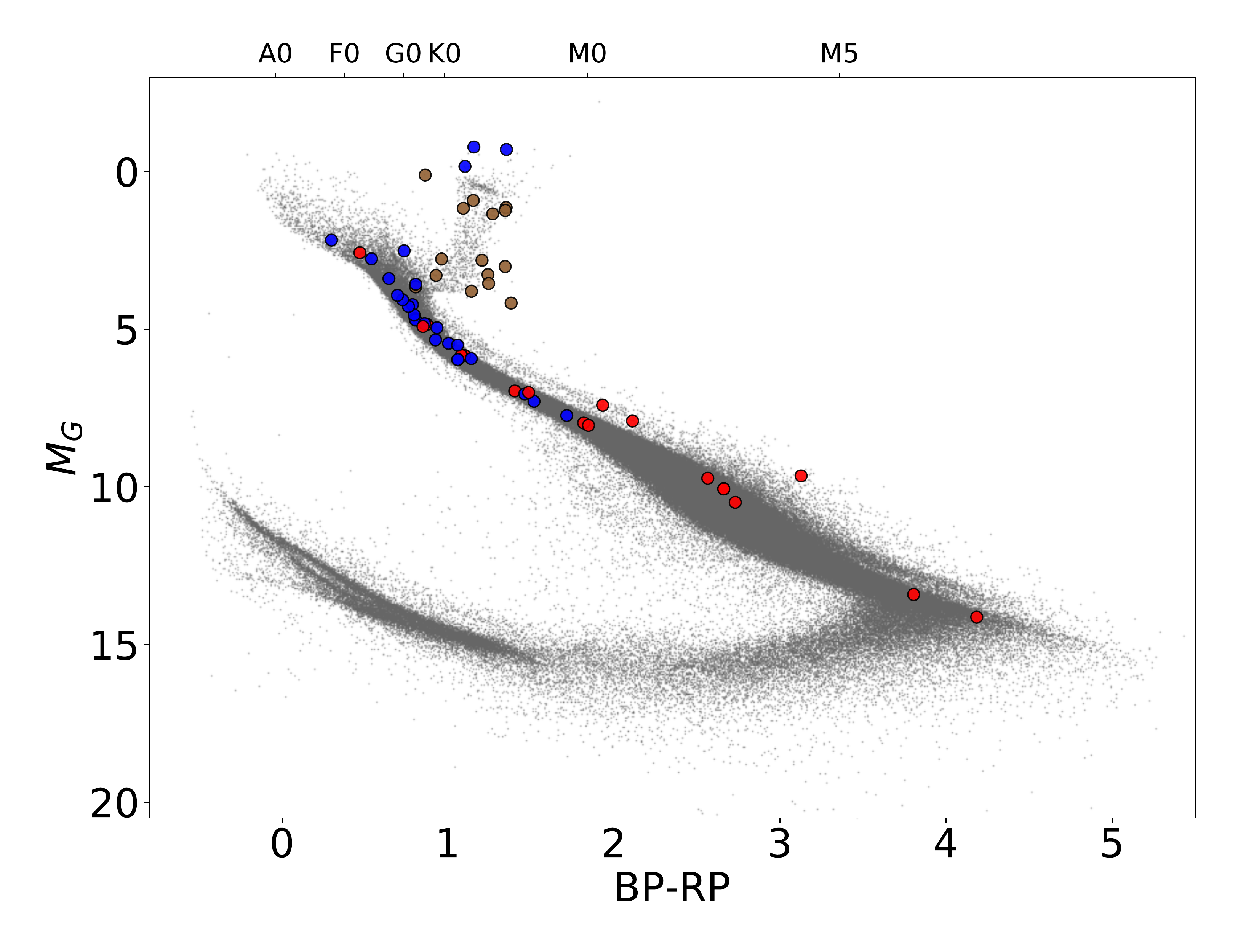}
\caption{Gaia EDR3 colour$-$magnitude diagram (within 100\,pc), showing the clusters from the k-means analysis from Fig.~\ref{fig:clustering}, using the same colour coding for the three clusters. Gaia magnitudes for stars brighter than $G=3^\mathrm{m}$ (e.g. the Sun) were estimated from $V$ and $I_C$ magnitudes, according to \citet{gaia_edr3}. The nearly equal mass M dwarf binaries YY\,Gem and AU\,Mic lie above the main sequence due to their binary nature. In case of AU\,Mic its youth also plays a role, the third star AT\,Mic is a PMS object, this triple is a member of the $\beta$\,Pic moving group.}
\label{fig:hrd_with_clusters}
\end{figure}

The \teff$-$FIP bias plots (see Fig.\,\ref{fig:fip_names} with the names of the stars indicated) show at least two distinct features, {\it i)} the blob of data points splits into two linear correlations, and {\it ii)} the part where M dwarfs are located seems to be separated. A question arises immediately, namely, is there a connection between the bimodal distribution of the FIP-bias values and the fact that as rotation increases, magnetic field increases and saturation can occur in the filling factor? Concerning the results of \citet{saar1996} and \citet{reiners2009}, it is possible that magnetic saturation occurs in the filling factor at higher rotation rates. According to \citet{baker2019} and \citet{laming2021}, higher filling factor leads to less volume for expansion with increasing altitude for the magnetic field, and IFIP is more likely to occur in conditions with minimal field expansion. A higher photospheric filling factor may also lead to an increase in subsurface magnetic reconnection, which appears to be linked to transient solar IFIP coronal abundances \citep{baker2019,2020ApJ...894...35B}. Figure\,\ref{fig:correlations_8}{\it e} shows that stars with higher FIP bias exhibit higher rotational velocities. However, this feature alone cannot explain the separation of the two branches in the diagrams.

The bimodality of the FIP diagram probably depicts the Vaughan$-$Preston gap, originally defined as a discontinuity in the activity index 'S' for FGK main sequence stars \citep{VPG1980}, i.e., medium activity stars are missing. The origin of this discontinuity is still actively debated.

Stellar rotation periods show bimodality in open clusters: the speed of the evolution for stars of similar age depends only on the mass of the stars; this is addressed by \citet{Barnes2010} depicting the rotational evolution of a star with age. The model finds a high speed transition from fast to slow rotation, thereby creating a void at medium rotational periods depending on the colour (temperature) and mass of the stars. As a consequence of the discontinuity of rotational periods and concerning the existing activity$-$rotation relations (faster rotation results in higher activity level), the bimodality occurs in the activity levels of open cluster members regarding their age, and the transition is not continuous, as observed by \citet{2009A&A...499L...9P}. However, this simple relation can be applied only to members of open clusters whose ages are known.

Unfortunately, we cannot use age as a parameter in finding the cause of the bimodality on the FIP bias diagrams. To derive the age of active field stars is complicated, if at all possible. The existing age estimates are usually quite uncertain due to the effect of strong and variable magnetism which is not taken into account in the stellar evolution models (for the evolved stars see more discussion below). Thus, only the stellar rotation remains as the parameter when characterizing the activity level through the known rotation$-$activity relations. In this work we apply the rotational velocities as a parameter, and for this we need to know the stellar radii as well. We use X-ray flux (surface units) as coronal activity measure. The bimodal distribution of the FIP-bias values is clear between the high and low rotational velocities, in line with the X-ray flux which is representative of the strength of magnetic activity; see Figs.\,\ref{fig:correlations_8}{\it e}-{\it f}.

\citet{2017NatCo...8..183B} have shown that the FIP-bias value of the Sun as a star changes with the solar cycle as much as 0.2. The magnetically active stars in the sample may have been at any point in their activity cycle at the time of measurement of their coronal abundances, which may increase the scatter in the \teff$-$FIP bias relation, and may narrow the gap between the two branches.

Our sample contains 17 evolved stars, most of them are members of close binaries, only four stars are single (see Table\,\ref{table:binarity}). From the evolved stars' sample 11 stars are active RS\,CVn binaries which have more complicated rotational evolution due to the gravitational and magnetic effects, making it impossible to get reliable age estimates. The active binaries with subgiant or giant primaries maintain fast rotation due to their binarity, and these stars make up the majority of the upper branch in the diagrams with fast rotating stars. The interested reader can find details about the theory of rotational evolution of low mass stars in close binaries, taking into account magnetic braking as well, by \citet{2019ApJ...881...88F} and references therein.

In Fig.\,\ref{fig:clustering} we present the result of a k-means clustering in the 4 dimensional parameter space (keeping only 4 PCs), projected into two dimensions. Based on the elbow method and the average silhouette score \citep[for an overview, see e.g.][]{kmeans_optimal_clusters}, the optimal number of k-means clusters is $k=3$, which resulted in physically interpretable groups. PC1 reflects mostly the contribution of the effects of the coronal parameters (FIP bias and X-ray flux), while PC2 depends mostly on the stellar parameters of radius and $\log g$, as displayed in Fig.\,\ref{fig:pca_explained_variance}. 
Figure\,\ref{fig:clustering} shows that the first cluster is formed by flare stars from late K to M type (light red background), the second one by MS stars from A to early K type (light blue background), and the third cluster by evolved stars (light brown background). The upper branch on the \teff$-$FIP bias diagram consists mainly of the stars from the third cluster, along with the fast rotating three K dwarfs and the fastest rotating solar-type stars (cf. Tables~\ref{table:binarity} and \ref{table:sample_params}), all of which situate close to the evolved stars in Fig.\,\ref{fig:clustering}. The three slowly rotating evolved stars ($\mu$\,Vel, Capella and $\beta$\,Cet) show up in the second cluster and lay in the lower branch in all of the \teff$-$FIP bias diagrams. To illustrate the three clusters, Fig.\,\ref{fig:hrd_with_clusters} shows the sample on the Gaia colour$-$magnitude diagram. We note that the same conclusion is reached with clustering in the original 12 dimensional parameter space.

\begin{figure}[th!]
\centering
\includegraphics[width=\columnwidth]{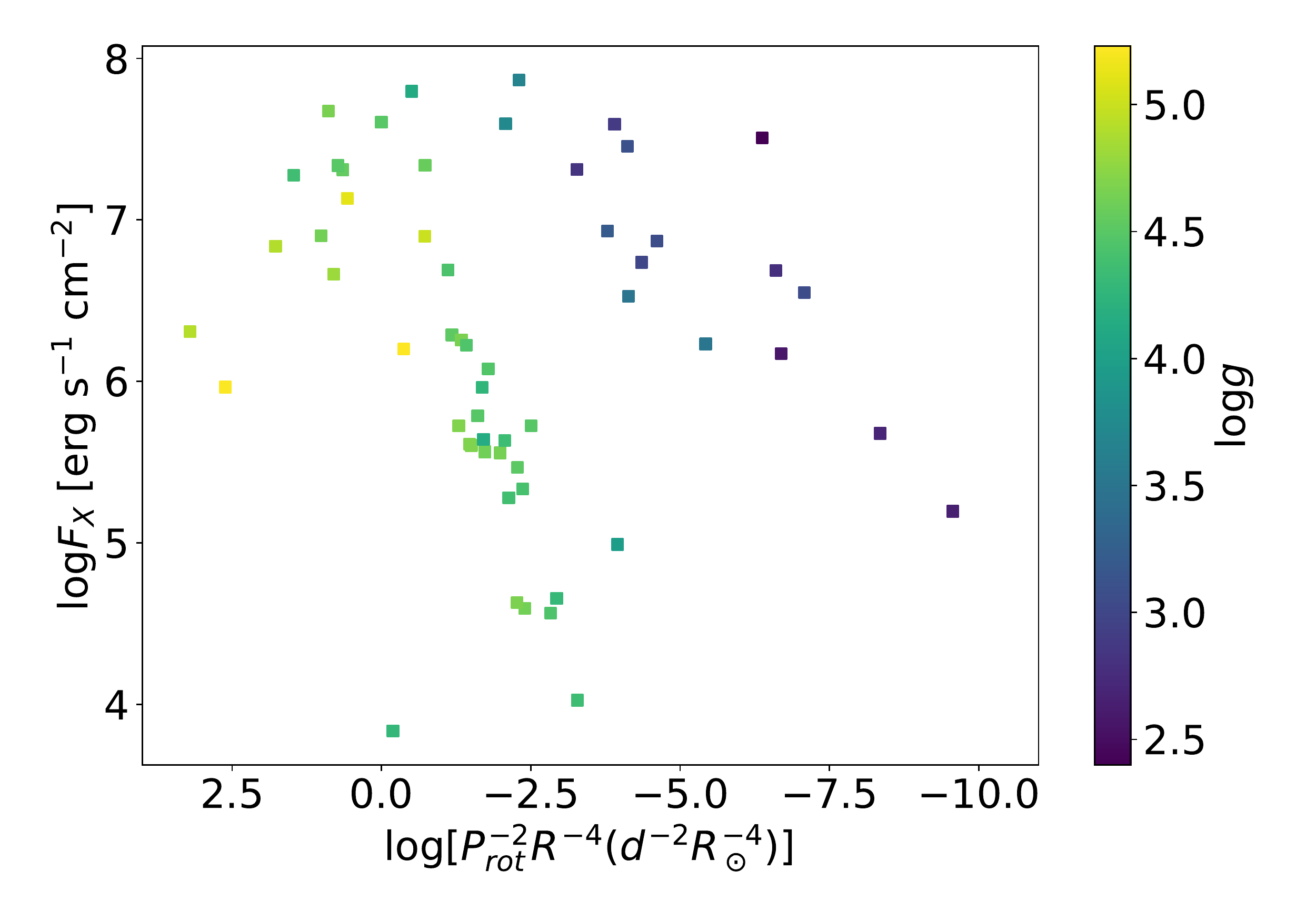}%

\includegraphics[width=\columnwidth]{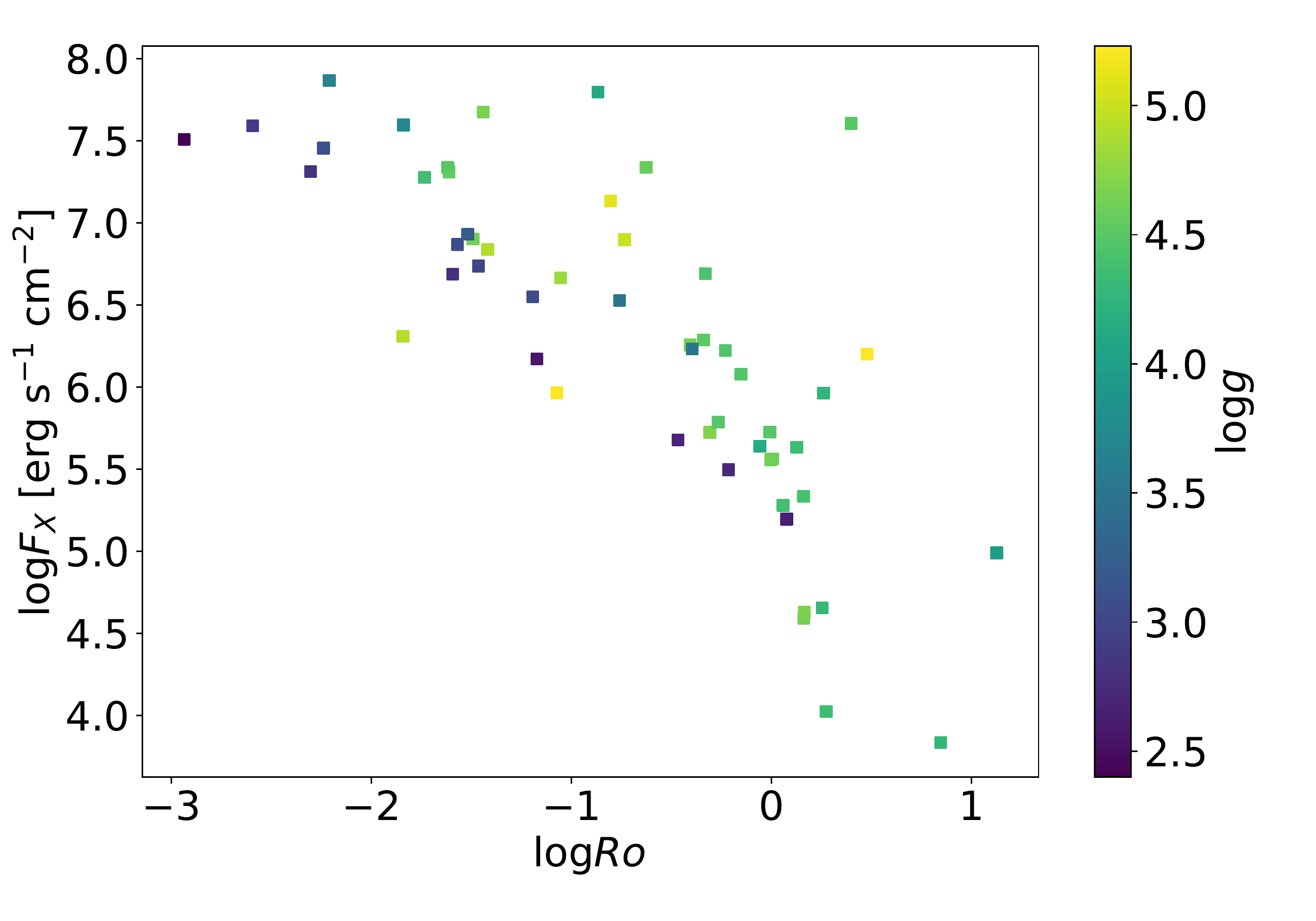}
\caption{\textit{Upper panel:} $\log F_X$ versus combined rotation period and stellar radius \textit{Lower panel:} $\log F_X$ versus Rossby number as in Figures 3b and 2a in \citet{2020NatAs...4..658L} but with $\log F_X$ instead of \logrhk. The colour bars in both panels indicate surface gravity.}
\label{fig:NatAst}
\end{figure}

\begin{figure}[th!]
\centering
\includegraphics[width=\columnwidth]{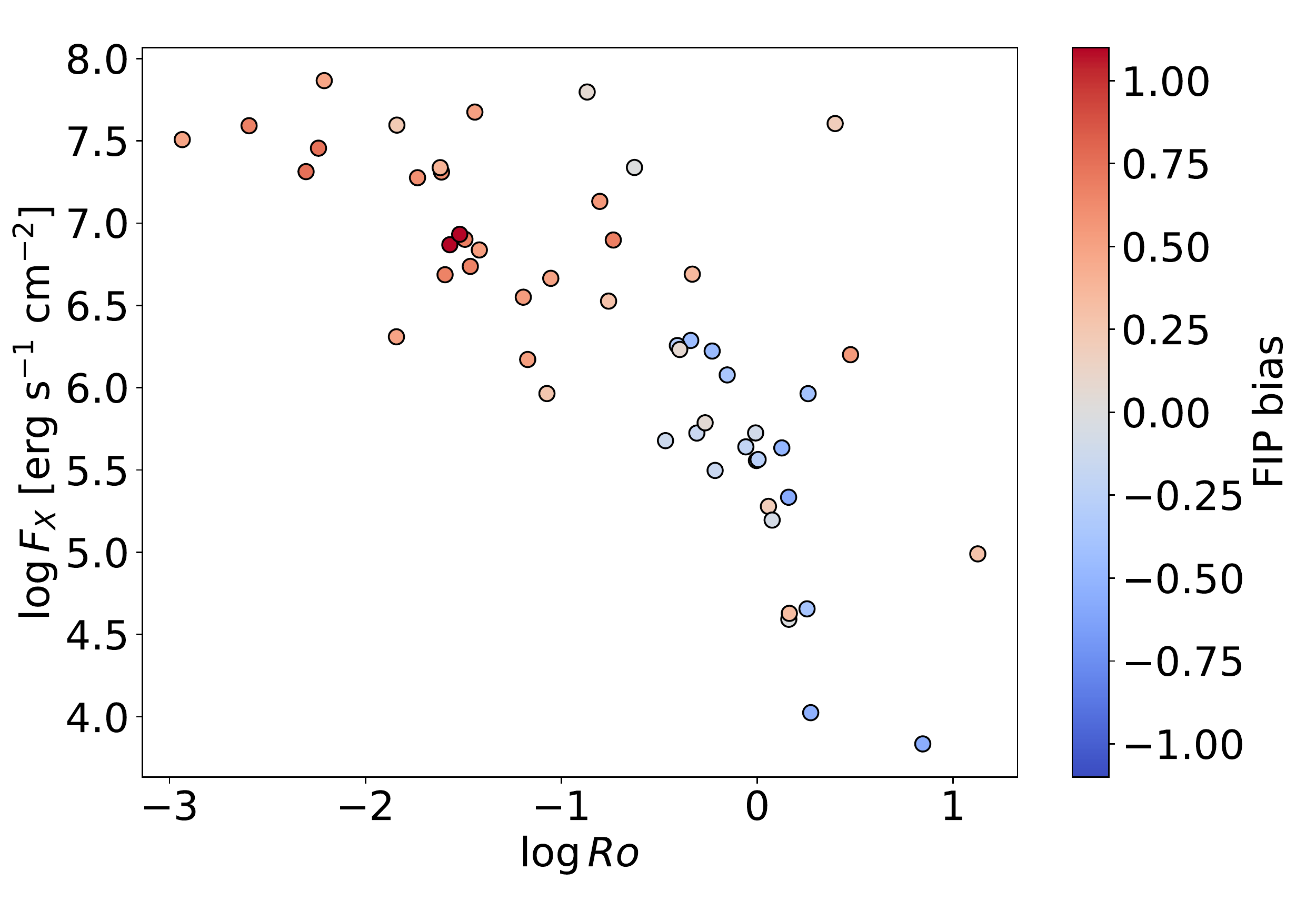}
\caption{X-ray flux versus Rossby number, coloured with the FIP bias.}
\label{fig:rossby_fip}
\end{figure}

\subsection{Clustering by different dynamo types?}\label{sect:disc_clustering}

The three clusters found in our heterogeneous active star sample shown in Fig.\,\ref{fig:clustering} may reflect the differences of the dynamo working inside these stars. Magnetic activity signatures can manifest in different ways on different parts of the Hertzsprung-Russell diagram. On the Sun, active regions are mostly concentrated to a $\pm 30^{\circ}$ latitude strip around the equator. On fast rotating solar-like stars, stellar spots tend to concentrate around the pole \citep[e.g., V1358\,Ori,][]{2019A&A...627A..52K}. Using surface flux transport models, \cite{2018A&A...620A.177I} found that this can indeed happen on faster rotating solar-like stars, due to the surface transport of emerging magnetic structures, which can be attributed to large scale surface flows (e.g., a strengthened meridional circulation) and the Coriolis-force, which can deflect the rising flux towards the poles.
On the other hand, fully convective M dwarfs can exhibit spot structures stable for a few thousand rotations \citep[e.g., V374\,Peg,][]{2016A&A...590A..11V}. Stars on the red giant branch can also have very different spot configurations, e.g., from time to time, polar spot structures of varying contrast were observed on $\zeta$\,And \citep{2007A&A...463.1071K,2012A&A...539A..50K,2016Natur.533..217R}. A similar RS\,CVn system, $\sigma$\,Gem also exhibits intermittent polar structures \citep[cf.][]{2015A&A...573A..98K,2021A&A...646A...6K}. Since spots are the most easily observed proxies of dynamo action, this may also mean that on different types of stars, we observe the manifestations of different types of dynamos.

The depths of the convection zones on the MS vary from the total stellar interior (full convection, low-mass M dwarfs) through different sizes of convection zone decreasing from fully convective to a very thin layer going from M, K, and G dwarfs to F-type stars, all with $\log g$ values of around 4.5$-$5. Both the depth of the convective region and speed of rotation have a bearing on differential rotation, which is crucial for the $\Omega$-effect, thus, this might point towards a more $\alpha$-effect dominated dynamo as the convective zone deepens. In case of evolved stars the convection zones are deep, but the surface gravity is much lower.

At first glance, one would expect the separation to be dependent on the Rossby number through activity, since it is connected to rotation, especially in light of the fact that $\log{L_\mathrm{X}}$ and $\log{F_\mathrm{X}}$ seem to play an important role here. However, it seems that the Rossby number does not. There could be a few reasons for that: the Rossby number was derived using different methods for MS and evolved stars using empirical relationships, and both rotation and the convective turnover time (e.g., through the depth of the convective envelope) can be drastically different on different parts of the Hertzsprung-Russell diagram. 

For young and evolved stars a common dynamo scaling has been proposed recently by \citet{2020NatAs...4..658L}. Following their approach, using the chromospheric activity index \logrhk\ we plotted the coronal activity index $\log F_X$ of the sample versus stellar rotation+radius and Rossby number in Fig.~\ref{fig:NatAst}. The result is very similar to that of \citet{2020NatAs...4..658L}: while there is a clear separation of $\log F_X$ between MS and evolved stars in the function of rotation combined with radius (free from $\tau_\mathrm{c}$), the use of Rossby number results in one relation regardless of evolutionary state of the stars. This result shows that the strength of coronal activity depends on the underlying dynamo.

The clustering of the stars in our sample points towards the possibility to further check the dynamos at work in the stars within the clusters. We note however, that the origin and working of the stellar dynamos are not well established theoretically to date, and has a lot of unknown effects due to various physical parameters. A thorough and detailed review of dynamos in active stars (including the Sun) is given by \citet{2017LRSP...14....4B} both from observational and theoretical points of view, and we direct the interested reader to consult this paper.

While the stars in the sample differ by dynamo type, the ways their coronae are heated could also be different (e.g., by waves, turbulence, braiding (nanoflares) and helicity conservation, see \citealt{coronal_heating}), which could have implications for the FIP effect.
Inverse FIP fractionation arises from upward propagating fast modes that are reflected/refracted back downwards (see e.g., \citealt{2017ApJ...844..153L}). As this happens in the chromosphere, the fractionation itself does not tell anything about how the corona is heated. Figure \ref{fig:rossby_fip} shows that as the Rossby number decreases, the coronal X-ray flux (depending on the strength of the magnetic field) increases, similar to the results of \cite{2015RSPTA.37340259T}. Stars with increasing magnetic strength/X-ray flux show FIP effect, which gradually turns to inverse FIP effect. The magnetic field eventually saturates. Although the stellar rotation/dynamo generates magnetic field, it is destroyed by reconnection before it can rise to the corona, and the waves produced by this reconnection give the inverse FIP fractionation.

\section{Summary and conclusions}

We present FIP-bias values for 59 stars extending the sample with active MS and evolved stars for which no FIP-bias values were available to date. The literature FIP-bias values are recalculated, and as a second set, new FIP-bias values are presented for all stars in the sample using homogeneously derived photospheric abundances based on LAMOST data. For a few stars we derived new photospheric abundances via spectral synthesis with SME. The main results can be summarized as follows:

\begin{itemize}
\item We did not find systematic difference among the FIP-bias values using the recalculated literature, KNN or SME photospheric abundances, except for the KNN values of the coolest stars due to the lack of such stars in the LAMOST training set.

\item We find that the \teff$-$FIP bias diagram is not a simple relation, but it now has two branches separated by $\sim$0.5 in FIP bias. The new (upper) branch consists of mostly evolved stars and additionally stars with high rotational velocity. The low-mass, flaring M dwarfs form a small blob in the continuation of the lower branch of the relation.

\item We suggest that the well-defined separation of the two branches in the \teff$-$FIP bias relation may be related to the bimodality in activity levels (the Vaughan-Preston gap), which in turn may be linked to a bimodality in rotation rates (periods) of stars of the same age but in different evolutionary stages due to different masses and/or binary evolution.

\item The separation of the two branches is observable through activity indicators like $v_\mathrm{rot}$, $\log{F_X}$, $Ro$ and \logrhk\ showing the importance of these parameters in the FIP bias of the stellar sample.

\item Statistical analysis of 12 stellar parameters reveals three clusters of stars in the sample. The evolved stars are well separated from the MS stars which are roughly divided by the sign of the FIP-bias values. The distribution of stars in the diagrams showing the clusters by means of distances between the stellar parameters and the linear discriminant analysis suggest that the FIP bias relates to the depths and structures of the convection zones of the stars indicated by their measured parameters. The properties (or the lack) of the convection zones have direct effect on the magnetic activity of the stars, and through this, on the FIP-bias values.
\end{itemize}

The results presented in this paper through investigating the effects of different stellar parameters widen our knowledge about the existing \teff$-$FIP bias relation. The extension of the sample to evolved stars may help to generalize this relation for different kinds of magnetically active stars in different evolutionary stages.

\begin{acknowledgements}
 We thank the referee for the useful suggestion concerning the connection between the (I)FIP and the heating of the stellar coronae. Konkoly Observatory, Budapest, Hungary hosted two workshops on Elemental Composition in Solar and Stellar Atmospheres (IFIPWS-1, 13-15 Feb, 2017 and IFIPWS-2, 27 Feb-1 Mar, 2018). The workshops have fostered collaboration by exploiting synergies in solar and stellar magnetic activity studies and exchanging experience and knowledge in both research fields. We thank G.\,Cs\"ornyei for the useful discussions and enlightening ideas regarding statistical methods. This work was supported by the Hungarian National Research, Development and Innovation Office grants NKFIH (OTKA) K-131508, KH-130526, NN129075, K129249 and by the NKFIH grant 2019-2.1.11-TÉT-2019-00056. Authors acknowledge the financial support of the Austrian-Hungarian Action Foundation (101\"ou13, 104\"ou2). LK acknowledges the financial support of the Hungarian National Research, Development and Innovation Office grant NKFIH PD-134784. LK is a Bolyai János Research Fellow.
JML was supported by the NASA Heliophysics Guest Investigator (80HQTR19T0029) and Supporting Research (80HQTR20T0076) programs, and by Basic Research FUnds of the Office of Naval Research.
DB is funded under STFC consolidated grant number ST/S000240/1 and LvDG is partially funded under the same grant.\\

\newline \textit{Software:} \texttt{python} \citep{python}, \texttt{matplotlib} \citep{matplotlib}, \texttt{numpy} \citep{numpy}, \texttt{scipy} \citep{scipy}, \texttt{pandas} \citep{pandas}, \texttt{sklearn} \citep{sklearn}.
\end{acknowledgements}

\bibliography{aa_fipforstars}

\clearpage

\begin{appendix} 

\onecolumn

\section{Tabulation of abundances for the FIP bias calculation}

\begin{table*}[h!]
\caption{Data for the calculation of the literature FIP bias.}\label{table:lit_abund_table}
\centering

\fontsize{7}{7}\selectfont

\begin{tabular}{c|c|c|c|c|c|c|c|c|c|c|c}
\hline\hline
\multicolumn{1}{c|}{Star} & 
\multicolumn{4}{c|}{Corona} & \multicolumn{4}{c|}{Photosphere} &
\multicolumn{1}{c|}{FIP bias} &
\multicolumn{1}{c|}{Corona} &
\multicolumn{1}{c}{Solar photosphere} \\

\multicolumn{1}{c|}{} &
\multicolumn{1}{c}{[C/Fe]} &
\multicolumn{1}{c}{[N/Fe]} &
\multicolumn{1}{c}{[O/Fe]} &
\multicolumn{1}{c|}{[Ne/Fe]} &
\multicolumn{1}{c}{[C/Fe]} &
\multicolumn{1}{c}{[N/Fe]} &
\multicolumn{1}{c}{[O/Fe]} &
\multicolumn{1}{c|}{[Ne/Fe]} &
\multicolumn{1}{c|}{} &
\multicolumn{1}{c|}{source} &
\multicolumn{1}{c}{source} \\
\hline
AD\,Leo & $1.45$ & $0.96$ & $1.64$ & $1.18$ & $0.93$ & $0.33$ & $1.19$ & $0.80$ & $0.49 \pm 0.11$ & (1) & (30) \\
CN\,Leo & $1.25$ & $0.64$ & $1.33$ & $0.76$ & $0.93$ & $0.33$ & $1.19$ & $0.80$ & $0.27 \pm 0.17$ & (2) Table 2 DEM1 & (24) \\
EQ\,Peg\,A & -- & $0.80$ & $1.68$ & $1.17$ & -- & $0.33$ & $1.19$ & $0.80$ & $0.53 \pm 0.07$ & (3) Table 7 Q & (25) \\
EQ\,Peg\,B & -- & $0.82$ & $1.61$ & $1.13$ & -- & $0.33$ & $1.19$ & $0.80$ & $0.49 \pm 0.08$ & (3) Table 7 Q & (25) \\
Prox\,Cen & -- & $1.13$ & $1.54$ & $1.04$ & -- & $0.33$ & $1.19$ & $0.80$ & $0.55 \pm 0.30$ & (3) Table 10 Q & (25) \\
EV\,Lac & -- & $0.93$ & $1.69$ & $1.10$ & -- & $0.33$ & $1.19$ & $0.80$ & $0.55 \pm 0.15$ & (3)Table 10 Q & (25) \\
YY\,Gem & -- & $1.12$ & $1.75$ & $1.26$ & -- & $0.33$ & $1.19$ & $0.80$ & $0.69 \pm 0.17$ & (3) Table 10 Q & (25) \\
AU\,Mic & $1.63$ & $1.19$ & $1.77$ & $1.39$ & $0.93$ & $0.33$ & $1.19$ & $0.80$ & $0.68 \pm 0.13$ & (1) & (30) \\
AT\,Mic & $1.42$ & $0.82$ & $1.55$ & $1.16$ & $0.93$ & $0.33$ & $1.19$ & $0.80$ & $0.51 \pm 0.07$ & (4) Table 4 3T model & (24), O, Fe (26) \\
AB\,Dor & $1.50$ & $1.06$ & $1.75$ & $1.35$ & $0.93$ & $0.33$ & $1.19$ & $0.80$ & $0.60 \pm 0.09$ & (1) & (30) \\
LO\,Peg & $1.50$ & $0.84$ & $1.61$ & $1.35$ & $0.93$ & $0.33$ & $1.19$ & $0.80$ & $0.59 \pm 0.07$ & (5) Table 1 Q & (26) \\
V471\,Tau & -- & -- & $1.57$ & $1.04$ & -- & -- & $1.19$ & $0.80$ & $0.39 \pm 0.10$ & (6) Table 4 & (26), O (27) \\
36\,Oph\,A & $0.57$ & $0.44$ & $1.19$ & $0.88$ & $0.83$ & $0.49$ & $1.35$ & $0.96$ & $-0.14 \pm 0.09$ & (1) & (30), ph. a. (33) \\
36\,Oph\,B & $0.44$ & $0.33$ & $1.10$ & $0.67$ & $0.83$ & $0.49$ & $1.35$ & $0.96$ & $-0.27 \pm 0.10$ & (1) & (30), ph. a. (33) \\
$\xi$ Boo\,B & $0.78$ & -- & $1.19$ & $0.97$ & $1.09$ & -- & $1.36$ & $0.97$ & $-0.16 \pm 0.15$ & (1) & (30), ph. a. (33) \\
61\,Cyg\,A & $0.95$ & $0.34$ & $1.12$ & $0.79$ & $0.93$ & $0.33$ & $1.19$ & $0.80$ & $-0.01 \pm 0.04$ & (1) & (30), ph. a. (35) \\
61\,Cyg\,B & $1.26$ & -- & $1.60$ & $1.06$ & $0.93$ & -- & $1.19$ & $0.80$ & $0.33 \pm 0.08$ & (1) & (30), ph. a. (35) \\
70\,Oph\,A & $0.64$ & $0.18$ & $0.92$ & $0.56$ & $0.88$ & $0.41$ & $1.27$ & $0.88$ & $-0.29 \pm 0.06$ & (1) & (30), ph. a. (33) \\
70\,Oph\,B & $0.98$ & $0.70$ & $1.35$ & $1.01$ & $0.88$ & $0.41$ & $1.27$ & $0.88$ & $0.15 \pm 0.10$ & (1) & (30), ph. a. (33) \\
$\epsilon$ Eri & $0.90$ & $0.42$ & $1.20$ & $0.84$ & $0.75$ & $0.35$ & $1.21$ & $0.82$ & $0.06 \pm 0.07$ & (1) & (30), ph. a. (33) \\
$\alpha$ Cen\,A & $0.72$ & $0.06$ & $0.35$ & $0.41$ & $0.99$ & $0.46$ & $1.32$ & $0.93$ & $-0.54 \pm 0.30$ & (1) low activity & (30), ph. a. (33) \\
$\alpha$ Cen\,B & $0.70$ & $0.14$ & $0.87$ & $0.33$ & $0.94$ & $0.43$ & $1.29$ & $0.90$ & $-0.38 \pm 0.15$ & (1) low activity & (30), ph. a. (33) \\
$\pi^1$ UMa & $0.54$ & $0.36$ & $0.84$ & $0.48$ & $1.07$ & $0.54$ & $1.40$ & $1.01$ & $-0.45 \pm 0.18$ & (1) & (30), ph. a. (33) \\
EK\,Dra & $0.98$ & $0.19$ & $1.12$ & $0.96$ & $0.93$ & $0.33$ & $1.19$ & $0.80$ & $-0.00 \pm 0.13$ & (1) & (30) \\
$\xi$ Boo\,A & $0.75$ & $0.20$ & $1.11$ & $0.69$ & $1.09$ & $0.50$ & $1.36$ & $0.97$ & $-0.29 \pm 0.04$ & (1) & (30), ph. a. (33) \\
$\chi^1$ Ori & $0.35$ & $-0.33$ & $0.64$ & $0.49$ & $0.99$ & $0.48$ & $1.34$ & $0.56$ & $-0.47 \pm 0.33$ & (7) & (24), Fe (28), ph. a. (29) \\
$\kappa$ Cet & $0.42$ & $-0.12$ & $0.67$ & $0.61$ & $1.03$ & $0.48$ & $1.34$ & $0.56$ & $-0.38 \pm 0.34$ & (7) & (24), Fe (28), ph. a. (29) \\
$\beta$ Com & $0.49$ & $-0.53$ & $0.59$ & $0.07$ & $0.96$ & $0.46$ & $1.32$ & $0.54$ & $-0.58 \pm 0.25$ & (7) & (24), Fe (28), ph. a. (29) \\
47\,Cas\,B & $0.83$ & $0.39$ & $1.03$ & $1.03$ & $0.93$ & $0.33$ & $1.19$ & $0.41$ & $0.19 \pm 0.36$ & (7) & (24), Fe (28) \\
$\iota$ Hor & $0.75$ & $0.43$ & $1.31$ & $0.79$ & $0.95$ & $0.30$ & $1.16$ & $0.38$ & $0.21 \pm 0.25$ & (8) & (30) \\
11\,LMi & $0.97$ & $0.11$ & $1.09$ & $0.33$ & $0.93$ & $0.33$ & $1.19$ & $0.80$ & $-0.10 \pm 0.22$ & (9) Table 3 & (30) \\
HR\,7291 & -- & -- & $0.68$ & $0.13$ & -- & -- & $1.19$ & $0.80$ & $-0.51 \pm 0.12$ & (9) Table 4 & (30) \\
$\sigma^2$ CrB & -- & $0.35$ & $1.17$ & $0.74$ & -- & $0.33$ & $1.19$ & $0.80$ & $0.06 \pm 0.04$ & (10) Table 5 Q & (26) \\
$\xi$ UMa & -- & -- & $1.42$ & $1.11$ & -- & -- & $1.19$ & $0.80$ & $0.35 \pm 0.05$ & (11) Table 3 & (26) \\
HR\,1099 & $1.50$ & $1.16$ & $1.87$ & $1.41$ & $0.93$ & $0.33$ & $1.19$ & $0.80$ & $0.76 \pm 0.12$ & (12) Table 4 APEC & (24), Fe (28) \\
UX\,Ari & $1.95$ & $1.72$ & $2.03$ & $1.72$ & $0.93$ & $0.33$ & $1.19$ & $0.80$ & $1.13 \pm 0.25$ & (12) Table 4 APEC & (24), Fe (28) \\
$\lambda$ And & $1.28$ & $0.74$ & $1.67$ & $1.31$ & $0.93$ & $0.33$ & $1.19$ & $0.80$ & $0.52 \pm 0.07$ & (12) Table 4 APEC & (24), Fe (28) \\
VY\,Ari & $1.57$ & $1.11$ & $1.78$ & $1.44$ & $0.93$ & $0.33$ & $1.19$ & $0.80$ & $0.74 \pm 0.09$ & (12) Table 4 APEC & (24), Fe (28) \\
Capella & $0.66$ & $0.40$ & $0.93$ & $0.40$ & $0.93$ & $0.33$ & $1.19$ & $0.80$ & $-0.13 \pm 0.20$ & (12) Table 4 APEC & (24), Fe (28) \\
$\sigma$ Gem & -- & $1.21$ & $1.61$ & $1.24$ & -- & $0.33$ & $1.19$ & $0.80$ & $0.66 \pm 0.26$ & (13) Table 4 & rel. to same solar \\
31\,Com & -- & -- & $1.31$ & $0.66$ & -- & -- & $1.19$ & $0.80$ & $0.07 \pm 0.19$ & (14) Table 5 & (25) \\
$\mu$ Vel & -- & -- & $1.06$ & $0.44$ & -- & -- & $1.19$ & $0.80$ & $-0.16 \pm 0.16$ & (14) Table 5, ObsID 3410 & (25) \\
$\beta$ Cet & -- & -- & $1.13$ & $0.54$ & -- & -- & $1.19$ & $0.80$ & $-0.08 \pm 0.14$ & (14) Table 5 & (25) \\
FK\,Com & -- & -- & $1.35$ & $1.06$ & -- & -- & $1.19$ & $0.80$ & $0.29 \pm 0.07$ & (15) Table 7 & (24) \\
YY\,Men & $1.14$ & $1.19$ & $1.41$ & $1.13$ & $0.93$ & $0.33$ & $1.19$ & $0.80$ & $0.49 \pm 0.31$ & (16) Table 2 XMM Method 1 & (26) \\
EI\,Eri & $0.10$ & $0.25$ & $0.15$ & $0.08$ & $0.00$ & $0.00$ & $0.00$ & $0.00$ & $0.23 \pm 0.08$ & (17) Figure 5 & (30) \\
V851\,Cen & $1.19$ & $1.05$ & $1.84$ & $1.50$ & $0.93$ & $0.33$ & $1.19$ & $0.80$ & $0.67 \pm 0.22$ & (18) Table 2 & (24) \\
AR\,Psc & $1.74$ & $1.00$ & $1.70$ & $1.18$ & $0.93$ & $0.33$ & $1.19$ & $0.80$ & $0.68 \pm 0.19$ & (19) Table 5 & (25) \\
AY\,Cet & $1.70$ & $1.01$ & $1.37$ & $0.96$ & $0.93$ & $0.33$ & $1.19$ & $0.80$ & $0.53 \pm 0.32$ & (19) Table 5 & (25) \\
II\,Peg & -- & $1.33$ & $2.47$ & $1.93$ & -- & $0.33$ & $1.19$ & $0.80$ & $1.22 \pm 0.14$ & (20) Table 4 Q & (24) \\
AR\,Lac & -- & $0.93$ & $1.51$ & $1.10$ & -- & $0.33$ & $1.19$ & $0.80$ & $0.49 \pm 0.17$ & (21) Table 3 & (24) \\
$\eta$ Lep & $0.78$ & -- & $1.09$ & $0.55$ & $1.03$ & -- & $1.35$ & $0.96$ & $-0.31 \pm 0.09$ & (1) & (30), ph. a. (32) \\
$\pi^3$ Ori & $0.79$ & $0.11$ & $0.92$ & $0.46$ & $1.12$ & $0.49$ & $1.35$ & $0.96$ & $-0.41 \pm 0.07$ & (1) & (30), ph. a. (33) \\
$\tau$ Boo\,A & $0.83$ & -- & $0.91$ & $0.50$ & $0.93$ & -- & $1.16$ & $0.77$ & $-0.21 \pm 0.09$ & (1) & (30), ph. a. (34) \\
Procyon & $1.20$ & $0.68$ & $1.43$ & $0.77$ & $0.93$ & $0.33$ & $1.19$ & $0.80$ & $0.29 \pm 0.17$ & (22) Table 4 & (24),  Fe=7.51 \\
Altair & $0.95$ & $0.32$ & $0.96$ & $0.26$ & $1.06$ & $0.89$ & $1.75$ & $1.36$ & $-0.56 \pm 0.42$ & (23) Table 2 MOS/RGS & (26), ph. a. (31) \\
\hline       
\end{tabular}
\tablefoot{Coronal abundances are taken in quiescent (Q) or low activity state, where available. The original assumed solar photospheric values are indicated in the last column, which were all converted to \cite{asplund2009} and Ne from \cite{drake_testa_neon}. If there were available stellar photospheric abundance (ph. a.) measurements for a star, the source is also given in the last column. The FIP bias and its uncertainty are calculated as the mean and standard deviation of [X/Fe]$_\mathrm{cor}-$[X/Fe]$_\mathrm{phot}$, where X can be C, N, O and Ne. The FIP bias column is corrected by +0.084 according to \cite{Wood2018}, except for the stars listing (1) as source. As discussed in Sect.~\ref{sect:data}, GJ\,338\,AB is omitted.\\
\textbf{References.} (1) \cite{Wood2018}, (2) \cite{Fuhrmeister2007}, (3) \cite{Liefke2008}, (4) \cite{Robrade2005}, (5) \cite{Lalitha2017}, (6) \cite{Garcia-Alvarez2005}, (7) \cite{telleschi2005}, (8) \cite{Sanz-Forcada2019}, (9) \cite{Peretz2015}, (10) \cite{Osten2003}, (11) \cite{corona_xi_uma}, (12) \cite{audard2003}, (13) \cite{huenemoerder2013}, (14) \cite{garcia-alvarez2006}, (15) \cite{Gondoin2002}, (16) \cite{Audard_etal_2004}, (17) \cite{corona_ei_eri}, (18) \cite{corona_v851_cen}, (19) \cite{corona_ar_psc_ay_cet}, (20) \cite{corona_ii_peg}, (21) \cite{corona_ar_lac}, (22) \cite{Raassen2002}, (23) \cite{Robrade2009}, (24) \cite{Anders_Grevesse1989}, (25) \cite{Asplund2005}, (26) \cite{Grevesse_Sauval1998}, (27) \cite{AllendePrieto2001}, (28) \cite{Grevesse_Sauval1999}, (29) \cite{AllendePrieto2006}, (30) \cite{asplund2009}, (31) \cite{Erspamer2003}, (32) \cite{Yuce2011}, (33) \cite{AllendePrieto2004}, (34) \cite{Gonzalez2007}, (35) \cite{Jofre2015}}
\end{table*}

\newpage
\begin{table*}[th!]
\caption{Data for the calculation of the KNN FIP bias.}\label{table:knn_abund_table}
\centering
\tiny
\begin{tabular}{c|c|c|c|c|c|c|c|c|c|c}
\hline\hline
\multicolumn{1}{c|}{Star} & 
\multicolumn{4}{c|}{Corona} & \multicolumn{4}{c|}{Photosphere} &
\multicolumn{1}{c|}{FIP bias} &
\multicolumn{1}{c}{Corona} \\

\multicolumn{1}{c|}{} &
\multicolumn{1}{c}{[C/Fe]} &
\multicolumn{1}{c}{[N/Fe]} &
\multicolumn{1}{c}{[O/Fe]} &
\multicolumn{1}{c|}{[Ne/Fe]} &
\multicolumn{1}{c}{[C/Fe]} &
\multicolumn{1}{c}{[N/Fe]} &
\multicolumn{1}{c}{[O/Fe]} &
\multicolumn{1}{c|}{[Ne/Fe]} &
\multicolumn{1}{c|}{} &
\multicolumn{1}{c}{source} \\
\hline
AD\,Leo & $1.45$ & $0.96$ & $1.64$ & $1.18$ & $0.58$ & $0.33$ & $0.87$ & $0.48$ & $0.74 \pm 0.10$ & (1) \\
CN\,Leo & $1.25$ & $0.64$ & $1.33$ & $0.76$ & $0.53$ & $0.32$ & $0.84$ & $0.80$ & $0.46 \pm 0.32$ & (2) Table 2 DEM1 \\
EQ\,Peg\,A & -- & $0.80$ & $1.68$ & $1.17$ & -- & $0.32$ & $0.84$ & $0.80$ & $0.65 \pm 0.25$ & (3) Table 7 Q \\
EQ\,Peg\,B & -- & $0.82$ & $1.61$ & $1.13$ & -- & $0.32$ & $0.84$ & $0.80$ & $0.61 \pm 0.22$ & (3) Table 7 Q \\
Prox\,Cen & -- & $1.13$ & $1.54$ & $1.04$ & -- & $0.32$ & $0.84$ & $0.80$ & $0.67 \pm 0.31$ & (3) Table 10 Q \\
EV\,Lac & -- & $0.93$ & $1.69$ & $1.10$ & -- & $0.32$ & $0.91$ & $0.80$ & $0.65 \pm 0.24$ & (3)Table 10 Q \\
YY\,Gem & -- & $1.12$ & $1.75$ & $1.26$ & -- & $0.32$ & $0.86$ & $0.80$ & $0.80 \pm 0.23$ & (3) Table 10 Q \\
AU\,Mic & $1.63$ & $1.19$ & $1.77$ & $1.39$ & $0.57$ & $0.32$ & $0.88$ & $0.49$ & $0.93 \pm 0.08$ & (1) \\
AT\,Mic & $1.42$ & $0.82$ & $1.55$ & $1.16$ & $0.52$ & $0.32$ & $0.84$ & $0.80$ & $0.70 \pm 0.24$ & (4) Table 4 3T model \\
AB\,Dor & $1.50$ & $1.06$ & $1.75$ & $1.35$ & $0.89$ & $0.36$ & $1.43$ & $1.04$ & $0.49 \pm 0.20$ & (1) \\
LO\,Peg & $1.50$ & $0.84$ & $1.61$ & $1.35$ & $1.07$ & $0.23$ & $1.69$ & $0.80$ & $0.46 \pm 0.31$ & (5) Table 1 Q \\
V471\,Tau & -- & -- & $1.57$ & $1.43$ & -- & -- & $1.25$ & $0.86$ & $0.53 \pm 0.17$ & (6) Table 4 \\
36\,Oph\,A & $0.57$ & $0.44$ & $1.19$ & $0.88$ & $0.90$ & $0.37$ & $1.47$ & $1.08$ & $-0.18 \pm 0.18$ & (1) \\
36\,Oph\,B & $0.44$ & $0.33$ & $1.10$ & $0.67$ & $0.87$ & $0.37$ & $1.41$ & $1.02$ & $-0.28 \pm 0.17$ & (1) \\
$\xi$ Boo\,B & $0.78$ & -- & $1.19$ & $0.97$ & $0.68$ & $0.31$ & $1.08$ & $0.70$ & $0.16 \pm 0.10$ & (1) \\
61\,Cyg\,A & $0.95$ & $0.34$ & $1.12$ & $0.79$ & $0.70$ & $0.29$ & $1.14$ & $0.76$ & $0.08 \pm 0.12$ & (1) \\
61\,Cyg\,B & $1.26$ & -- & $1.60$ & $1.06$ & $0.61$ & $0.33$ & $1.04$ & $0.66$ & $0.54 \pm 0.12$ & (1) \\
70\,Oph\,A & $0.64$ & $0.18$ & $0.92$ & $0.56$ & $0.80$ & $0.41$ & $1.30$ & $0.92$ & $-0.28 \pm 0.11$ & (1) \\
70\,Oph\,B & $0.98$ & $0.70$ & $1.35$ & $1.01$ & $0.69$ & $0.35$ & $1.08$ & $0.70$ & $0.31 \pm 0.04$ & (1) \\
$\epsilon$ Eri & $0.90$ & $0.42$ & $1.20$ & $0.84$ & $0.83$ & $0.39$ & $1.34$ & $0.96$ & $-0.04 \pm 0.11$ & (1) \\
$\alpha$ Cen\,A & $0.72$ & $0.06$ & $0.35$ & $0.41$ & $0.83$ & $0.52$ & $1.22$ & $0.84$ & $-0.47 \pm 0.31$ & (1) low activity \\
$\alpha$ Cen\,B & $0.70$ & $0.14$ & $0.87$ & $0.33$ & $0.79$ & $0.54$ & $1.21$ & $0.83$ & $-0.33 \pm 0.17$ & (1) low activity \\
$\pi^1$ UMa & $0.54$ & $0.36$ & $0.84$ & $0.48$ & $0.91$ & $0.37$ & $1.30$ & $0.91$ & $-0.32 \pm 0.21$ & (1) \\
EK\,Dra & $0.98$ & $0.19$ & $1.12$ & $0.96$ & $0.84$ & $0.39$ & $1.29$ & $0.90$ & $-0.04 \pm 0.17$ & (1) \\
$\xi$ Boo\,A & $0.75$ & $0.20$ & $1.11$ & $0.69$ & $0.85$ & $0.38$ & $1.34$ & $0.95$ & $-0.19 \pm 0.07$ & (1) \\
$\chi^1$ Ori & $0.35$ & $-0.33$ & $0.64$ & $0.88$ & $0.91$ & $0.36$ & $1.28$ & $0.89$ & $-0.39 \pm 0.31$ & (7) \\
$\kappa$ Cet & $0.42$ & $-0.12$ & $0.67$ & $1.00$ & $0.84$ & $0.39$ & $1.29$ & $0.91$ & $-0.28 \pm 0.32$ & (7) \\
$\beta$ Com & $0.49$ & $-0.53$ & $0.59$ & $0.46$ & $0.92$ & $0.36$ & $1.27$ & $0.88$ & $-0.52 \pm 0.23$ & (7) \\
47\,Cas\,B & $0.83$ & $0.39$ & $1.03$ & $1.42$ & $0.86$ & $0.39$ & $1.29$ & $0.90$ & $0.14 \pm 0.33$ & (7) \\
$\iota$ Hor & $0.75$ & $0.43$ & $1.31$ & $1.18$ & $0.88$ & $0.44$ & $1.25$ & $0.86$ & $0.15 \pm 0.19$ & (8) \\
11\,LMi & $0.97$ & $0.11$ & $1.09$ & $0.33$ & $0.78$ & $0.58$ & $1.18$ & $0.80$ & $-0.12 \pm 0.32$ & (9) Table 3 \\
HR\,7291 & -- & -- & $0.68$ & $0.13$ & -- & -- & $1.25$ & $0.80$ & $-0.53 \pm 0.08$ & (9) Table 4 \\
$\sigma^2$ CrB & -- & $0.35$ & $1.17$ & $0.74$ & -- & $0.40$ & $1.28$ & $0.80$ & $0.01 \pm 0.03$ & (10) Table 5 Q \\
$\xi$ UMa & -- & -- & $1.42$ & $1.50$ & $0.94$ & $0.34$ & $1.34$ & $0.96$ & $0.39 \pm 0.33$ & (11) Table 3 \\
HR\,1099 & $1.50$ & $1.16$ & $1.87$ & $1.41$ & $0.78$ & $0.45$ & $1.19$ & $0.80$ & $0.76 \pm 0.05$ & (12) Table 4 APEC \\
UX\,Ari & $1.95$ & $1.72$ & $2.03$ & $1.72$ & $0.80$ & $0.45$ & $1.21$ & $0.80$ & $1.12 \pm 0.21$ & (12) Table 4 APEC \\
$\lambda$ And & $1.28$ & $0.74$ & $1.67$ & $1.31$ & $0.87$ & $0.37$ & $1.38$ & $0.80$ & $0.48 \pm 0.09$ & (12) Table 4 APEC \\
VY\,Ari & $1.57$ & $1.11$ & $1.78$ & $1.44$ & $0.80$ & $0.43$ & $1.22$ & $0.80$ & $0.75 \pm 0.09$ & (12) Table 4 APEC \\
Capella & $0.66$ & $0.40$ & $0.93$ & $0.40$ & $0.62$ & $0.64$ & $1.09$ & $0.80$ & $-0.11 \pm 0.19$ & (12) Table 4 APEC \\
$\sigma$ Gem & -- & $1.21$ & $1.61$ & $1.24$ & -- & $0.48$ & $1.22$ & $0.80$ & $0.60 \pm 0.19$ & (13) Table 4 \\
31\,Com & -- & -- & $1.31$ & $0.66$ & -- & -- & $1.24$ & $0.80$ & $0.05 \pm 0.15$ & (14) Table 5 \\
$\mu$ Vel & -- & -- & $1.06$ & $0.44$ & -- & -- & $1.16$ & $0.80$ & $-0.15 \pm 0.18$ & (14) Table 5, ObsID 3410 \\
$\beta$ Cet & -- & -- & $1.13$ & $0.54$ & -- & -- & $1.10$ & $0.80$ & $-0.03 \pm 0.21$ & (14) Table 5 \\
FK\,Com & -- & -- & $1.35$ & $1.06$ & -- & -- & $1.17$ & $0.80$ & $0.30 \pm 0.06$ & (15) Table 7 \\
YY\,Men & $1.14$ & $1.19$ & $1.41$ & $1.13$ & $0.81$ & $0.45$ & $1.26$ & $0.80$ & $0.47 \pm 0.25$ & (16) Table 2 XMM Method 1 \\
EI\,Eri & $1.03$ & $0.58$ & $1.34$ & $1.27$ & $0.77$ & $0.46$ & $1.21$ & $0.82$ & $0.32 \pm 0.15$ & (17) Figure 5 \\
V851\,Cen & $1.19$ & $1.05$ & $1.84$ & $1.89$ & $0.80$ & $0.44$ & $1.24$ & $0.85$ & $0.74 \pm 0.27$ & (18) Table 2 \\
AR\,Psc & $1.74$ & $1.00$ & $1.70$ & $1.57$ & $0.67$ & $0.54$ & $1.09$ & $0.70$ & $0.83 \pm 0.27$ & (19) Table 5 \\
AY\,Cet & $1.70$ & $1.01$ & $1.37$ & $1.35$ & $0.76$ & $0.51$ & $1.27$ & $0.88$ & $0.59 \pm 0.34$ & (19) Table 5 \\
II\,Peg & -- & $1.33$ & $2.47$ & $2.32$ & $0.88$ & $0.34$ & $1.38$ & $0.99$ & $1.22 \pm 0.17$ & (20) Table 4 Q \\
AR\,Lac & -- & $0.93$ & $1.51$ & $1.48$ & $0.94$ & $0.34$ & $1.38$ & $1.00$ & $0.49 \pm 0.25$ & (21) Table 3 \\
$\eta$ Lep & $0.78$ & -- & $1.09$ & $0.55$ & $1.03$ & $0.33$ & $1.35$ & $0.96$ & $-0.30 \pm 0.09$ & (1) \\
$\pi^3$ Ori & $0.79$ & $0.11$ & $0.92$ & $0.46$ & $0.99$ & $0.33$ & $1.30$ & $0.91$ & $-0.31 \pm 0.12$ & (1) \\
$\tau$ Boo\,A & $0.83$ & -- & $0.91$ & $0.50$ & $0.95$ & $0.41$ & $1.24$ & $0.85$ & $-0.27 \pm 0.13$ & (1) \\
Procyon & $1.20$ & $0.68$ & $1.43$ & $0.77$ & $0.98$ & $0.35$ & $1.29$ & $0.80$ & $0.25 \pm 0.15$ & (22) Table 4 \\
Altair & $0.95$ & $0.32$ & $0.96$ & $0.64$ & $1.07$ & $0.30$ & $1.42$ & $1.42$ & $-0.25 \pm 0.36$ & (23) Table 2 MOS/RGS \\
\hline
\end{tabular}
\tablefoot{Coronal abundances are taken in quiescent (Q) or low activity state, where available. The photospheric abundances are predicted with the KNN method using data from \cite{lamost}, and are relative to \cite{asplund2009} solar composition and Ne from \cite{drake_testa_neon}. As there were no Ne abundances in the LAMOST dataset, the photospheric [Ne/Fe] column was calculated as [Ne/Fe]$_\mathrm{phot}=$[O/Fe]$_\mathrm{phot} + \log_{10}{0.41}$. The FIP bias and its uncertainty are calculated as the mean and standard deviation of [X/Fe]$_\mathrm{cor}-$[X/Fe]$_\mathrm{phot}$, where X can be C, N, O and Ne. The FIP bias column is corrected by +0.084 according to \cite{Wood2018}, except for the stars listing (1) as source. As discussed in Sect.~\ref{sect:data}, GJ\,338\,AB is omitted.\\
\textbf{References.} (1) \cite{Wood2018}, (2) \cite{Fuhrmeister2007}, (3) \cite{Liefke2008}, (4) \cite{Robrade2005}, (5) \cite{Lalitha2017}, (6) \cite{Garcia-Alvarez2005}, (7) \cite{telleschi2005}, (8) \cite{Sanz-Forcada2019}, (9) \cite{Peretz2015}, (10) \cite{Osten2003}, (11) \cite{corona_xi_uma}, (12) \cite{audard2003}, (13) \cite{huenemoerder2013}, (14) \cite{garcia-alvarez2006}, (15) \cite{Gondoin2002}, (16) \cite{Audard_etal_2004}, (17) \cite{corona_ei_eri}, (18) \cite{corona_v851_cen}, (19) \cite{corona_ar_psc_ay_cet}, (20) \cite{corona_ii_peg}, (21) \cite{corona_ar_lac}, (22) \cite{Raassen2002}, (23) \cite{Robrade2009}}
\end{table*}

\newpage
\section{Results of the spectral synthesis}

\begin{table*}[th!]
\tiny
\caption{Results from the SME fit of CFHT-ESpaDoNS and TBL-NARVAL spectra.}
\label{table:sme_results}
\centering          
\begin{tabular}{c|c|c|c|c|c|c|c|c|c}
\hline\hline 
star & \teff & $\log{g}$ & [Fe/H] & $v_\mathrm{mic}$ & $v \sin{i}$ & A(C) & A(O)$_\mathrm{NLTE}$ & A(Fe) & FIP bias \\
 & [K] & [cgs] &  & [km s$^{-1}$] & [km s$^{-1}$] &  & & & \\
\hline
70\,Oph A & 5320 $\pm$ 40 & 4.52 $\pm$ 0.06 & -0.02 $\pm$ 0.07 & 1.12 $\pm$ 0.02 & 3.7 $\pm$ 0.05 & 8.49 & 8.90 & 7.42 & -0.49 $\pm$ 0.07$^a$ \\
70\,Oph B & 4400 & 4.47 & -0.02 & 1.85 & 6.4 & 8.79 & 8.81 & 7.60 & 0.12 $\pm$ 0.23$^a$ \\
$\beta$\,Com & 6090 $\pm$ 60 & 4.41 $\pm$ 0.05 & -0.03 $\pm$ 0.05 & 1.28 $\pm$ 0.1 & 5.1 $\pm$ 0.05 & 8.35 & 8.61 & 7.45 & -0.54 $\pm$ 0.30$^b$ \\
EK\,Dra & 5840 $\pm$ 100 & 4.57 $\pm$ 0.20 & -0.01 $\pm$ 0.10 & 1.6 $\pm$ 0.05 & 18.3 $\pm$ 0.1 & 8.50 & 8.87 & 7.53 & -0.13 $\pm$ 0.15$^{a,b}$ \\
$\epsilon$\,Eri & 5050 $\pm$ 10 & 4.48 $\pm$ 0.05 & -0.11 $\pm$ 0.01 & 0.99 $\pm$ 0.01 & 3.7 $\pm$ 0.2 & 8.53 & 8.97 & 7.49 & -0.22 $\pm$ 0.06$^a$ \\
$\eta$\,Lep & 6920 $\pm$ 70 & 4.19 $\pm$ 0.02 & -0.09 $\pm$ 0.05 & 1.83 $\pm$ 0.2 & 19 $\pm$ 0.2 & 8.27 & 9.40 & 7.45 & -0.63 $\pm$ 0.5$^a$ \\
$\kappa$\,Cet & 5740 $\pm$ 20 & 4.48 $\pm$ 0.01 & -0.02 $\pm$ 0.06 & 1.25 $\pm$ 0.03 & 5.7 $\pm$ 0.1 & 8.45 & 8.80 & 7.45 & -0.46 $\pm$ 0.07$^b$ \\
$\chi^1$\,Ori & 6020 $\pm$ 10 & 4.45 $\pm$ 0.03 & -0.06 $\pm$ 0.05 & 1.21 $\pm$ 0.05 & 9.7 $\pm$ 0.1 & 8.35 & 8.62 & 7.47 & -0.38 $\pm$ 0.09$^b$ \\
$\xi$\,Boo A & 5550 $\pm$ 100 & 4.66 $\pm$ 0.20 & -0.19 $\pm$ 0.10 & 1.29 $\pm$ 0.1 & 5.5 $\pm$ 0.2 & 8.50 & 8.90 & 7.48 & -0.29 $\pm$ 0.03$^a$ \\
Sun (Moon) & 5840 $\pm$ 60 & 4.44 $\pm$ 0.10 & -0.06 $\pm$ 0.10 & 1.13 $\pm$ 0.05 & 2.7 $\pm$ 0.1 & 8.44 & 8.65 & 7.44 & -- \\
$\pi^1$\,UMa & 5950 $\pm$ 70 & 4.53 $\pm$ 0.10 & -0.12 $\pm$ 0.10 & 1.4 $\pm$ 0.05 & 10 $\pm$ 0.1 & 8.36 & 8.69 & 7.47 & -0.27 $\pm$ 0.18$^{a,b}$ \\
$\pi^3$\,Ori & 6430 $\pm$ 40 & 4.25 $\pm$ 0.01 & -0.05 $\pm$ 0.01 & 1.37 $\pm$ 0.1 & 17.8 $\pm$ 0.2 & 8.18 & 8.67 & 7.43 & -0.23 $\pm$ 0.19$^a$ \\
$\tau$\,Boo A & 6370 $\pm$ 30 & 4.14 $\pm$ 0.04 & 0.16 $\pm$ 0.10 & 1.34 $\pm$ 0.2 & 16.2 $\pm$ 0.2 & 8.20 & 8.64 & 7.45 & -0.17 $\pm$ 0.20$^a$ \\
$\lambda$\,And & 4510 & 2.57 & -0.56 & 1.26 & 9.2 & 8.42 & 10.19 & 7.33 & -1.02 $\pm$ 0.69$^c$ \\
$\beta$\,Cet & 4720 & 2.65 & -0.15 & 1.58 & 6.3 & 8.22 & 9.16 & 7.35 & -0.79 $\pm$ 0.46$^d$ \\
$\sigma$\,Gem & 4630 & 2.79 & -0.10 & 1.62 & 27.4 & 8.37 & 9.69 & 7.34 &  -0.58 $\pm$ 0.26$^e$ \\
\hline
\end{tabular}
\tablefoot{Error bars (except for the FIP bias) are calculated as the standard deviation of the results from multiple spectra, and are omitted when only a single spectrum was available. This represents the internal precision of the method, not the true accuracy. The oxygen abundances already include the NLTE correction from \citet{NLTE_MPIA}.\\
\textbf{Sources of coronal abundances.} $^a$~\citet{Wood2018}, $^b$~\citet{telleschi2005}, $^c$~\citet{audard2003}, $^d$~\citet{Garcia-Alvarez_2006}, $^e$~\citet{Huenemoerder_etal_2013}
}
\end{table*}

\section{References for the individual stars listed in Table\,\ref{table:sample_params}}\label{A2}

\noindent{\large{\it{M-type flare stars}}}\vspace{-0.5cm}

\begin{quote}
\item{\bf AD\,Leo}
\teff, [Fe/H]: \citet{Rojas-Ayala2012}, $\log g$, $R$: \citet{TICv8}, $P_\mathrm{rot}$: \citet{Morin_etal.2008}, \logrhk: \citet{2017A&A...600A..13A}, $\log L_X$: \citet[][their Table 1]{Wood2018}, age (mean): \citet{Shkolnik2009}

\item{\bf CN\,Leo}
\teff: \citet{2014A&A...563A..35S}, $\log g$, $R$: \citet{TICv8}, [Fe/H]: \citet{Rojas-Ayala2012}, $P_\mathrm{rot}$: \citet{Newton_etal_2018}, $\log L_X$: \citet{2004A&A...417..651S}, age (mean): \citet{2006A&A...447..709P}

\item{\bf EQ\,Peg~A}
\teff: \citet{Yee2017}, $\log g$: calculated from mass,  [Fe/H]: \citet{Yee2017}, $P_\mathrm{rot}$: \citet{Morin_etal.2008}, \logrhk: \citet{Gagneetal.2016}, $R$, $\log L_X$: \citet[][their Table 6]{Wood2018}, mass: 0.36$M_{\odot}$ from the Multiple Star Catalog (\texttt{http://www.ctio.noao.edu/$\sim$atokovin/stars/stars.php}), age: \citet{Zuckermanetal.2013}

\item{\bf EQ\,Peg~B}
\teff: \citet{Houdebine2019}, $\log g$: calculated from mass, [Fe/H] (as EQ\,Peg~A): \citet{Yee2017}, $P_\mathrm{rot}$: \citet{Morin_etal.2008}, $R$, $\log L_X$: \citet[][their Table 1]{Wood2018}, mass: 0.19$M_{\odot}$ from the Multiple Star Catalog (\texttt{http://www.ctio.noao.edu/$\sim$atokovin/stars/stars.php}), age: \citet{Zuckermanetal.2013}

\item{\bf Proxima\,Cen}
\teff, $\log g$: \citet{Zhaoetal.2018}, [Fe/H] (as $\alpha$\,Cen~A,~B): \citet{Zhaoetal.2018}, $P_\mathrm{rot}$: \citet{Klein_etal2020}, $R$: \citet{Zhaoetal.2018},  \logrhk: \citet{2017A&A...600A..13A}, $\log L_X$: \citet[][their Table 6]{Wood2018}, age (as $\alpha$\,Cen~A,~B): \citet{Zhaoetal.2018}

\item{\bf EV\,Lac}
\teff, $\log g$, [Fe/H]: \citet{2019A&A...625A..68S}, $P_\mathrm{rot}$: \citet{1995A&A...300..819C}, $R$: \citet{2019A&A...625A..68S}, \logrhk: \citet{Gagneetal.2016}, $\log L_X$: \citet[][their Table 6]{Wood2018}, age (mean): \citet{Shkolnik2009}

\item{\bf YY\,Gem}
\teff, $\log g$ (calculated from mass), $R$, mass: \citet{MacDonald_Mullan2017}, $P_\mathrm{rot}$: \citet{2019ApJ...873...69K}, [Fe/H], $\log L_X$, age:  \citet{MacDonald_Mullan2014}

\item{\bf AU\,Mic}
\teff: \citet{Houdebine2019}, $\log g$: \citet{2020A&A...642A.115C}, [Fe/H]: \citet{Houdebine2009}, $P_\mathrm{rot}$: \citet{2017A&A...600A..83M},
\logrhk: \citet{IbanezBustosetal.2019}, $R$, $\log L_X$: \citet[][their Table 1]{Wood2018}, age: \citet{Gagne2018(BANYANXI.)}, 

\item{\bf AT\,Mic}
\teff: \citet{BANYANIV.}, $\log g$: calculated from mass,  [Fe/H]: \citet{Houdebine2009}, $P_\mathrm{rot}$, $R$: \citet{2017A&A...600A..83M}, 
$\log L_X$: \citet{2015A&A...581A..28P}, mass: 0.61$M_{\odot}$ from the Multiple Star Catalog (\texttt{http://www.ctio.noao.edu/$\sim$atokovin/stars/stars.php}), age: \citet{Gagne2018(BANYANXI.)}, 

\end{quote}

\noindent{\large{\it{Fast rotating K dwarfs with flares}}}\vspace{-0.5cm}

\begin{quote}

\item{\bf AB\,Dor}
\teff: \citet{Close_etal_2007}, $\log g$: calculated from mass, [Fe/H]: \citet{1997A&AS..124..299C}, $P_\mathrm{rot}$: \citet{2009A&ARv..17..251S}, $R$: \citet{2011A&A...533A.106G},
\logrhk: \citet{2018A&A...616A.108B},  $\log L_X$: \citet[][their Table 1]{Wood2018},
mass: 0.86$M_{\odot}$ from \citet{2011A&A...533A.106G}, age: \citet{Gagne2018(BANYANXI.)}, 

\item{\bf LO\,Peg}
\teff, $\log g$: \citet{Folsom2016}, [Fe/H]: \citet{1994MNRAS.270..153J}, $P_\mathrm{rot}$: \citet{Dal&Tas2003}, $R$: \citet{Folsom2016},
\logrhk: \citet{Plavchan2009},  $\log L_X$: \citet{Lalitha2017}, age: \citet{Gagne2018(BANYANXI.)}, 

\item{\bf V471\,Tau}
\teff, $\log g$, [Fe/H], $P_\mathrm{rot}$, $R$, $\log L_X$, $\log F_X$: \citet{2021arXiv210302041K}, age: \citet{Gagne2018(BANYANXI.)} 

\end{quote}

\noindent{\large{\it{K dwarfs}}}\vspace{-0.5cm}
\begin{quote}

\item{\bf GJ\,338~A}
\teff, $\log g$ [Fe/H], $R$: \citet{2019A&A...625A..68S},  $P_\mathrm{rot}$: \citet{2020A&A...637A..93G}, 
\logrhk: \citet{Gagneetal.2016}, $\log L_X$: \citet[][their Table 6]{Wood2018}, age (upper limit, $(B-V)_{\rm SIMBAD}$=1$\fm$41): \citet{Barnes2007}, 

\item{\bf GJ\,338~B}
\teff, $\log g$ [Fe/H], $R$: \citet{2019A&A...625A..68S},  $P_\mathrm{rot}$:  \citet{2020A&A...637A..93G}
\logrhk: \citet{Gagneetal.2016}, $\log L_X$: \citet[][their Table 6]{Wood2018}, age (upper limit, $(B-V)_{\rm SIMBAD}$=1$\fm$42): \citet{Barnes2007}, 

\item{\bf 36\,Oph~A}
\teff, $\log g$: \citet{Luck2017}, [Fe/H]: \citet{1989A&A...225..369C}, $P_\mathrm{rot}$, \logrhk, age: \citet{Barnes2007}, $R$, $\log L_X$: \citet[][their Table 1]{Wood2018}

\item{\bf 36\,Oph~B}
\teff, $\log g$: \citet{Luck2017}, [Fe/H]: \citet{1989A&A...225..369C}, $P_\mathrm{rot}$, \logrhk, age: \citet{Barnes2007}, $R$, $\log L_X$: \citet[][their Table 1]{Wood2018}

\item{\bf $\boldsymbol\xi$\,Boo~B}
\teff, $\log g$, [Fe/H]: \citet{Aleo2017}, $P_\mathrm{rot}$, \logrhk, age: \citet{Barnes2007}, $R$, $\log L_X$: \citet[][their Table 1]{Wood2018}

\item{\bf 61\,Cyg~A}
\teff, $\log g$: \citet{2015A&A...582A..49H}, [Fe/H]: \citet{2005AJ....129.1063L}, $P_\mathrm{rot}$, \logrhk, age: \citet{Barnes2007}, $R$, $\log L_X$: \citet[][their Table 1]{Wood2018}
 
\item{\bf 61\,Cyg~B}
\teff, $\log g$: \citet{2015A&A...582A..49H}, [Fe/H]: \citet{2005AJ....129.1063L}, $P_\mathrm{rot}$, \logrhk, age: \citet{Barnes2007}, $R$, $\log L_X$: \citet[][their Table 1]{Wood2018}

\item{\bf 70\,Oph~A}
\teff, $\log g$, [Fe/H]: from this paper,  $P_\mathrm{rot}$, \logrhk, age: \citet{Barnes2007}, $R$, $\log L_X$: \citet[][their Table 1]{Wood2018}

\item{\bf 70\,Oph~B}
\teff, $\log g$, [Fe/H]: from this paper, \logrhk (as 70\,Oph~A), age: \citet{Barnes2007}, $R$, $\log L_X$: \citet[][their Table 1]{Wood2018}

\item{\bf ${\boldsymbol\epsilon}$\,Eri}
\teff, $\log g$, [Fe/H]: from this paper,  $P_\mathrm{rot}$, \logrhk, age: \citet{Barnes2007}, $R$, $\log L_X$: \citet[][their Table 1]{Wood2018}

\end{quote}

\noindent{\large{\it{Solar type stars}}}\vspace{-0.5cm}
\begin{quote}

\item{\bf Sun} 
        
\item{\bf ${\boldsymbol\alpha}$\,Cen~A}
\teff, $\log g$, [Fe/H]: \citet{2018A&A...615A.172M}, $P_\mathrm{rot}$: \citet{Angus_etal_2015}, \logrhk: \citet{Mamajek2008}, $R$, $\log L_X$: \citet[][their Table 1]{Wood2018}, age (mean): \citet{Zhaoetal.2018}

\item{\bf $\boldsymbol\alpha$\,Cen~B}
\teff, $\log g$, [Fe/H]: \citet{2018A&A...615A.172M}, $P_\mathrm{rot}$: \citet{Angus_etal_2015}, \logrhk: \citet{Mamajek2008}, $R$, $\log L_X$: \citet[][their Table 1]{Wood2018}, age (mean): \citet{Zhaoetal.2018}

\item{\bf $\boldsymbol\pi^1$\,UMa}
\teff, $\log g$, [Fe/H]: from this paper, $P_\mathrm{rot}$, \logrhk, age: \citet{Barnes2007}, $R$, $\log L_X$: \citet[][their Table 1]{Wood2018}

\item{\bf EK\,Dra}
\teff, $\log g$, [Fe/H]: from this paper, $P_\mathrm{rot}$: \citet{2002A&A...393..225M},
\logrhk: \citet{Barnes2007}, $R$, $\log L_X$: \citet[][their Table 1]{Wood2018}, age: \citet{Gagne2018(BANYANXI.)}, 

\item{\bf $\boldsymbol\xi$\,Boo~A}
\teff, $\log g$, [Fe/H]: from this paper, $P_\mathrm{rot}$, \logrhk, age: \citet{Barnes2007}, $R$, $\log L_X$: \citet[][their Table 1]{Wood2018}

\item{\bf $\boldsymbol\chi^1$\,Ori}
\teff, $\log g$, [Fe/H]: from this paper, $P_\mathrm{rot}$, \logrhk, age: \citet{Barnes2007}, $R$, $\log L_X$: \citet[][their Table 6]{Wood2018}

\item{\bf $\boldsymbol\kappa$\,Cet}
\teff, $\log g$, [Fe/H]: from this paper, $P_\mathrm{rot}$, \logrhk, age: \citet{Barnes2007}, $R$, $\log L_X$: \citet[][their Table 6]{Wood2018}

\item{\bf $\boldsymbol\beta$\,Com}
\teff, $\log g$, [Fe/H]: from this paper, $P_\mathrm{rot}$, \logrhk, age: \citet{Barnes2007}, $R$, $\log L_X$: \citet[][their Table 6]{Wood2018}

\item{\bf 47\,Cas~B}
\teff: \citet{2004A&A...427..667N}, [Fe/H]: \citet{2011A&A...530A.138C}, $P_\mathrm{rot}$, $\log L_X$, age: \citet{telleschi2005}, $R$: \citet{2015A&A...578A.129J}

\item{\bf $\boldsymbol\iota$\,Hor}
\teff, $\log g$, [Fe/H], $R$: \citet{Fuhrmann2017}, $P_\mathrm{rot}$: \citet{Metcalfe2010},
\logrhk: \citet{2018A&A...616A.108B}, $\log L_X$: \citet{Peretz2015}, age: \citet{Gagne2018(BANYANXI.)}, 

\item{\bf 11\,LMi}
\teff: \citet{Yee2017}, $\log g$, [Fe/H], $P_\mathrm{rot}$, \logrhk: \citet{2018A&A...619A...6O}, $R$: \citet{Boyajian2013}, $\log L_X$: \citet{Peretz2015}, age: \citet{2016A&A...593A..51M}

\item{\bf HR\,7291}
\teff: \citet[][their Table 4]{2019A&A...624A..10G}, $\log g$, [Fe/H], $P_\mathrm{rot}$, $R$: \citet{Fares2013}, age: \citet{Takeda2007}, \logrhk: \citet{Wright2004}, $\log L_X$: \citet{Peretz2015}

\item{\bf $\boldsymbol\sigma^2$\,CrB}
\teff, $R$, [Fe/H], age: \citet{Raghavan_etal_2009}, $\log g$: \citet{Luck2017}, $P_\mathrm{rot}$: \citet{2003A&A...399..315S}, \logrhk: \citet{Gray2003}, $\log L_X$: \citet{2015A&A...581A..28P}

\item{\bf $\boldsymbol\xi$\,UMa~B}
\teff, $\log g$, [Fe/H]: \citet{Zhao_etal2002}, $P_\mathrm{rot}$ (=$P_\mathrm{orb}$), $R$: \citet{2003ApJS..145..147S}, $\log L_X$: \citet{DL1993}

\end{quote}

\noindent{\large{\it{Evolved stars}}}\vspace{-0.5cm}

\begin{quote}

\item{\bf HR\,1099}
\teff, $\log g$, \logrhk: \citet{2020A&A...644A..67S}, [Fe/H]: \citet{Luck2017}, $P_\mathrm{rot}$: \citet{2006A&A...455..595L}, $R$: \citet{Fekel_1983}, $\log L_X$: \citet{Huenemoerder_etal_2013}

\item{\bf UX\,Ari}
\teff, $\log g$, [Fe/H], $P_\mathrm{rot}$, $R$, age: \citet{Hummel_etal_2017}, $\log L_X$: \citet{Makarov_2003}

\item{\bf $\boldsymbol\lambda$\,And}
\teff, $\log g$, [Fe/H]: from this paper, $P_\mathrm{rot}$: \citet{1993A&AS..100..173S}, $R$, age: \citet{2016A&A...593A..51M}, \logrhk: \citet{Gray2003}, $\log L_X$: \citet{1992ApJS...82..311D}

\item{\bf VY\,Ari}
\teff, $\log g$, [Fe/H]: \citet{1998A&A...338..661O}, $P_\mathrm{rot}$: \citet{1997A&AS..125...11S}, $R$: \citet{2018A&A...616A...1G}, $\log L_X$: \citet{Makarov_2003}

\item{\bf Capella}
\teff, $\log g$, [Fe/H], $P_\mathrm{rot}$, $R$, age: \citet{Torres_etal_2015}, $\log L_X$: \citet{Makarov_2003}

\item{\bf $\boldsymbol\sigma$\,Gem}
\teff, $\log g$, [Fe/H]: from this paper, $P_\mathrm{rot}$, $R$, age: \citet{Roettenbacher_etal_2015}, $\log L_X$: \citet{Huenemoerder_etal_2013}

\item{\bf 31\,Com}
\teff, $\log g$, [Fe/H], $P_\mathrm{rot}$, $R$, age: \citet{2010A&A...520A..52S}, $\log L_X$: \citet{Garcia-Alvarez_2006}

\item{\bf $\boldsymbol\mu$\,Vel}
\teff: \citet{Ayres_etal_2007}, $\log g$: calculated from mass, $R$: \citet{Mullan_etal_2006}, $\log L_X$: \citet{Makarov_2003}, mass: 3.3$M_{\odot}$ from \citet{2003A&A...409..251M}

\item{\bf $\boldsymbol\beta$\,Cet}
\teff, $\log g$, [Fe/H]: from this paper, $P_\mathrm{rot}$: \citet{2013A&A...556A..43T}, $R$: \citet{2011A&A...535A..59B}, \logrhk: \citet{Rich_etal_2017}, $\log L_X$: \citet{Makarov_2003}, age: \citet{2018A&A...616A..33S}

\item{\bf FK\,Com}
\teff, $\log g$, $P_\mathrm{rot}$: \citet{2004AN....325..402K}, [Fe/H]: \citet{2019A&A...628A..94A}, $R$: \citet{2018A&A...616A...1G}, $\log L_X$: \citet{2016ApJS..223....5A}

\item{\bf YY\,Men}
\teff, $\log g$: \citet{1993A&A...273..194R}, [Fe/H], $P_\mathrm{rot}$, $R$, $\log L_X$: \citet{Audard_etal_2004}

\item{\bf EI\,Eri}
\teff, $\log g$, $R$, $P_\mathrm{rot}$, age: \citet{Washuettletal.2009}, [Fe/H]: \citet{1998A&A...338..661O}, $\log L_X$: \citet{Ostenetal.2002}

\item{\bf V851\,Cen}
\teff, $\log g$, [Fe/H]: \citet{2003A&A...397..747K}, $P_\mathrm{rot}$: \citet{Kiraga2012(ASAS)}, $R$: \citet{Stawikowski-Glebocki1994a}, $\log L_X$: \citet{DL1993}

\item{\bf AR\,Psc}
\teff, $\log g$, [Fe/H]: \citet{Shan_etal2006}, $P_\mathrm{rot}$: \citet{Matranga_etal2010}, $R$: \citet{Stawikowski-Glebocki1994b}, $\log L_X$: \citet{DL1993}, age: \citet{Fekel1996}

\item{\bf AY\,Cet}
\teff, $\log g$, [Fe/H], age: \citet{2018A&A...615A..31D}, $P_\mathrm{rot}$: \citet{Strassmeier_etal1990}, $R$: \citet{Stawikowski-Glebocki1994b}, $\log L_X$: \citet{2003ApJS..145..147S}

\item{\bf II\,Peg}
\teff, $\log g$, [Fe/H], $R$: \citet{1998A&A...334..863B}, $P_\mathrm{rot}$: \citet{Strassmeier_etal1990}, $\log L_X$: \citet{DL1993}

\item{\bf AR\,Lac}
\teff, $\log g$, $P_\mathrm{rot}$ (=$P_\mathrm{orb}$): \citet{1998A&A...332..541L}, [Fe/H]: \citet{1992A&AS...95..273C}, $R$: \citet{Zboril_etal2005}, $\log L_X$: \citet{DL1993}

\end{quote}

\noindent{\large{\it{F stars, hotter than 6300\,K}}}\vspace{-0.5cm}

\begin{quote}

\item{\bf $\boldsymbol\eta$\,Lep}
\teff, $\log g$, [Fe/H]: from this paper, \logrhk: \citet{2018A&A...616A.108B}, $R$, $\log L_X$: \citet[][their Table 1]{Wood2018}, age: \citet{2011A&A...530A.138C}

\item{\bf $\boldsymbol\pi^3$\,Ori}
\teff, $\log g$, [Fe/H]: from this paper, 
\logrhk: \citet{Wright2004}, $R$, $\log L_X$: \citet[][their Table 1]{Wood2018}, age: \citet{Wright2004}

\item{\bf $\boldsymbol\tau$\,Boo~A}
\teff, $\log g$, [Fe/H]: from this paper, $P_\mathrm{rot}$: \citet{Fares_etal2009}, \logrhk, age: \citet{Mamajek2008}, $R$, $\log L_X$: \citet[][their Table 1]{Wood2018}

\item{\bf Procyon}
\teff, $\log g$, [Fe/H]: \citet{2004A&A...413..251K}, $P_\mathrm{rot}$: \citet{Ayres1991}, $R$: \citet{Aufdenberg_2005}, $\log L_X$: \citet{2002A&A...394..911N}, age: \citet{Liebert_etal_2013}

\item{\bf Altair}
\teff (mean), [Fe/H], $R$ (mean): \citet{Monnier_etal2007}, $\log g$: \citet{1990A&AS...85.1015M}, $P_\mathrm{rot}$: \citet{Peterson_etal2006}, $\log L_X$, age: \citet{Robrade2009}

\end{quote}

\end{appendix}

\end{document}